\documentclass[preprint]{aastex}

\usepackage{rotating}

\slugcomment{The Astrophysical Journal, in press}

\shorttitle{Dust in Galactic planetary nebulae}
\shortauthors{Stanghellini et~al. }

\begin{document}

\title{
The nature of dust in compact Galactic planetary nebulae from {\it Spitzer} spectra\footnote{based on observations made with the {\it Spitzer Space Telescope},
which is operated by the Jet Propulsion Laboratory, California Institute of Technology, under a contract with NASA.}.}

\author{Letizia Stanghellini}
\affil{National Optical Astronomy Observatory, Tucson, AZ 85719; lstanghellini@noao.edu}
\author{D. A. Garc\'{\i}a-Hern\'andez}
\affil{Instituto de Astrof\'{\i}sica de Canarias, V\'{\i}a L\'actea s/n, La Laguna, E-38200 Tenerife, 
Spain; affiliated to Departamento de Astrof\'{\i}sica, Universidad de La Laguna; agarcia@iac.es}
\author{Pedro Garc\'\i a-Lario}
\affil{Herschel Science Centre, European Space Astronomy Centre, Research and Scientific Support Department of ESA, 
Villafranca del Castillo, P.O. Box 50727. E-28080 Madrid, Spain; Pedro.Garcia-Lario@sciops.esa.int}
\author{James E. Davies}
\affil{\it{Spitzer} Science Center, Infrared Processing and Analysis Center, California Institute of Technology, 1200 East California Boulevard, Pasadena, CA 91125, USA; jdavies@ipac.caltech.edu}
\author{Richard A.~Shaw}
\affil{National Optical Astronomy Observatory, 950 N. Cherry Av.,
Tucson, AZ  85719; shaw@noao.edu}
\author{Eva Villaver}
\affil{Departamento de F\'{\i}sica Te\'orica C-XI, Universidad Aut\'onoma de Madrid, E-28049 Madrid, Spain; eva.villaver@uam.es}
\author{Arturo Manchado}
\affil{Instituto de Astrof\'isica de Canarias, V\'{\i}a L\'actea s/n, La Laguna, E-38200 Tenerife, 
Spain; affiliated to affiliated to Departamento de Astro\'{\i}sica, Universidad
de La Laguna and CSIC, Spain; amt@iac.es}
\author{Jose V.~Perea-Calder\'on}
\affil{European Space Astronomy Centre, INSA S.~A., P.O. Box 50727. E-28080 Madrid, Spain; Jose.Perea@sciops.esa.int}

\begin{abstract}
We present the Spitzer/IRS spectra of 157 compact Galactic planetary nebulae (PNe).
These young PNe provide insight on the effects of dust in early post-AGB evolution, before much of
the dust is altered or destroyed by the hardening stellar radiation field. Most of the selected targets have
PN-type IRS spectra, while a few turned out to be misclassified stars. We inspected the
group properties of the PN spectra and classified them based on the different dust classes
(featureless, or F; carbon-rich dust, or CRD; oxygen-rich dust, or ORD; mixed-chemistry dust, or MCD) and
subclasses (aromatic and aliphatic; crystalline and amorphous). All PNe are characterized by
dust continuum and more than $80\%$ of the sample shows solid state features above the
continuum, in contrast with the Magellanic Cloud sample where only $\sim40\%$ of the entire
sample displays solid state features; this is an indication of the strong link between
dust properties and metallicity. The Galactic PNe that show solid state features are almost
equally divided among the CRD, ORD, and MCD. We analyzed  dust properties together with
other PN properties and found that (i) there is an enhancement of MCD PNe toward the Galactic center, in
agreement with studies of Galactic bulge PNe; (ii) CRD PNe could be seen as
defining an evolutionary sequence, contrary to the ORD and MCD PNe, which are scattered in all evolutionary diagrams;
(iii) carbon-rich and oxygen-rich grains retain different equilibrium temperatures, as expected from models; 
(iv) ORD PNe are highly asymmetric, i.~e. bipolar or bipolar-core, and
CRD PNe highly symmetric, i.e., round or elliptical; point-symmetry is statistically more common in MCD than
in other dust class PNe. By comparing the sample of this paper to that of Magellanic Cloud
PNe we find that the latter sample does not include MCD PNe, and the other dust classes are differently
populated, with continuity of the fraction of F, CRD, ORD, and MCD population from high to low metallicity environments. 
We also find  similar sequences for CRD PNe in the Galactic disk and the Magellanic Clouds, except that the Magellanic
Cloud PNe seem to attain higher dust temperatures at similar evolutionary stages, in agreement with the observational findings of 
smaller dust grains (i.e, lower radiation efficiency) in low metallicity interstellar environments.
\end{abstract}


\section{Introduction}

Planetary nebulae (PNe) are the gas and dust envelopes ejected toward the end of the
evolution of low- and intermediate-mass stars (LIMS, 1-8 M$_{\odot}$), at the tip of the 
thermally-pulsing asymptotic giant  branch (TP-AGB; e.~g., Herwig 2005) phase. The
stellar ejecta carries both the products of nucleosynthesis, such as carbon and nitrogen, 
and the $\alpha$-elements, such as oxygen, neon, argon, and sulfur,
whose net yields tend to be near zero in this mass range. 

Dust may be fundamental to PN formation, since radiation pressure on the dust grains 
formed at the surface of AGB stars may be important in triggering the envelope
ejection. Planetary nebulae are ideal dust probes during their early evolutionary stages,
since sputtering of the dust grains is expected to  affect their dusty nature in their
lifetime. Dust particles in PNe are typically cool (50-150 K) and radiate in the mid infrared, 
producing a near-blackbody continuum spectrum (Cohen \& Barlow 1974) that
contributes to $\sim$ 40$\%$ of  the total emergent flux (Zhang \& Kwok 1991), and that
peaks between 25 and 60 $\mu$m. Spitzer IRS spectroscopy has been very successful in 
tracing this dust component in PNe, as well as in a variety of other stellar targets including
the AGB stars.  Previous systematic PN studies based on IRS spectra include Galactic bulge
(GO program 3633, Bobrowsky), halo  (GO program 20049, Kwitter), disk (GTO program
40035, Bernard-Salas), and  Magellanic Cloud (GTO program 103, Houck, and GO program 20443, Stanghellini) PNe.  These
data collectively show PN spectra with thermal dust  continua, nebular emission lines, and a
variety of dust signatures, with characteristics of carbon-rich and oxygen-rich  compounds
and different types of grains. From the comparative study of the different samples some
trends have emerged: it appears that oxygen dust features, such as crystalline silicates,
are more common in the bulge population  (Gutenkunst et al. 2008; Perea-Calder\'on et al.
2009) than in the Galactic or Magellanic Cloud PNe (Stanghellini et al. 2007,
hereafter S07; Bernard-Salas et al. 2009). In particular, in the Magellanic Cloud PNe the
oxygen and carbon dust features are never observed in the same spectrum. It is apparent that
the galaxy metallicity has a strong effect on the nature of dust (S07; Bernard-Salas et al. 2009), 
and, from the Magellanic Cloud sample, it is apparent that planetary nebula bipolarity correlates strongly with oxygen-rich dust
PNe (S07). Finally, gas and dust chemistry in Magellanic Cloud PNe are strongly correlated
(S07). 

What was lacking from the Spitzer database was a large, homogeneous sample of Galactic PN spectra,
especially for young,  unevolved PNe, to study the early dust features and their evolution
across the Galactic disk. In this paper we present the study of 157 Galactic angularly small targets classified as 
PNe, observed during the last cryogenic cycle of {\it Spitzer}. We collected the present sample to
gain insight on the impact of dust and metallicity   in stellar evolution, a fundamental
question in astrophysics. The extensive target list is essential to explore the different
dust types with statistical significance, and  to study them across the Galactic disk, and
in relation to the Galactic populations. The large sample presented here fills an important
gap in the Spitzer program. Our compact Galactic PN dataset
is also complemented by a WFC3/{\it HST} imaging survey of about one-third of the targets
presented here. The morphological analysis, and that of the central stars (CS) from the WFC3
data will complement the dust analysis presented here, allowing a much more detailed view on
how dust affects the post-AGB and PN evolution in our galaxy compared to other environments.

Systematic comparison of IRS spectra of PNe belonging to different populations such as the
Galactic disk, halo, bulge, and the satellite galaxies, can address many key questions that
have opened up in recent years regarding AGB stars as well. First, the absence of heavily
obscured AGB stars in the Magellanic Clouds (Trams et al. 1999; Groenewegen et al. 2000; 
Garc\'{\i}a-Hern\'andez et al. 2009), in contrast with their Galactic counterparts
(e.~g., Garc\'{\i}a-Hern\'andez et al. 2006, 2007a), seems to indicate that, on average, lower
metallicity environments such as those of the Magellanic Clouds are less favorable to dust
production. The relative number of C-rich versus O-rich AGB stars in galaxies of the local
group increases with decreasing metallicity (Cioni \& Habing 2003; Cioni et al. 2003; Schultheis et al.
2004), which shows that mass-loss efficiency depends on metallicity (H{\"o}fner 2011).

In this paper we present the 5--40 $\mu$m IRS spectra from Spitzer program GO
50261. In $\S$2 we describe the target selection, exposures, and  observing
strategy. The data  analysis, including spectral extraction, classification of
the dust types, and continuum fitting, are presented in $\S$3.  In $\S$4 we
explore the relation of the dust characteristics with respect to other physical
properties of the PNe in our sample, and in $\S$5 we extend the comparison to
PNe in the Magellanic Clouds.  The summary and conclusions are in $\S$6.

\begin{figure}
\plotone{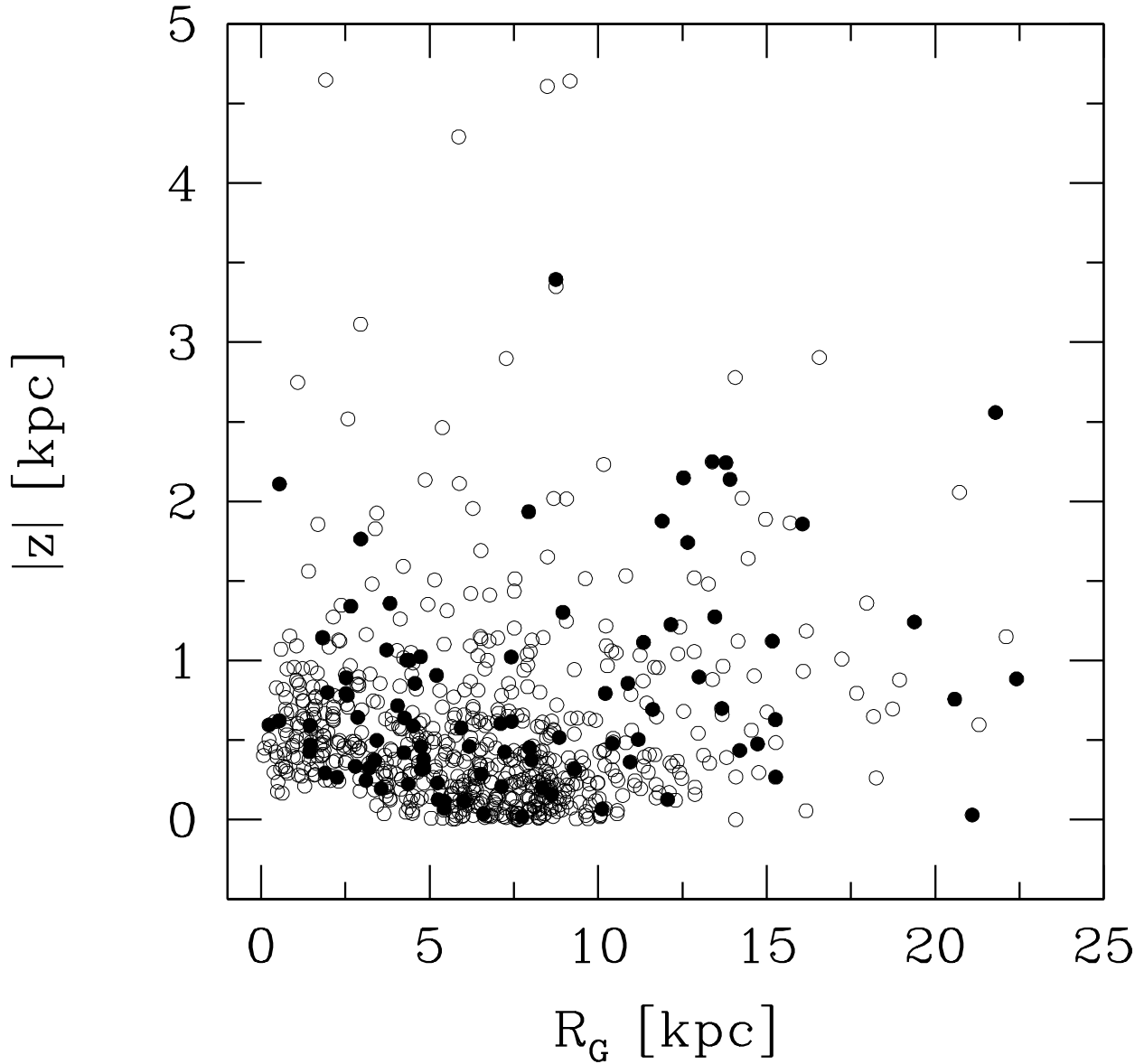}
\caption{The spatial distribution of our targets (filled circles) compared to the general PN population in the Galaxy (open
circles). This plot is limited to those PNe whose statistical distance could be calculated, thus it represents subsamples of both our
target sample and the general Galactic PN population.}
\label{Fig. 1}
\end{figure}

\section{Observations}

The aim of the target selection is to cover a homogenous and as much as possible complete set
of compact Galactic PNe, most of which should be young. To this end, we selected PNe smaller than 4\arcsec~ in
apparent size. Galactic PNe with diameters $<$ 4\arcsec~ should be younger than $\sim 2.0\times 10^3$
yr (Villaver et al. 2002a, b) if they have heliocentric distances smaller than 6 kpc and if their expansion velocities are 
typical (20-40 km s$^{-1}$).
Statistical distances (Stanghellini et al. 2008) help us  distinguish which PNe are nearby and young from those that are 
distant and evolved. The population of spectroscopically confirmed Galactic PNe is
listed either in the Strasbourg--ESO catalog of  Galactic PNe (Acker et al.~1992) or in the
MASH survey (Parker et al.~2006). Of the 1143 PN in the Strasbourg-ESO catalog, 143 are point
sources, and 86 have $\theta<4$\arcsec. The MASH survey gives another 2 PN with
$\theta<4$\arcsec ~(but no point sources).  From the ~230 PNe thus selected we explicitly
eliminated the PNe already observed with Spitzer/IRS.  Most of the exclusions are Galactic
bulge and halo PNe, only 10 compact PNe in the Galactic disk had been observed with {\it Spitzer} before with
similar observing configuration. Finally, we also exclude a dozen targets previously observed with IRAS or
MSX.  This filtering yielded a final sample of 157 compact Galactic PNe, representing for the
most part the disk population. 

Twenty-six PNe in our sample may actually be bulge PNe. The fact that a PN belongs to the bulge rather than the disk is
controversial, if nothing else for the reason that the Galactic PN distances are not known
with high precision, thus finding a PN in the direction of the bulge might not be sufficient
to exclude it form the disk population. Stanghellini \& Haywood (2010) made this selection
for all Galactic PNe based on the best distance scale available, and included a brightness criterion, but uncertainties in the
separation of the bulge to disk PN samples are of course possible. Furthermore, many of the PNe of our
sample do not have a distance determined at all, given the lack of measurable angular
diameter.  Knowing the radial velocities of the targets could help, unfortunately at this time velocities are available only for a minority
of the sample, and none for  the PNe whose distances are not known.

Three PNe in our sample could belong to the Galactic halo. In
Table 1 we list the observed targets with their IAU designation (column 1), their common name
(column 2), and their equatorial coordinates (equinox 2000, columns 3 and 4). In
columns 5 through 7 of Table 1 we list the IRS campaign number, the observing mode, and the exposure time respectively. 
In the observing mode column,  "0" is for the combination of SL (short-low, 5-14 $\mu$m) and LL (long-low, 14-40 $\mu$m) modules, and "1" for 
the combination of the SL, SH (short-high, 10-20 $\mu$m) and LH (long-high, 20-40 $\mu$m). Nearly all targets were observed with the SL module 
and "0" and "1" denote respectively whether the longer wavelength part of the spectra where observed at low or high resolution.

\begin{figure}
\plotone{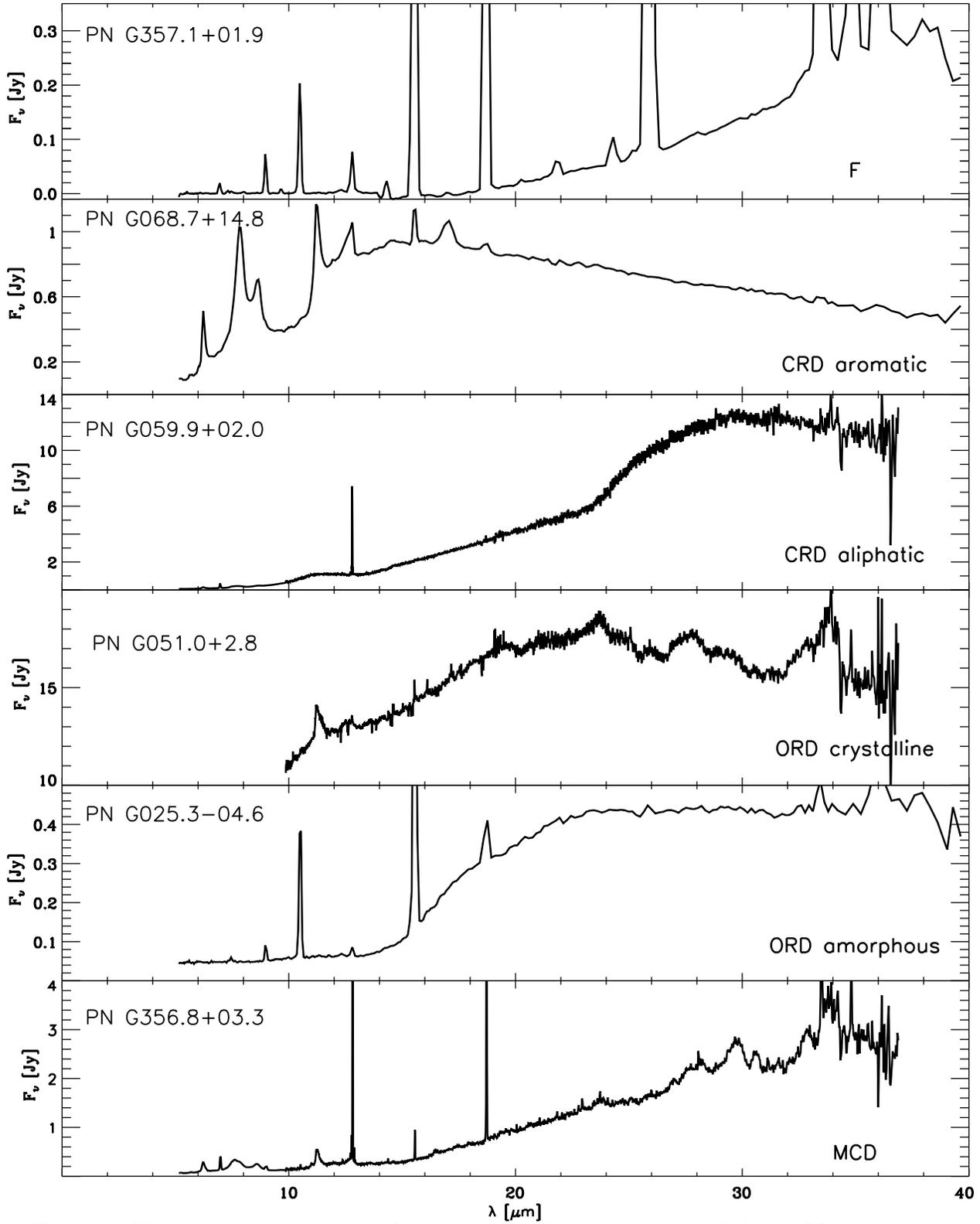}
\caption{The spectral templates of dust types found in our compact Galactic PN sample.}
\label{Fig.2}
\end{figure}

In Figure 1 we show the space distribution of our targets whose distance is known (92 PNe)
compared with the general PN population (Stanghellini \& Haywood 2010), where their distance
from the Galactic plane is plotted against their radial distance from the Galactic center,
and where we assume that the Galactic center is 8 kpc from the sun. It is  clear from this
plot that the selected sample well represents the Galactic disk population of PNe. 
Most of the PNe in this study are actually young, being selected to be compact in apparent size. 

We acquired, reduced, and analyzed the Galactic PN IRS/Spitzer spectra in a similar manner to that 
for Magellanic Cloud PNe (S07).  Our AOR were built with the following requirements: (1) we aimed at
obtaining  5--40 $\mu$m spectra, with different modes depending on the estimated PN
brightness; (2) we estimated the fluxes of our targets  based on the upper limit provided by
IRAS (Zhang \& Kwok 1991). We chose the observing mode, computed the exposure time
and the number of cycles needed for each mode using Spitzer IRS PET, the
saturation levels for all modes, and a conservative choice for the fluxes. For the PN with
very low fluxes, based on the IRAS upper limits and our experience with IRS spectra of
Magellanic Cloud PN, we observed in SL and LL modes, with 3 integrations of respectively 14
and 30 seconds each. Peak up observations were not necessary, since we derive all coordinates
form the astrometric catalog by Kerber et al. (2003), where the precision is better than
0.35$\arcsec$ . 

\section{Data analysis: calibration, spectral extraction, and dust classification.}

Raw data were retrieved from the Spitzer archive together with the latest pipeline calibration
files. Initial file preparation included removing  {\it Bowing} of spectral orders  in the LH
data using DARK SETTLE; cleaning rogue pixels  using IRSCLEAN MASK, where we used custom
masks for rogue pixels cleaning for all our data;  and subtracting the two nod positions for the SL and LL data 
to obtain sky-subtracted spectra. All low- resolution spectral images were checked free from background source contamination.
We did not subtract the sky in the high resolution data, but the sky
background is very low for these bright sources. We then extracted the 1D spectra from the co-added 2D spectral 
images using SPICE version 2.2..

The spectral merging, averaging, and continuum fitting was done with the package SMART (Higdon et al. 2004). The
final spectrum for each dataset  includes merging, cleaning and subtracting any spurious
jump, and averaging the nod positions and orders.  We subtracted a spurious
pedestal above 14 $\mu$m from the spectrum of PN~G012.5-09.8, so both final reduced spectrum and the
fitted black body curve have been modified accordingly.  We also subtracted a pedestal along the whole spectrum of PN~G041.8+04.4 
and PN~G044.1+05.8 to bring it down to a level of 0.0 Jy at 5 $\mu$m.  

Three distinct components are apparent in the spectral energy distributions: the dust continuum, the nebular emission in the form of collisionally excited lines of atomic gas and, in the majority of the targets, solid-state dust emission features. An analysis of the abundances in the nebular atomic gas will be presented in a future paper; for now we focus on the spectral signatures of the dust. We fit the dust continuum in each target with the black-body 
fitting routine described in the package SMART/IDEA in order to determine a characteristic dust temperature and IR luminosity. We performed a continuum fit on each spectrum after masking the nebular and solid state emission features and the low-signal region  long-ward of $\sim37 \mu$m where the system sensitivity drops rapidly. The temperature and luminosity were varied to minimize the RMS deviation, and the IR luminosity was derived from the integral of the Planck function at that temperature over all wavelengths. In most cases the continuum was well sampled over the bulk of the observed spectrum, through the whole SED, and we constrained the fitted function to pass through or below the average continuum. 

The form of the fitting included an emissivity term, $\tau \propto \lambda ^{-\alpha}$, ($\alpha$ is given in column (5) of table 3 for the converging fits) and the fitting routine finds a solution for both alpha and the proportionality constant (see also Hony et al. 2002). 
The black body fits were robustly obtained for about half of the PN sample, and in these cases the formal uncertainties of the fit to temperature and IR luminosity were small, owing to the very slow change in the Planck function with wavelength.  For cases where the fits did not converge with the proposed method (flagged as {\it N} in Table 3, column 4), we did not use the derived dust temperatures in the tables or Figures of this paper. We plan to use more sophisticated fitting models in a future analysis. For the PNe whose black body fit converge, we have validated them further by comparing the fits against the the 65 $\mu$m fluxes from the Akari data archive, and the 60 $\mu$m fluxes
 from IRAS archive\footnote{In Table 3, column (6) we give the satellite fluxes we have used to asses the validity of the fits, where we used the Akari 65 $\mu$m flux (Yamamura et al., 2010) when available, otherwise the 60 $\mu$m IRAS flux (given in Acker et al. 1992). Errors are given when available from the Akari data archives; flux errors are not available for IRAS, only flux quality, where a low quality indicates an upper limit to the 60 $\mu$m flux, which is still valid, in our case, to constraint the continuum fit.} in order to determine whether the IRS spectra do
sample their flux maxima (i.e., the 60 [or 65] $\mu$m flux is lower than the maximum flux of the IRS spectra).  In these cases we also have checked whether the 60-65 $\mu$m fluxes in the literature were not lower that our continuum fit extrapolations at the same wavelength.  The majority of the (converging) fitted continua are perfectly compatible with the long wavelength photometry where available (55 PNe, flagged with {\it A} in Table 3, column 4); the IR flux long-ward of 60 $\mu$m shows that the observed SEDs sample the functional maximum of the fits. Furthermore, our extrapolated fluxes and those in the literature correlate very well with one another, with a global (linear) correlation coefficient of 0.82. We use the {\it A} fits with confidence in our group analysis and figures. 

Another 14 PNe are below the limit of detectability of the Akari and IRAS satellites at 60-65 $\mu$m, although their continuum fits seem to be excellent; they are flagged with {\it B} in column (4) of Table 3. 

Finally, 10 PNe have (converging) black body fits which are badly constrained by the 60-65 $\mu$m archival flux (they are flagged with {\it C} in Table 3, column 4). What happens in these cases is that the 60 and/or 65 $\mu$m fluxes are considerably higher than the IRS flux maxima.  
This  discrepancy can have several causes, including (a) a bad fit of the continuum, (2) the presence of other sources, or (3) high background at 60-65 $\mu$m. We have inspected these sources and found 4 possible bulge PNe among them, which could imply higher than average  background, and one PN (PN~G356.5-03.6) whose WISE data archive inspection within the Spitzer beam disclosed multiple sources which could cause the mismatch between the extrapolated fits and the 60-65 $\mu$m archival fluxes. For the remaining  PNe there is not a good explanation for the flux mismatch, likely an unexplained high background at long wavelength, and they will be studied in more detail in the future. To be very conservative with our conclusions we will not use the black body fits that do not comply with the available 60-65 $\mu$m photometry in the Figures (29-36) or analysis (Section 4) of this paper.

While these black-body fits reflect the physical origin of the continuum shape, this characterization is only broadly representative of the thermal dust. Indeed, the dust within a given nebula may be composed of a variety of chemical species, each with their own emissivity, at varying distances from the source of excitation. Thus, there may in reality be an ensemble of thermal emission within the nebula which is not perfectly represented by a single black-body temperature. We revisit this point and the impact it could have in the analysis presented in Section 4.2. In addition, the residuals in the black-body subtracted continuum may complicate the interpretation of very broad, weak, solid-state features. 
For those, one needs to perform polynomial fits and subtractions (see, for example, Garc\'{\i}a-Hern\'andez et al. 2010).

\begin{figure}
\plotone{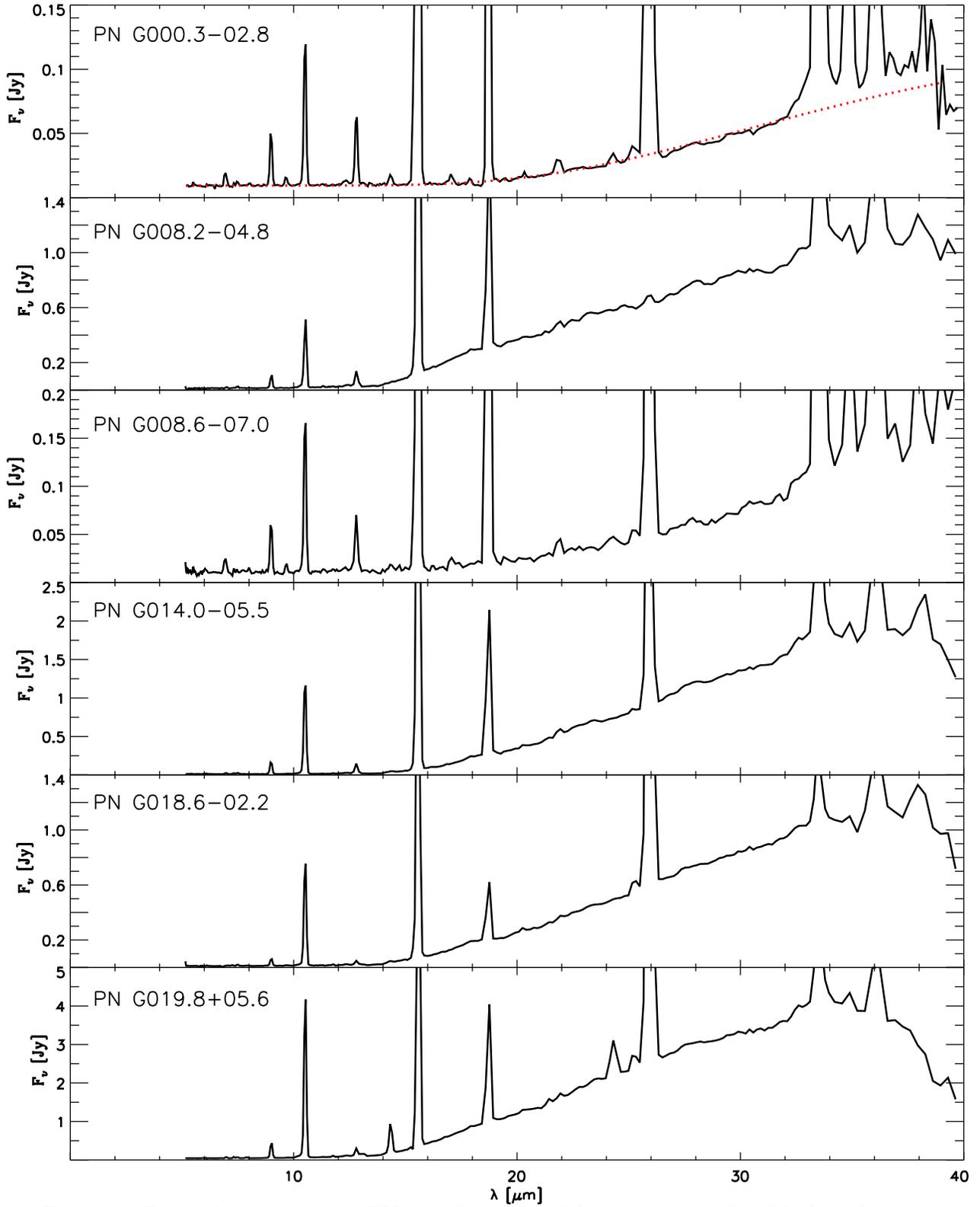}
\caption{Featureless spectra, in PN~G order. Dotted lines correspond to black body continuum fits 
of type {\it A} or {\it B} (see text, and Table 3).}
\label{}
\end{figure}

\begin{figure}
\plotone{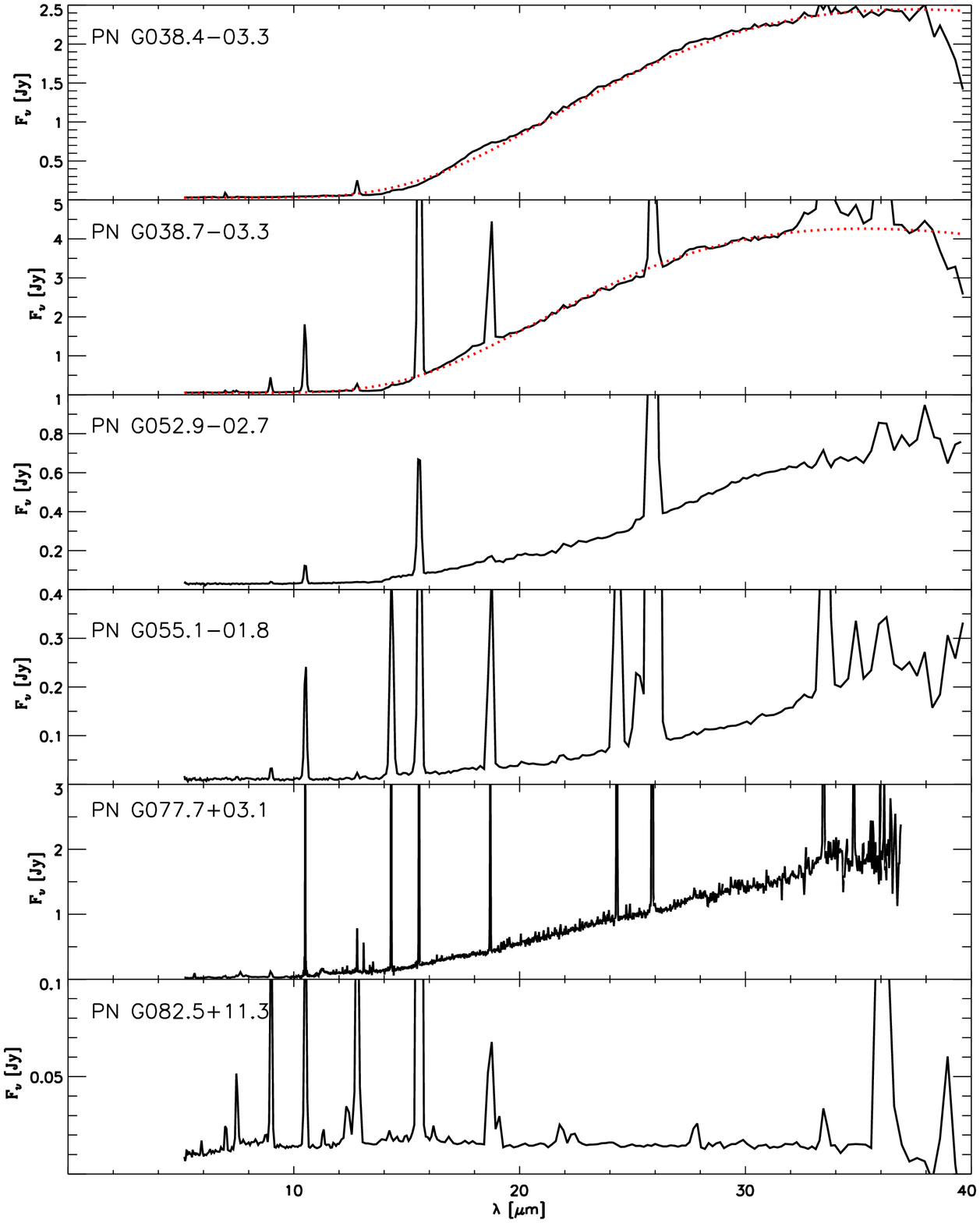}
\caption{Featureless spectra, cont.}
\label{}
\end{figure}

\begin{figure}
\plotone{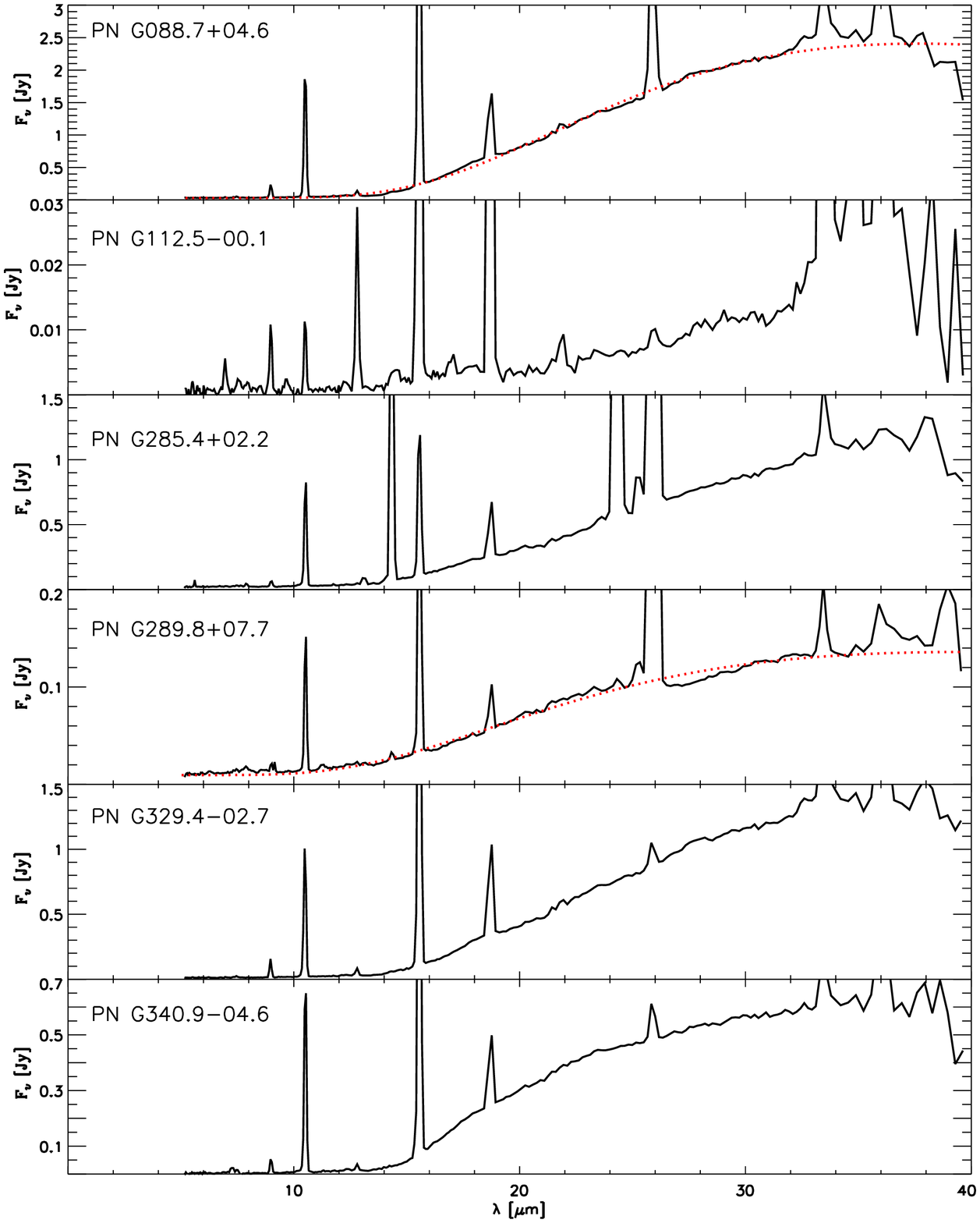}
\caption{Featureless spectra, cont.}
\label{}
\end{figure}

\begin{figure}
\plotone{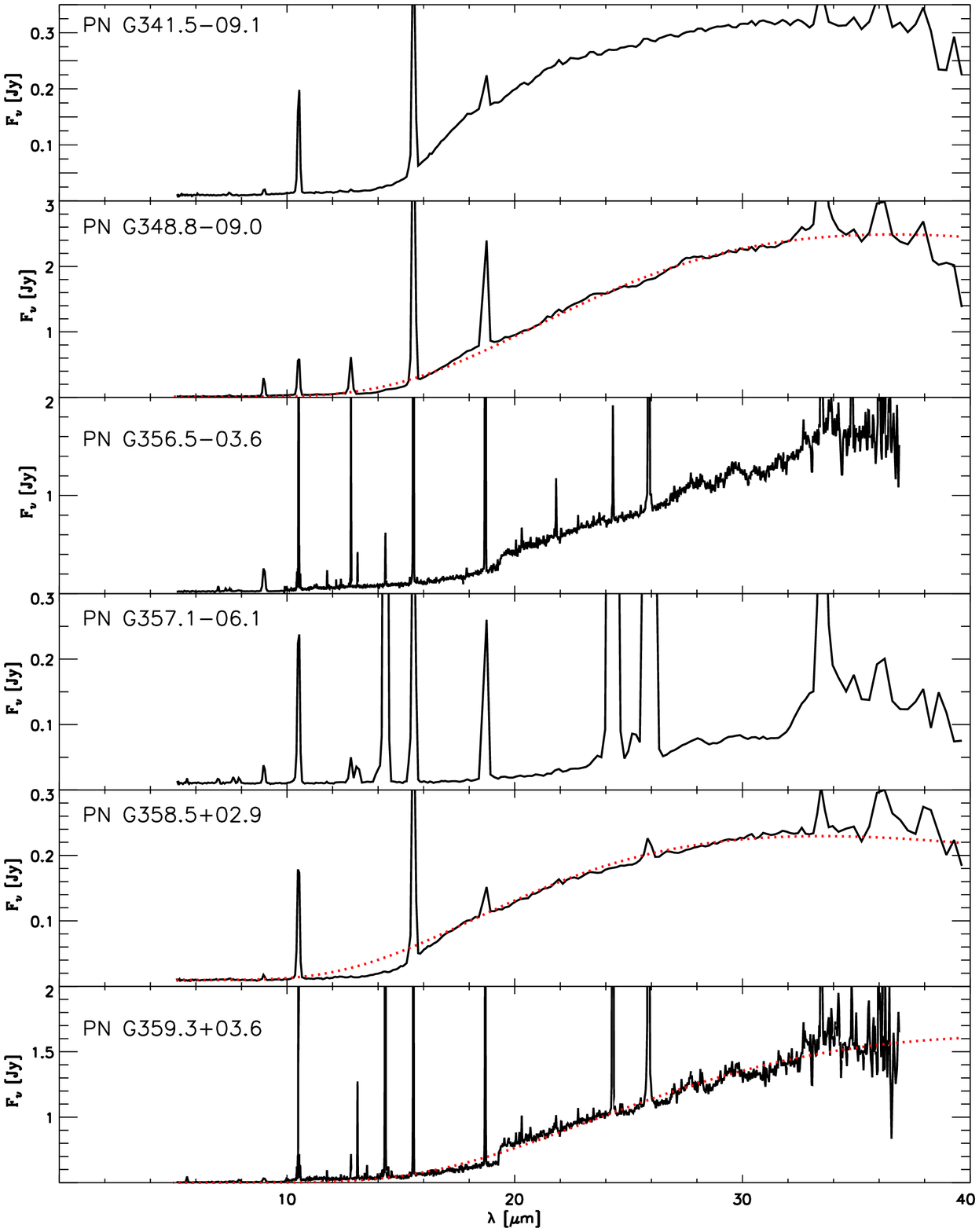}
\caption{Featureless spectra, cont.}
\label{}
\end{figure}

In Table 2 we give a description of our dust classification scheme. We define
therein the dust classes, and also several subclasses that we encounter in the analysis of
our large sample of compact Galactic PNe. As for the Magellanic Cloud sample, we classify the
spectra based on their dust features. A detailed analysis of the atomic emission lines and elemental
abundances will be published elsewhere. We distinguish among (1) the absence of prominent
dust features, (2) the presence of carbon-rich dust features, and (3) the presence of
oxygen-rich dust features. PNe lacking prominent molecular/dust emission features and with very low continuum are defined as
{\it F}. Those with carbon-rich features are in the {\it CRD} class, with subclasses
determining weather they show aromatic and/or aliphatic features. Similarly for the
oxygen-rich dust features, or {\it ORD}, where the subclasses indicate the presence of
crystalline and/or amorphous dust grains. Finally in many cases the spectra show both 
CRD and ORD characteristics, and we call these dual-chemistry PNe as {\it MCD} (or mixed-chemistry dust) PNe. 
In Figure 2 we
present the array of template spectra according to dust class and subclass. Table 3 lists, for all targets, the dust class (column
2), and subclass (column 3).

In Figures 3 through 26 we plot all IRS PN spectra from our program, and we overplot (as dotted lines) the continuum fits for {\it A} and {\it B} fit types (see Table 3).
There are a total of 25 PNe with continuum-only (F) spectra; one is shown in Figure 2 (top panel)
and the remaining 24 in Figures 3 through 6. None of these spectra shows very obvious molecular/dust emission
features above the continuum, and the continua are very faint. In a couple of cases (PN~G038.4-0.3. and PN~G038.7-03.3,
both in Fig. 4) the dust profile has a different shape than the other F PNe, in particular
there is a suggestion of a bump in the $\sim$20-30 $\mu$m spectral region, thus the F
designation might be uncertain. Actually, there may be some ORD or CRD PNe among the
subsample of objects with featureless spectra but the weakness of the dust features prevents
to confirm the dominant chemistry (ORD or CRD) of their circumstellar shells. It is worth
noting that two of the F spectra (PN~G356.5-03.6 and PN~G359.3+03.6, both in Fig. 6) show a
jump between the SH and LH modules ($\sim \lambda < 19.5 \mu m$) that, in principle, may
be attributed to an extended nature of these sources at the longer wavelengths. However, due
to the compact nature ($<$4") of our PN sample and because we do not see any mismatch
between the SL and SH modules, we suspect that this LH flux excess is possibly due to
 a known systematic effect that may affects some LH observations
(see e.g., Garc\'{\i}a-Hern\'andez et al. 2007b) or, less likely, to background emission. In general the background is very faint
relative to the source in these LH observations, but there is really no way to check if this is indeed the case.
In any case, this slight mismatch between
the SH and LH observations does not affect any of our results. In these case, we have scaled the continuum for the final fits.

\begin{figure}
\plotone{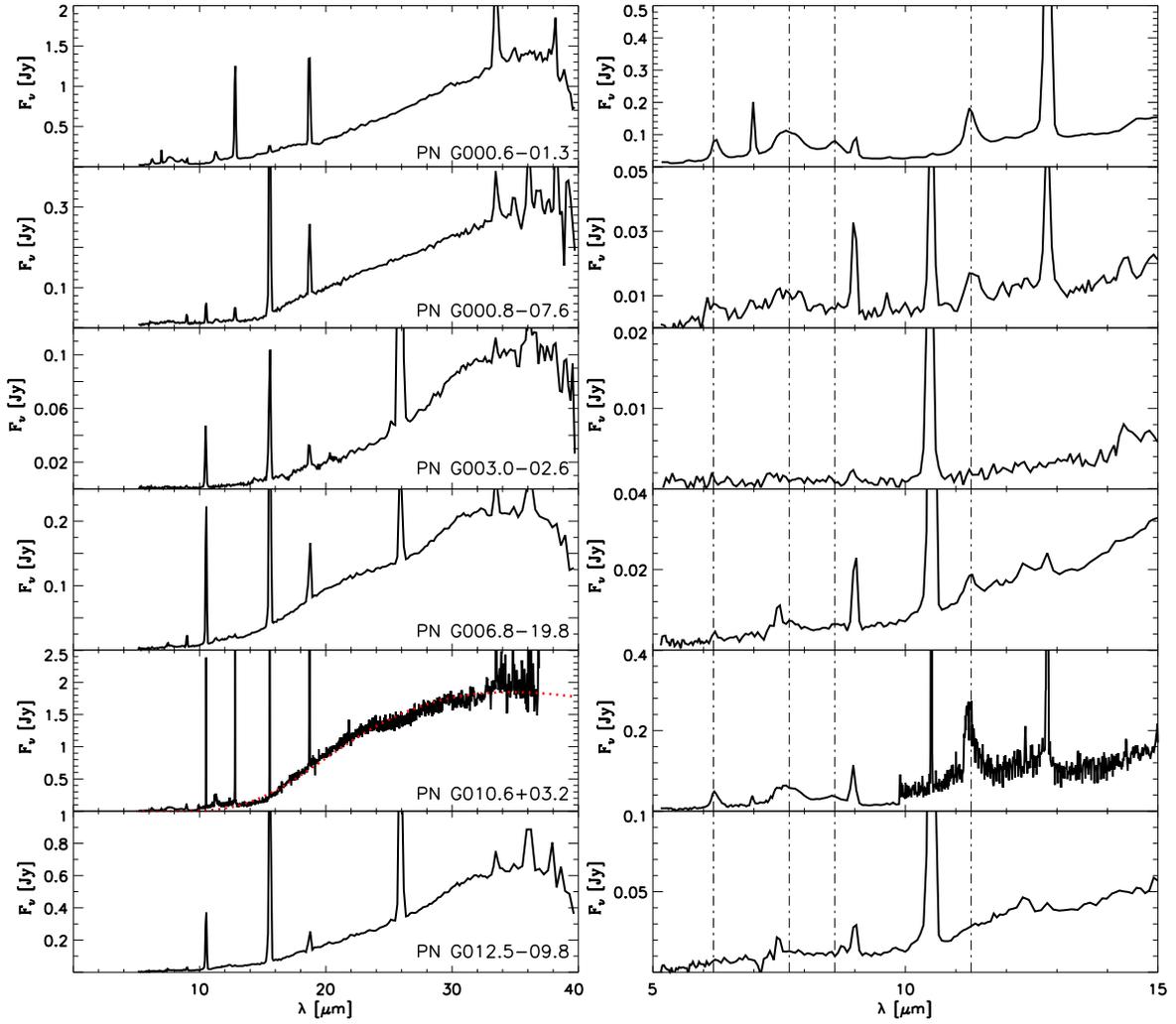}
\caption{CRD spectra, in PN~G order. Both aromatic, aliphatic, and aromatic/aliphatic spectra are shown here, see Table 3 for complete list. On the left panels we show the complete spectra, while right panels show the 5-12 $\mu$m sections of the spectra, where the 6.2, 7.7, 8.6, and 11.2 $\mu$m AIB band positions have been indicated with vertical lines. Dotted lines have the same meaning than in Fig. 3}
\label{}
\end{figure}

\begin{figure}
\plotone{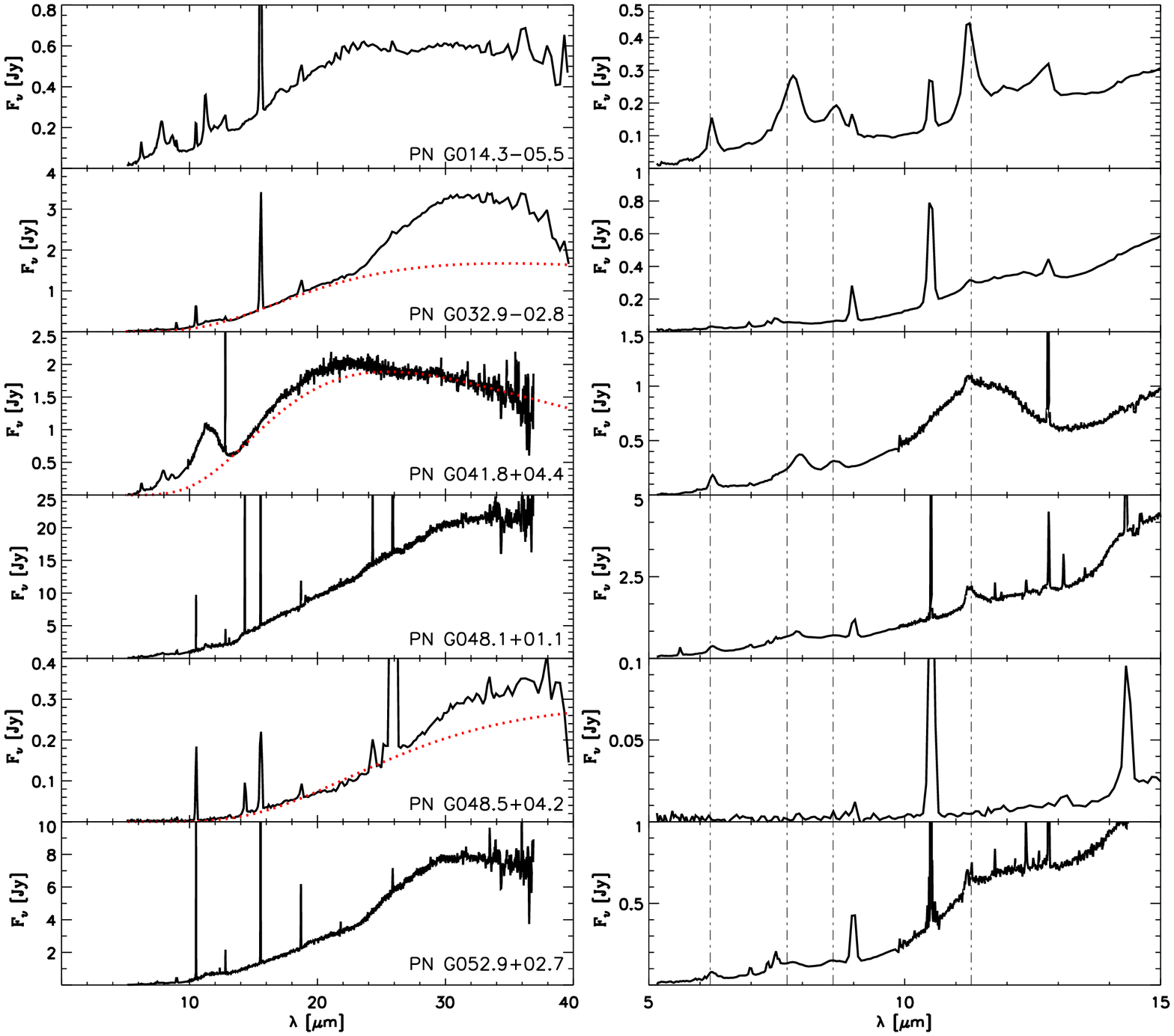}
\caption{CRD spectra, cont.}
\label{}
\end{figure}

\begin{figure}
\plotone{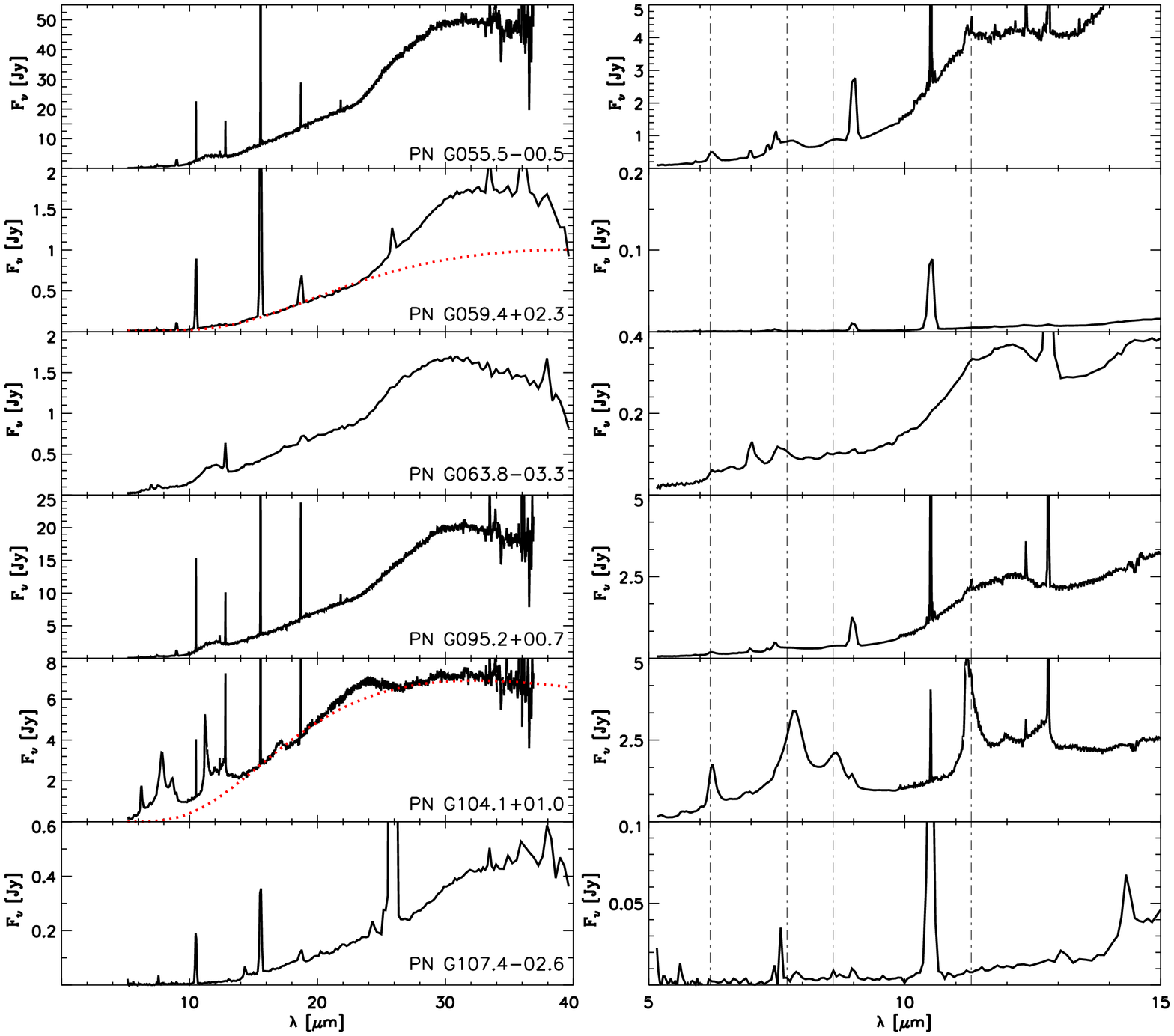}
\caption{CRD spectra, cont.}
\label{}
\end{figure}

\begin{figure}
\plotone{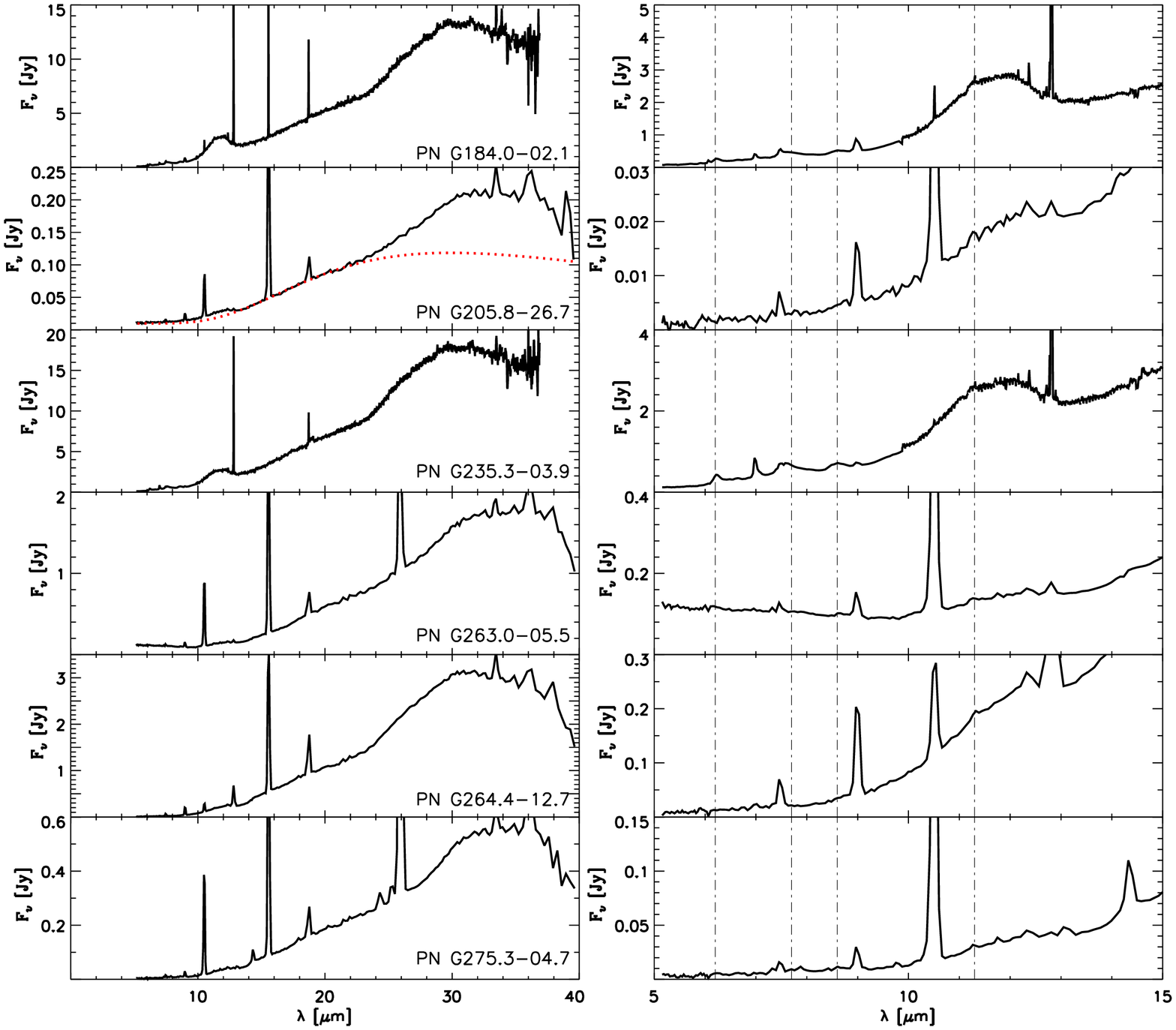}
\caption{CRD spectra, cont.}
\label{}
\end{figure}

\begin{figure}
\plotone{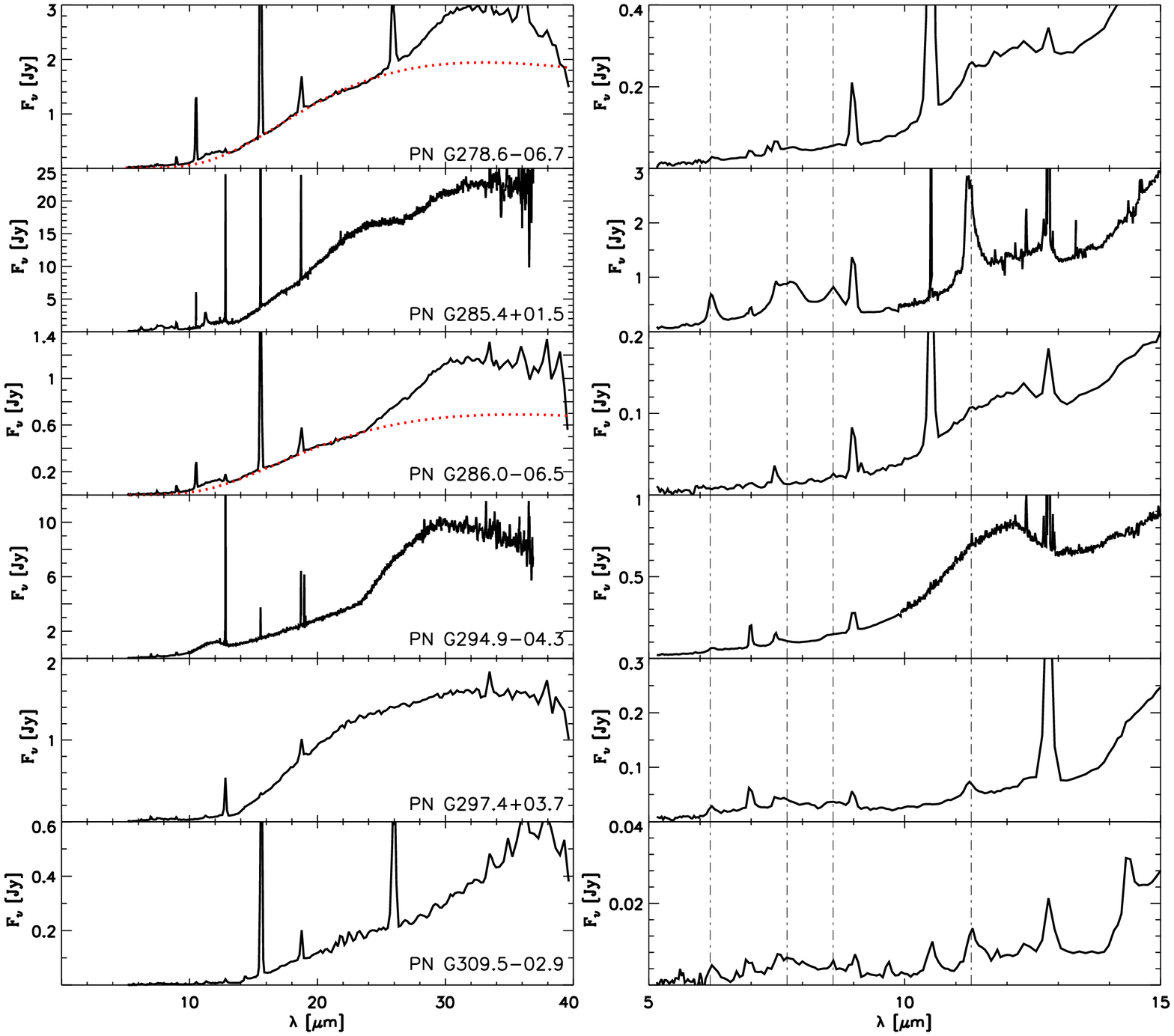}
\caption{CRD spectra, cont.}
\label{}
\end{figure}

\begin{figure}
\plotone{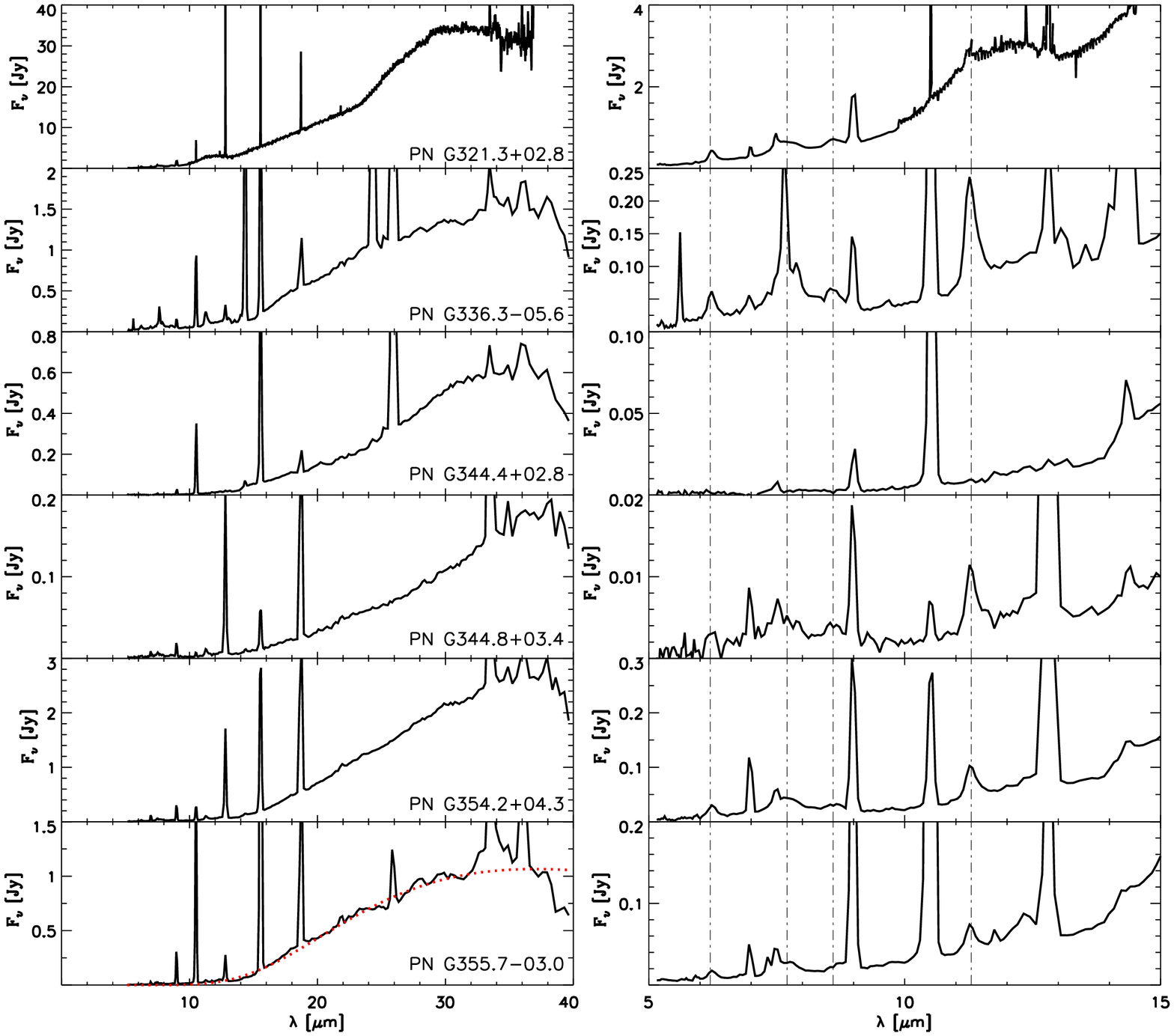}
\caption{CRD spectra, cont.}
\label{}
\end{figure}

Carbon-rich dust features can be aromatic or aliphatic. The aromatic features are
represented by the well-known family of infrared features at 6.2, 7.7, 8.6, and 11.2 $\mu$m
usually attributed to polycyclic aromatic hydrocarbons (PAHs; e.g., Leger \& Puget 1984), or,
as more appropriately defined by Kwok \&  Zhang (2011), to aromatic infrared bands (heretofore, AIBs). 
The AIBs have been observed in several types of astronomical sources.  It is worth noting that
in several cases a single (and usually very weak) 11.2 $\mu$m feature in the spectra,
especially if the feature is narrower than common AIB features, is due to interstellar rather
than circumstellar emission, thus the nebula would not be classified as CRD solely on this emission feature. 
In Figure
2 we show a typical example of obvious AIB features in PN~G068.7+14.8, where all 4 bands
are clearly present. PNe with spectra showing any of these bands are classified as CRD;
our sample includes 12 additional such spectra. 

A larger number of the CRD spectra show other emission features that are likely related to carbon-rich dust as well,
but of different types. The 6-9 $\mu$m feature 
can be attributed to hydrogenated amorphous carbons (HACs), large PAH clusters, or very small carbon grains. 
On the other hand, 
the broad 10-15 $\mu$m emission (centered at about 11.5 $\mu$m) and the 25-35 $\mu$m emission (the so-called 
30$\mu$m feature) are usually attributed to SiC (e.g., Speck et al. 2009) and MgS 
(e.g., Hony et al. 2002), respectively. However, the observed variation of these broad features 
is quite consistent with the variety of properties of HACs, which are manifest in a wide range of different spectra 
depending on their physical and chemical properties 
(e.g., size, shape, hydrogenation; Scott \& Duley 1996; Scott et al. 1997; Grishko et al. 2001). 
In particular, Grishko et al. 2001 showed that HACs can explain the 21, 26, and 30 $\mu$m features.
Zhang et al. (2009) also argue that MgS is very unlikely the carrier of the aliphatic 30 $\mu$m emission 
seen in C-rich evolved stars. Thus, it is very unlikely that the carriers of the broad 11.5 um and 30 $\mu$m 
are related with SiC and MgS, respectively, as it was 
suggested in the past. We tentatively define all these features as aliphatic dust features, found in 22 PNe of our sample.

Three of the CRD spectra in our sample show both aromatic and aliphatic
features, thus the total number of  CRD PNe in our sample amounts to 38, or $\sim$25$\%$ of
the PN spectra. All CRD spectra are shown in Figures 7 through 12, where we show complete
spectra in the left panels of each figure, and the 5-15 $\mu$m section of the spectra in the
right panels, to characterize the presence (or absence) of AIB features. Most of the CRD
spectra are predominantly aromatic or aliphatic, as noted in column (3) of Table 3. In a few
cases both sets of features are present, and we mark the PN as aromatic/aliphatic type. In a
minority of cases the CRD dust features are very faint, and had the spectra been shallower we would have
classified them as F. Their dust type in column (3) of Table
3 is then flagged as uncertain. Interestingly, two CRD aliphatic PNe, PN~G063.8-03.3 and 
PN~G235.3-03.9, display strong C$_{60}$ fullerene features (Garc\'{\i}a-Hern\'andez et al.
2010), and are part of a small group of Galactic and Magellanic Cloud PNe that show the
infrared emission bands of this complex molecule (Cami et al. 2010; Garc\'{\i}a-Hern\'andez et al. 2010, 2011b).
PNe that contain fullerenes are typically low-excitation and low-mass C-rich objects displaying
the broad 11.5 and 30 $\mu$m aliphatic features. The detection of fullerenes in the H-rich
circumstellar shells of these PNe (see also Garc\'{\i}a-Hern\'andez et al. 2011a) has
challenged our understanding of the formation of these complex fullerene molecules in space,
indicating that fullerenes may be more abundant than previously believed and demonstrating
that normal C-rich PNe are important factories of complex organic compounds.

Carbon-rich dust features have been found in 42 additional PNe, where crystalline
silicate bands (e.g., Waters et al. 1998) are also present. These objects, mixed-chemistry
dust (or MCD) PNe, showing both AIBs (C-rich) and crystalline silicates (O-rich) in their
shells (e.g., Perea-Calder\'on et al. 2009), have not been found in the Magellanic Clouds to
date (S07; Bernard-Salas et al. 2009), but are very common in the Galactic Bulge
(Gutenkunst et al. 2008; Perea-Calder\'on et al. 2009). A template spectrum of MCD type PNe is shown in Figure 2,
while all other MCD PN spectra are presented in Figures 13 to 19, where we show complete
spectra in left panels, and the $5 < \lambda< 15 \mu m$ spectra in the right panels, as in the CRD PN
spectra. In the MCD PN plots we also indicate the prominent crystalline silicate
features at 23.5, 27.5, and 33.8 $\mu$m - usually attributed to olivine and pyroxenes - with
vertical dotted lines. Crystalline silicates and other oxygen-rich dust features (e.g., the amorphous silicate
bands at $\sim$10 and 18 $\mu$m), not associated with carbon-rich dust, have been detected
in 45 PNe. Of these, 16 have only crystalline silicate features, 24 show possible
amorphous oxygen features, and five  show possible signatures of both types (amorphous and crystalline) of
oxygen-rich dust. We call all these objects ORD  (oxygen-rich
dust) PNe. The details of their subtypes are given in Table 2, and their spectra are shown in
Figures 20 to 26. 

It turns out that 7 of the 157 targets acquired by us are misclassified as PNe, thus their
spectral analysis will be published elsewhere.  The IRS spectra of the remaining 150 Galactic PNe indicate
that many more show dust features than their
Magellanic Cloud counterparts, as expected, given their relative metallicities (S07). Only
$\sim$17$\%$ of the Galactic sample does not show molecular/dust features, while the featureless
PNe in the Clouds represented $>$41$\%$ of the samples (S07; Bernard-Salas et al. 2009). 
The PNe spectra with molecular/dust features in the
Galaxy are distributed evenly among the ORD, CRD, and MCD  types, respectively representing
approximately 30, 25, and 28 $\%$ of the sample; in the Magellanic Clouds, on the other
hand, most of the dust-rich PN spectra are CRD, and only very few PNe show ORD features,
while MCD PNe have not been detected in the Magellanic Clouds. A more detailed analysis
(e.g., integrated fluxes, positions, etc.) of the organic (AIBs) and solid-state features
(aliphatic dust, amorphous and crystalline silicates) seen in our Galactic Disk PN sample in
comparison Ä those of the Magellanic Clouds will be presented in a forthcoming paper
(Garc\'{\i}a-Hern\'andez et al. in preparation).

\begin{figure}
\plotone{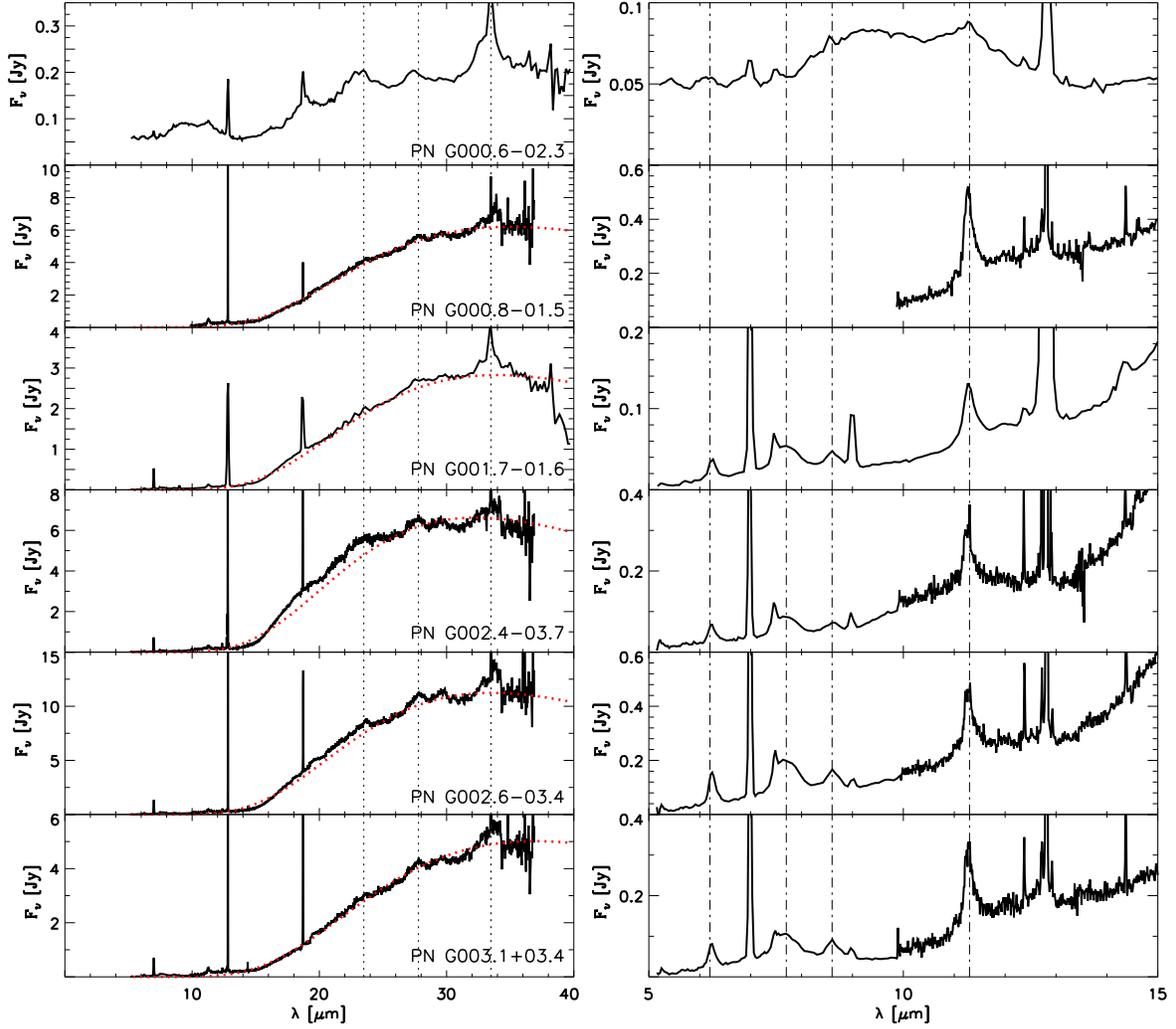}
\caption{MCD spectra, in PN~G order.  On the left panels we show the complete spectra, where the crystalline silicate bands at 23.5, 27.5, and 33.8 $\mu$m have been marked with vertical dotted lines; right panels show the 5-12 $\mu$m sections of the spectra, where the 6.2, 7.7, 8.6, and 11.2 $\mu$m AIB band positions have been indicated with vertical lines. Dotted lines have the same meaning as in Fig. 3}
\label{}
\end{figure}

\begin{figure}
\plotone{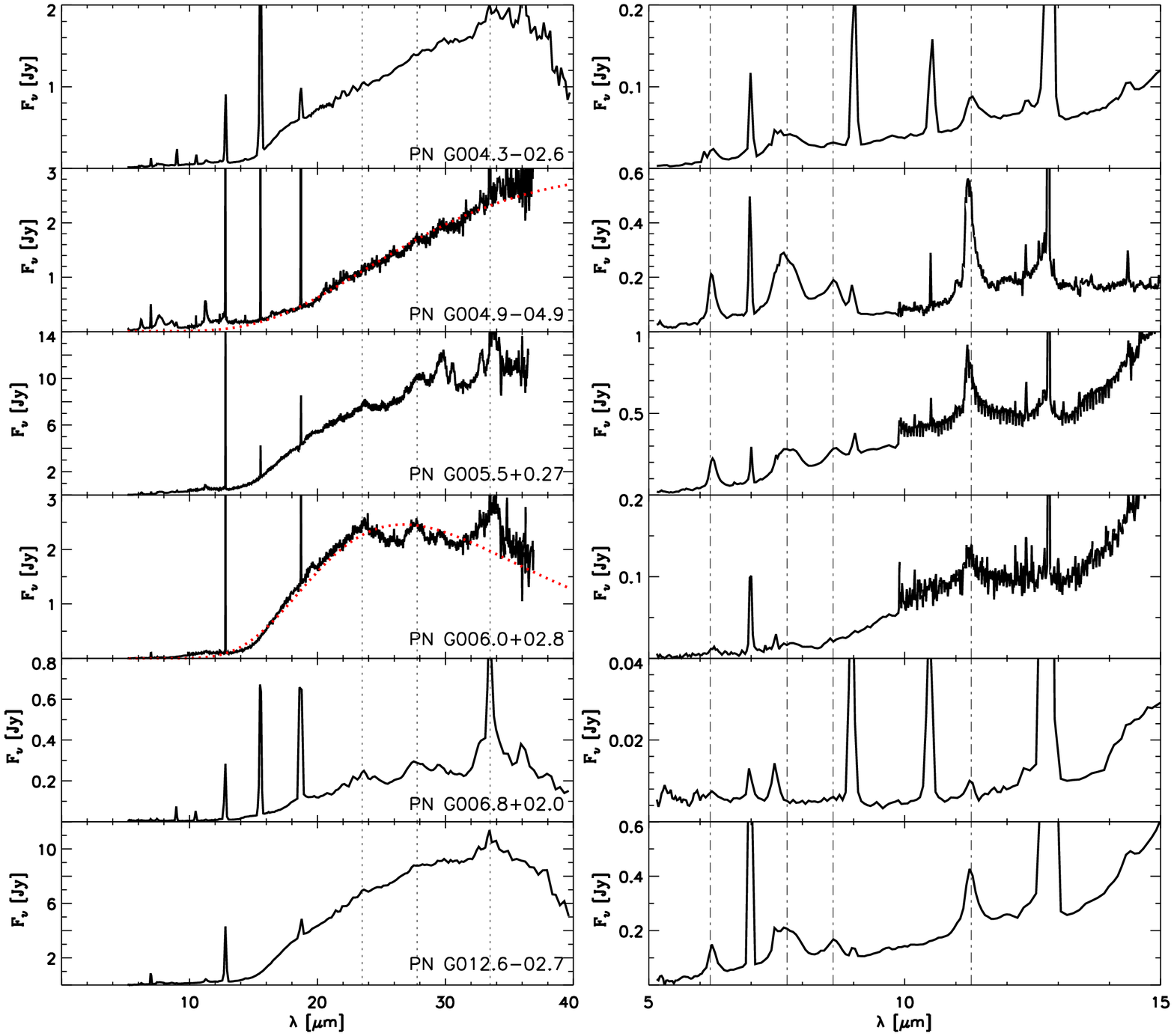}
\caption{MCD spectra, cont.}
\label{}
\end{figure}

\begin{figure}
\plotone{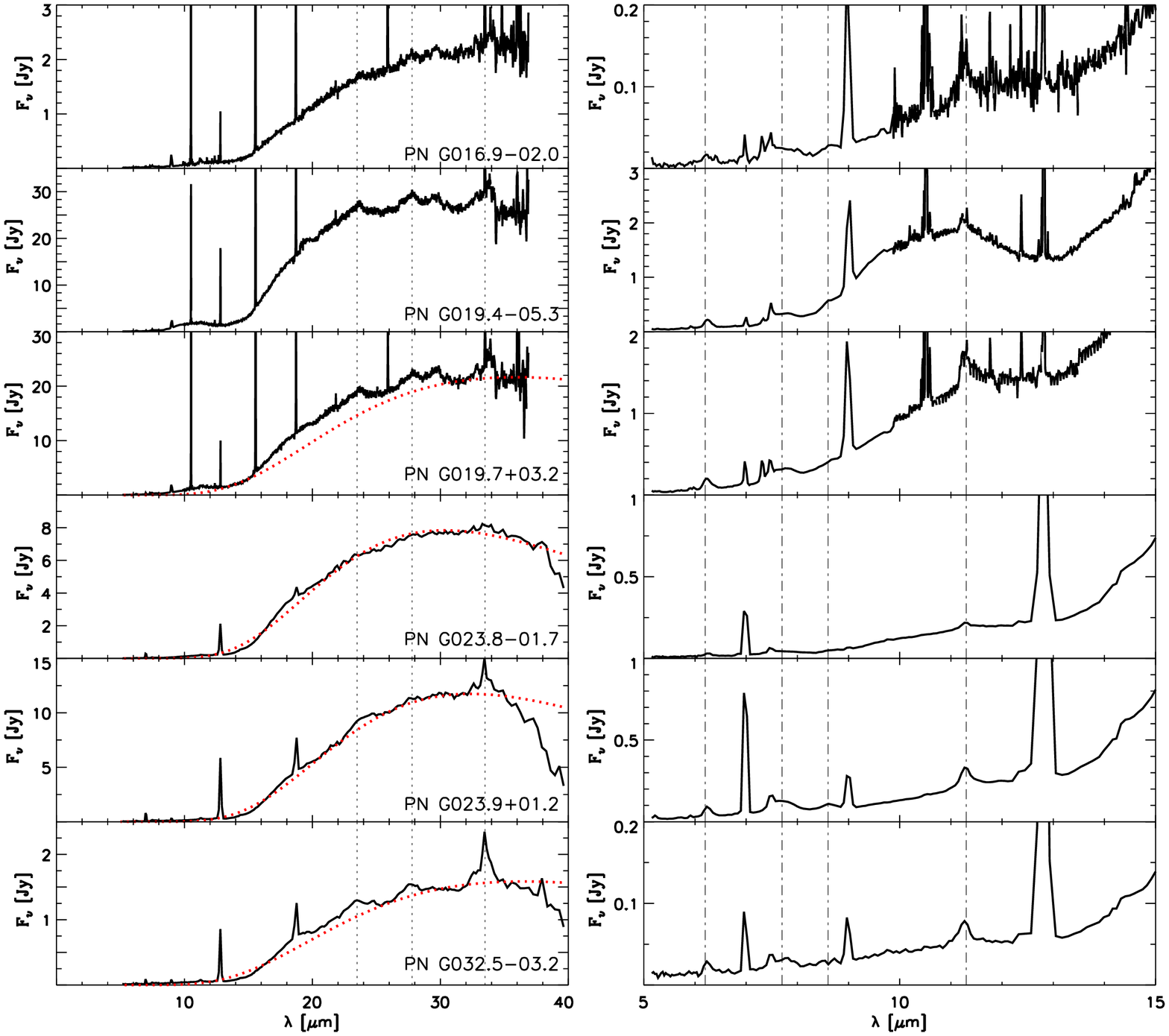}
\caption{MCD spectra, cont.}
\label{}
\end{figure}

\begin{figure}
\plotone{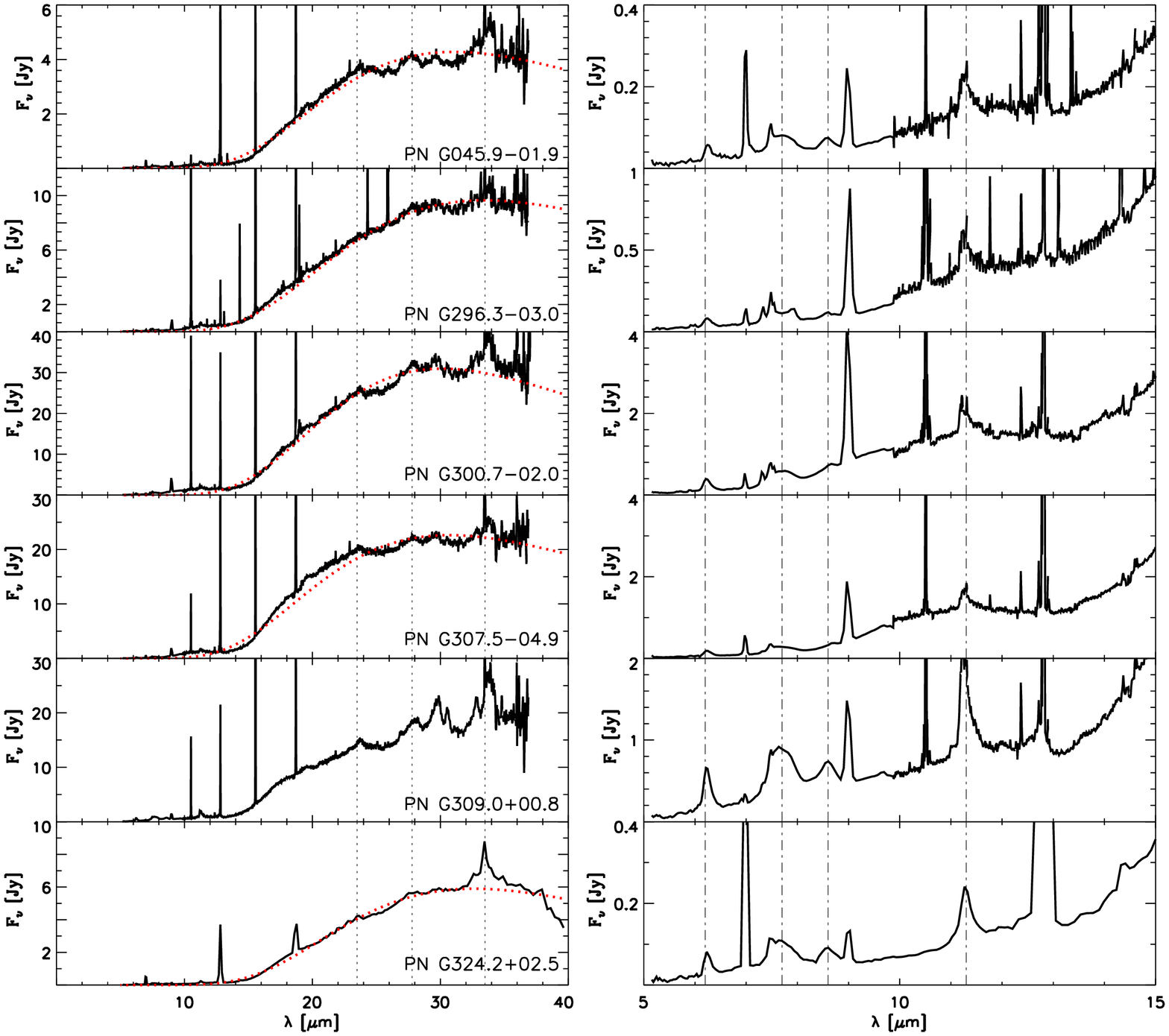}
\caption{MCD spectra, cont.}
\label{}
\end{figure}

\begin{figure}
\plotone{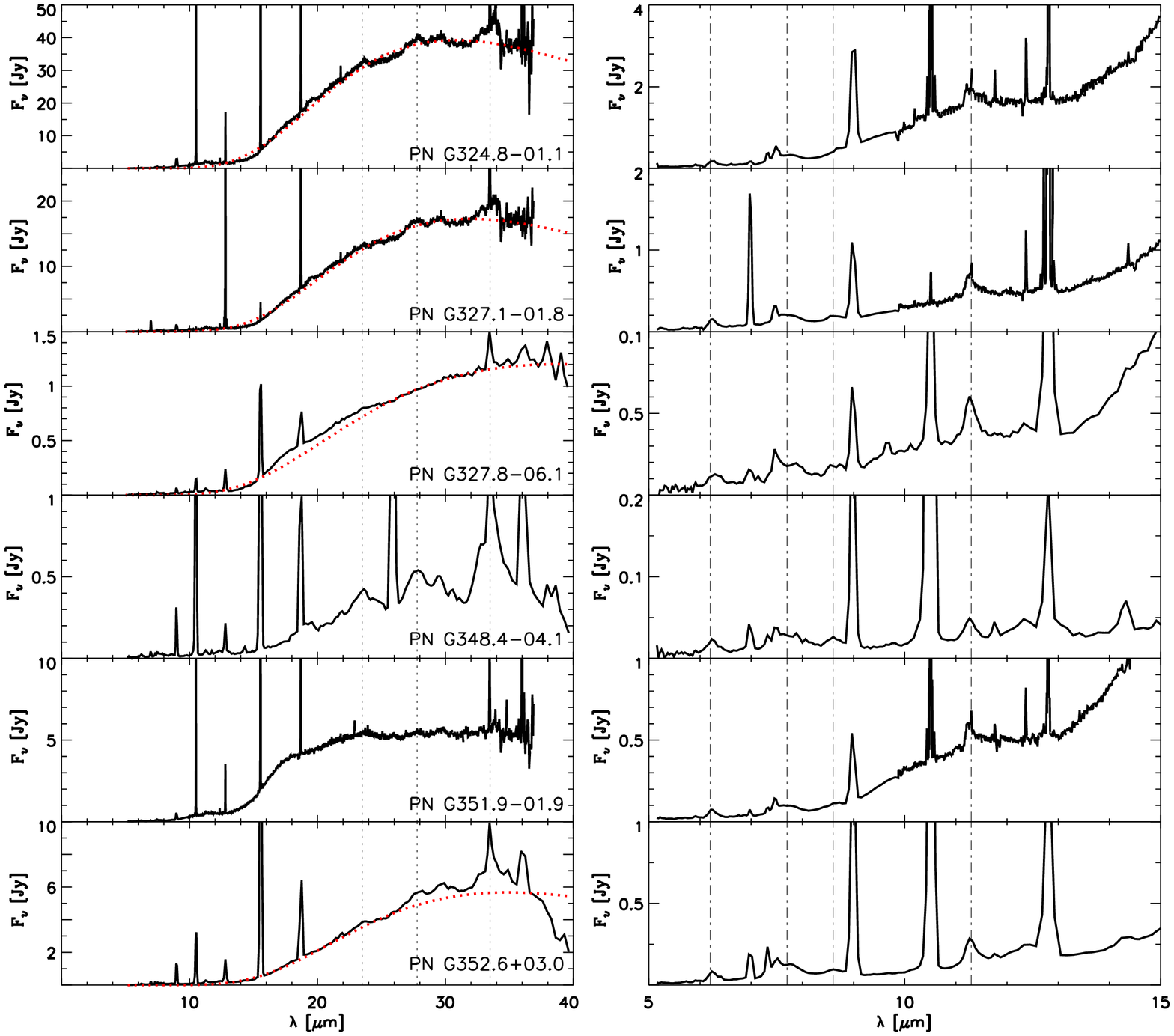}
\caption{MCD spectra, cont.}
\label{}
\end{figure}

\begin{figure}
\plotone{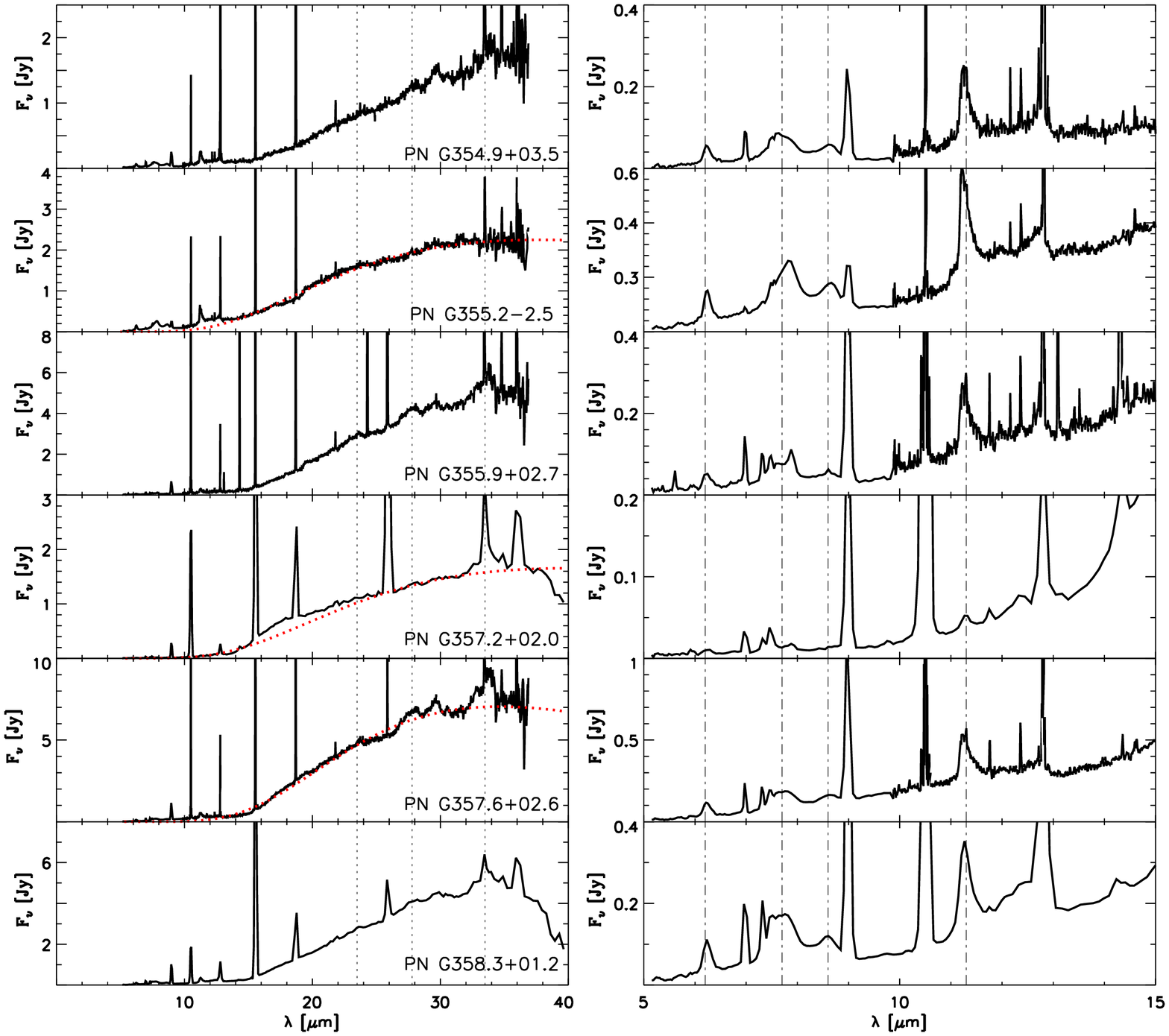}
\caption{MCD spectra, cont.}
\label{}
\end{figure}

\begin{figure}
\plotone{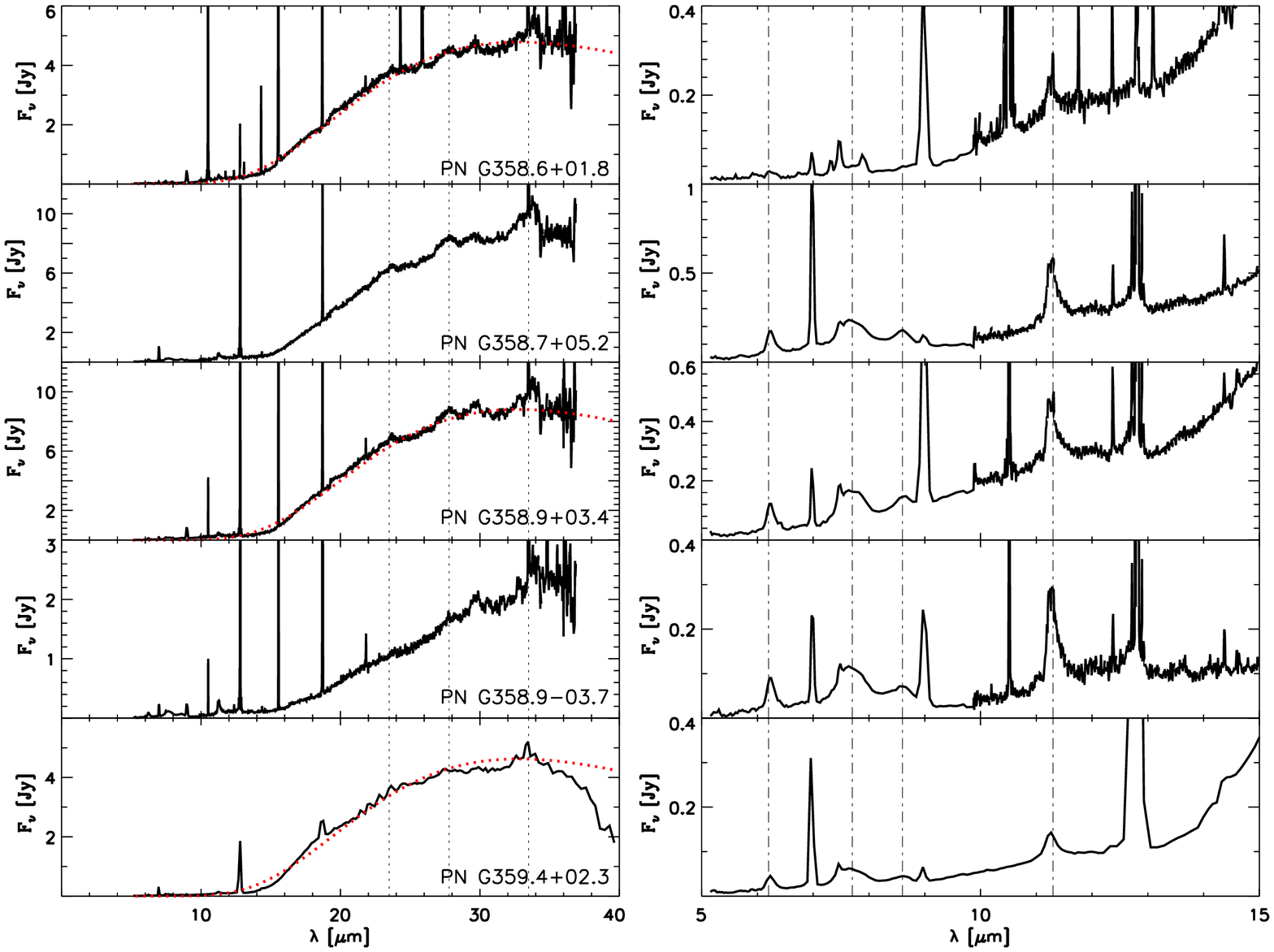}
\caption{MCD spectra, cont.}
\label{}
\end{figure}

\section{Correlations between IR spectral characteristics and other nebular and
stellar properties}

\subsection{Ancillary data from the literature}

From the IRS analysis in this paper we have acquired the knowledge of several IR-related
properties of the PNe. In Table 3, together with the class and subclass of the dust spectra,
we give the IRS-spectra related parameters, namely, the fit type (column 4), the emissivity exponent (column 5),
the 60-65 $\mu$m flux from the Akari or IRAS surveys, used to assess the validity of the continuum fit  (column 6)
the dust temperature and its uncertainty
 (column 7), the
total IR luminosity (column 8), and the infrared excess, derived classically as in Pottasch (1984)  (column 9).  
Dust temperatures are given only in the cases where the black body fit converge, either with simple black-body or with an 
emissivity term. It is worth noting that the different dust types show a difference in emissivity, as expected, with the average of the 
exponent $\alpha$ in CRD PNe being 0.52$\pm$0.73, compared to 
an average value of 3.16$\pm$2.06 in ORD PNe.
The infrared luminosity was determined by integrating the flux under the black body fit of the IRS dust continua, 
then scaled to absolute luminosity by using the statistical distances.
The PNe whose fits converge may have {\it A, B} or {\it C} fit type (column 4), as described in section 3.
In Figures 29 and beyond we plot T$_{\rm dust}$ and L$_{\rm IR}$ only in the case of an {\it A} or {\it B} black body fit type.

Other properties of the PNe that we have used in this study have been listed in Table 4. The
most important parameter to gain astrophysical insight  from the analysis of PN properties is
of course their distances. Individual heliocentric distances determined with reliable
methods, such as cluster membership, are not available for most of the PNe in our sample. We
then use the best statistical distance scale available, based on the Stanghellini et al.  (2008)
Magellanic Cloud calibration. In order to get statistical distances on this calibrating scale
we need the angular diameters, and the 5 GHz (or H$\beta$) fluxes. The fluxes, and the upper limits to angular diameters,
 are available
for all targets, but actual angular diameters are known only for 50 of the 150 confirmed PNe in our IRS
sample, thus their distances are readily available only for approximately a third of the sample (see also
Stanghellini \& Haywood, 2010).  An additional 42 PNe have been observed by us with WFC3
(Shaw et al. in preparation), and from these space-based images we were able to derive the
angular sizes, thus calculate their distances with the statistical scheme described above.
Distances from the Galactic center, and the PN sizes derived as described in Stanghellini \&
Haywood (2010), are listed in Table 4. 

Plasma diagnostics for the observed PN based on optical spectra are scarce in the literature,
with only a dozen PNe having adequate electron density and elemental abundance. We have used
the [SII] $\lambda$6717-6731 emission lines measured from low resolution spectroscopy and
published by Acker et al. (1992) to derive an estimate of the nebular densities, by assuming
electron temperatures of 10,000 K and by using the {\it nebular} routines in {\it
IRAF/stsdas} (Shaw \& Dufour 1995). The densities calculated are listed in column (4) of Table 4. There is not
enough information to determine oxygen or other abundances for most of the compact PNe in our
sample. The abundance analysis based on the IR lines in the IRS spectra will be the subject
of a future paper.

For all other nebular parameters, we used the Stanghellini \& Haywood (2010) recent, complete, and uniform
collection of Galactic PN data.

\begin{figure}
\plotone{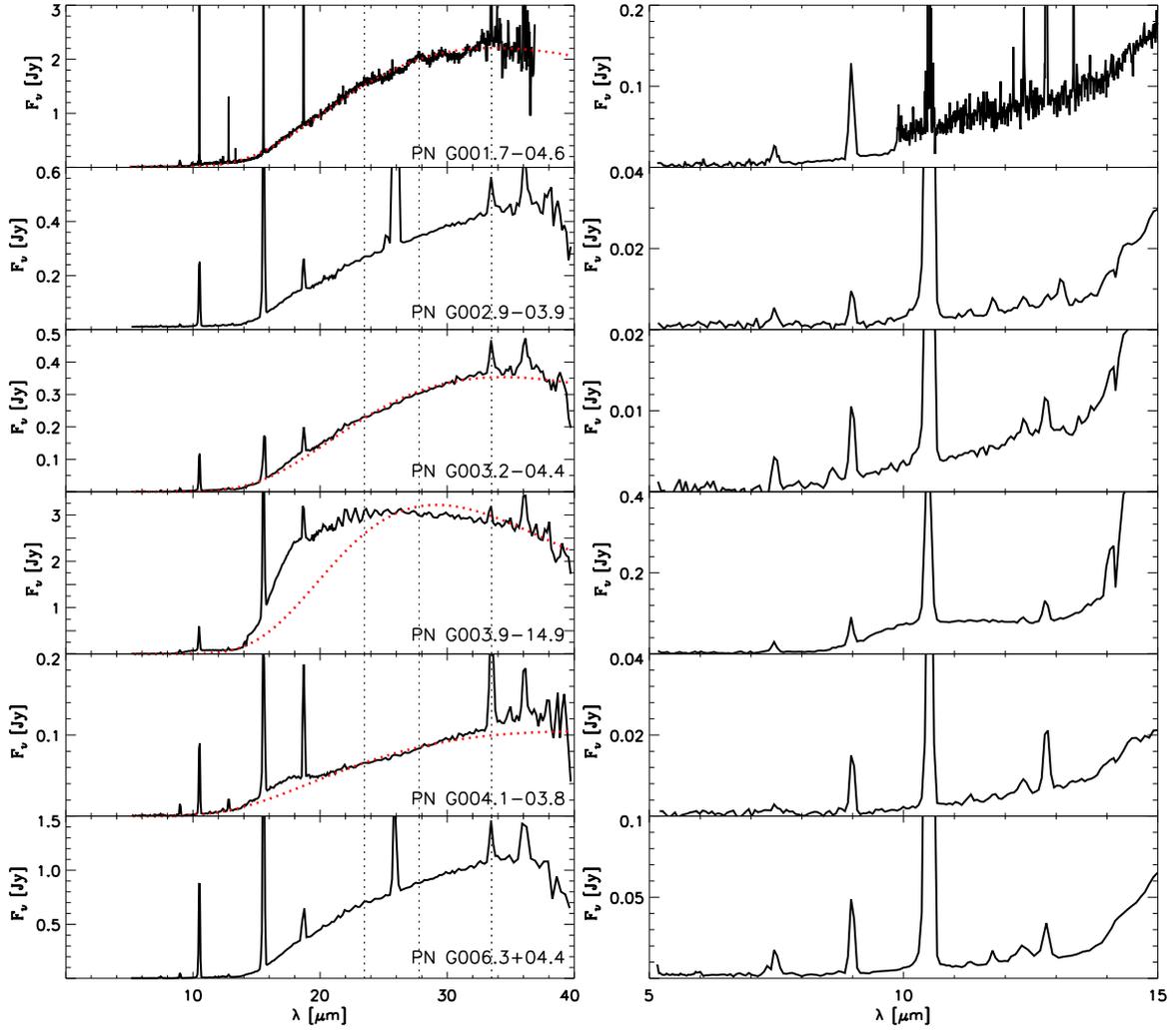}
\caption{ORD spectra, in PN~G order.  On the left panels we show the complete spectra, where the crystalline silicate bands at 23.5, 27.5, and 33.8 $\mu$m have been marked with vertical dotted lines; right panels show the 5-12 $\mu$m sections of the spectra. Dotted lines as in Fig. 3}
\label{}
\end{figure}

\begin{figure}
\plotone{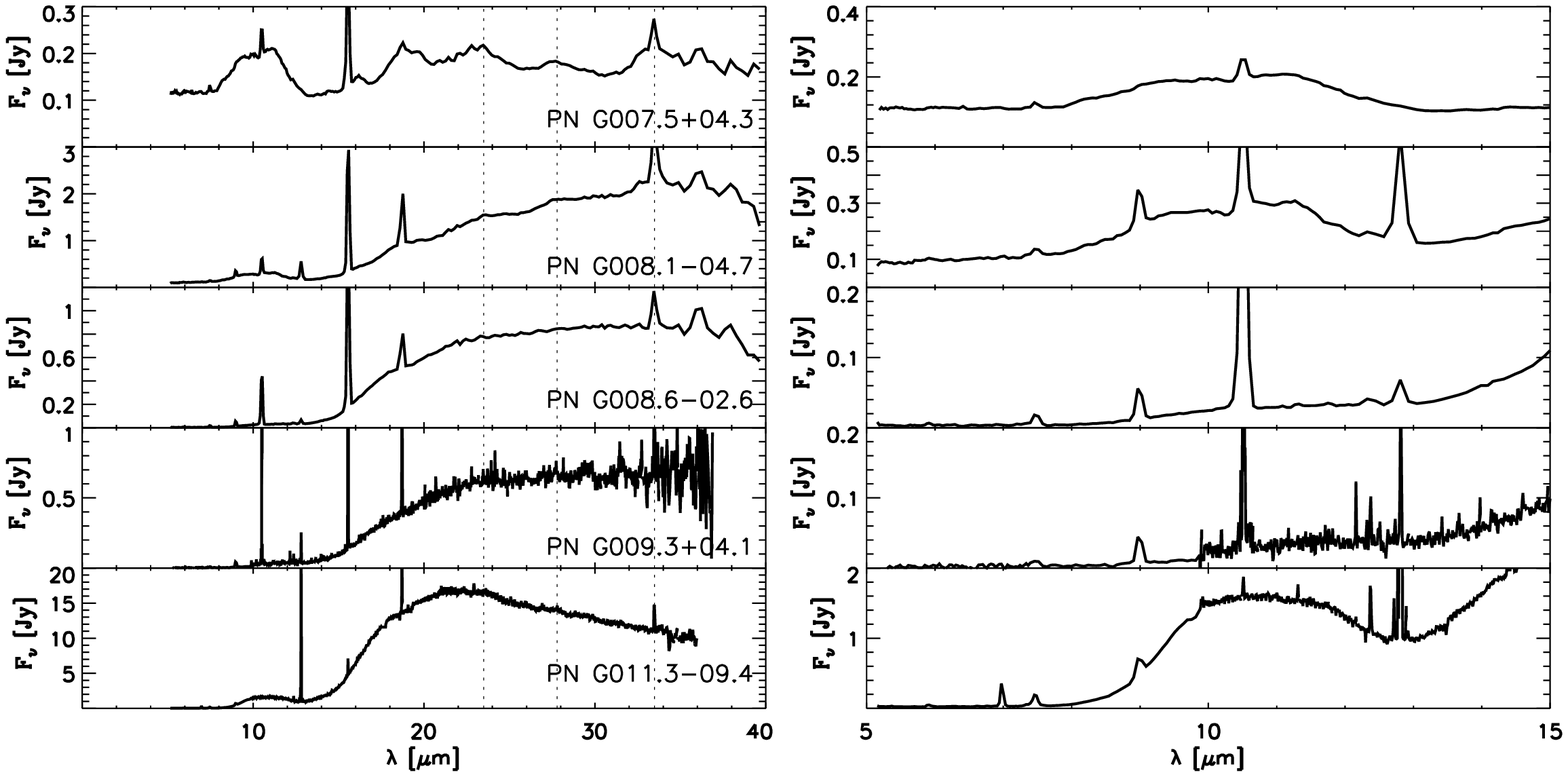}
\caption{ORD spectra, cont.}
\label{}
\end{figure}

\begin{figure}
\plotone{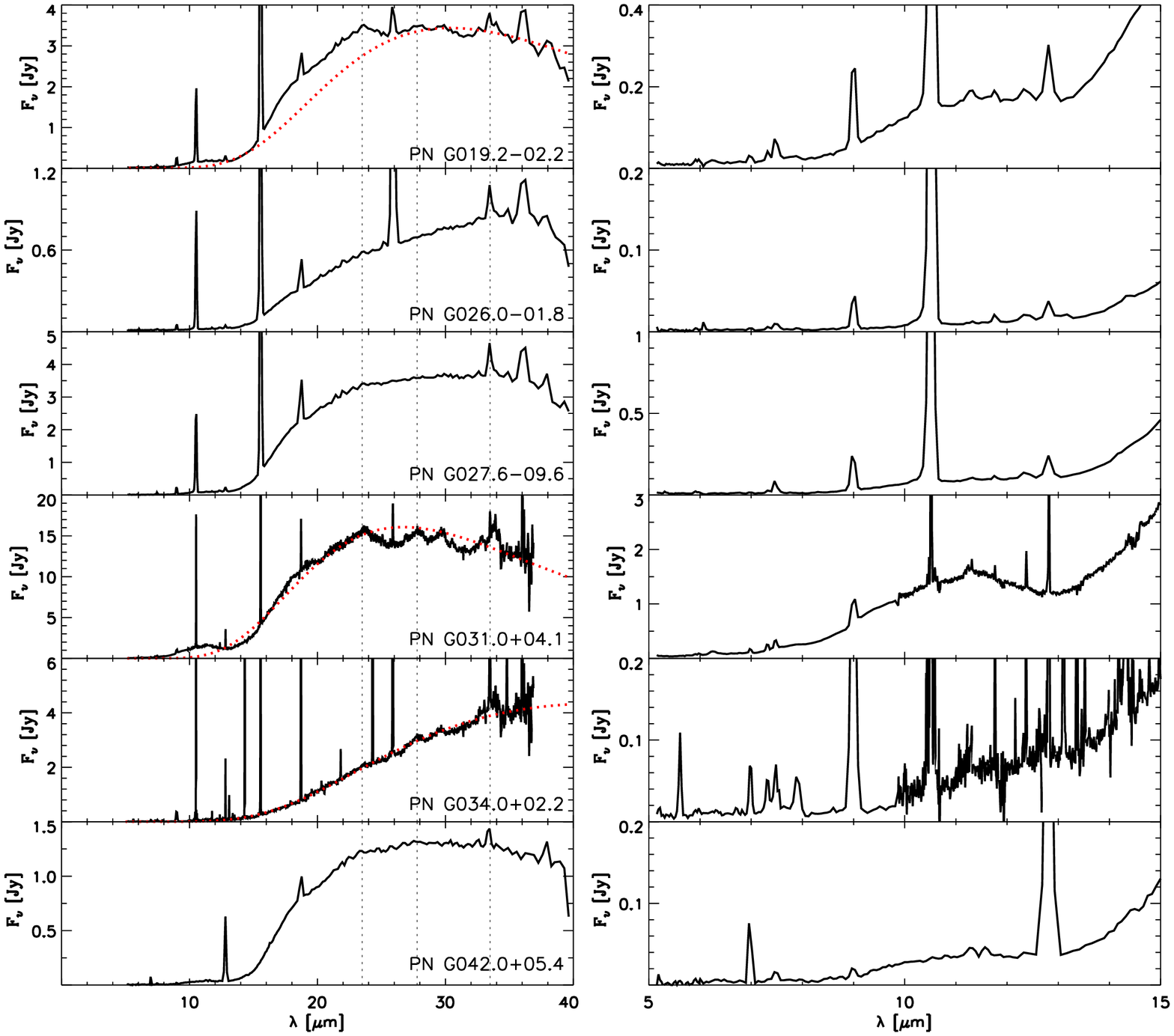}
\caption{ORD spectra, cont.}
\label{}
\end{figure}

\begin{figure}
\plotone{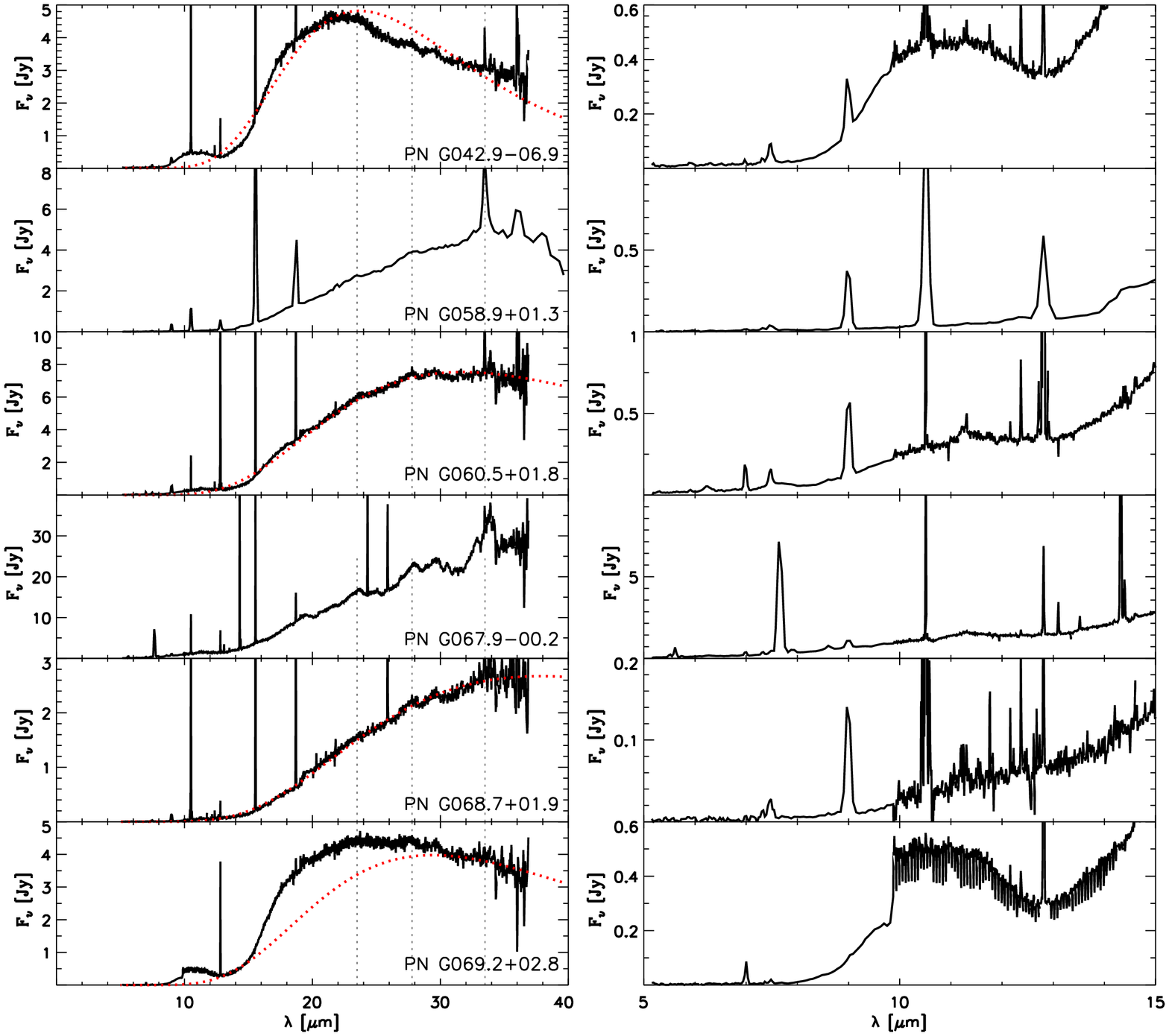}
\caption{ORD spectra, cont.}
\label{}
\end{figure}

\begin{figure}
\plotone{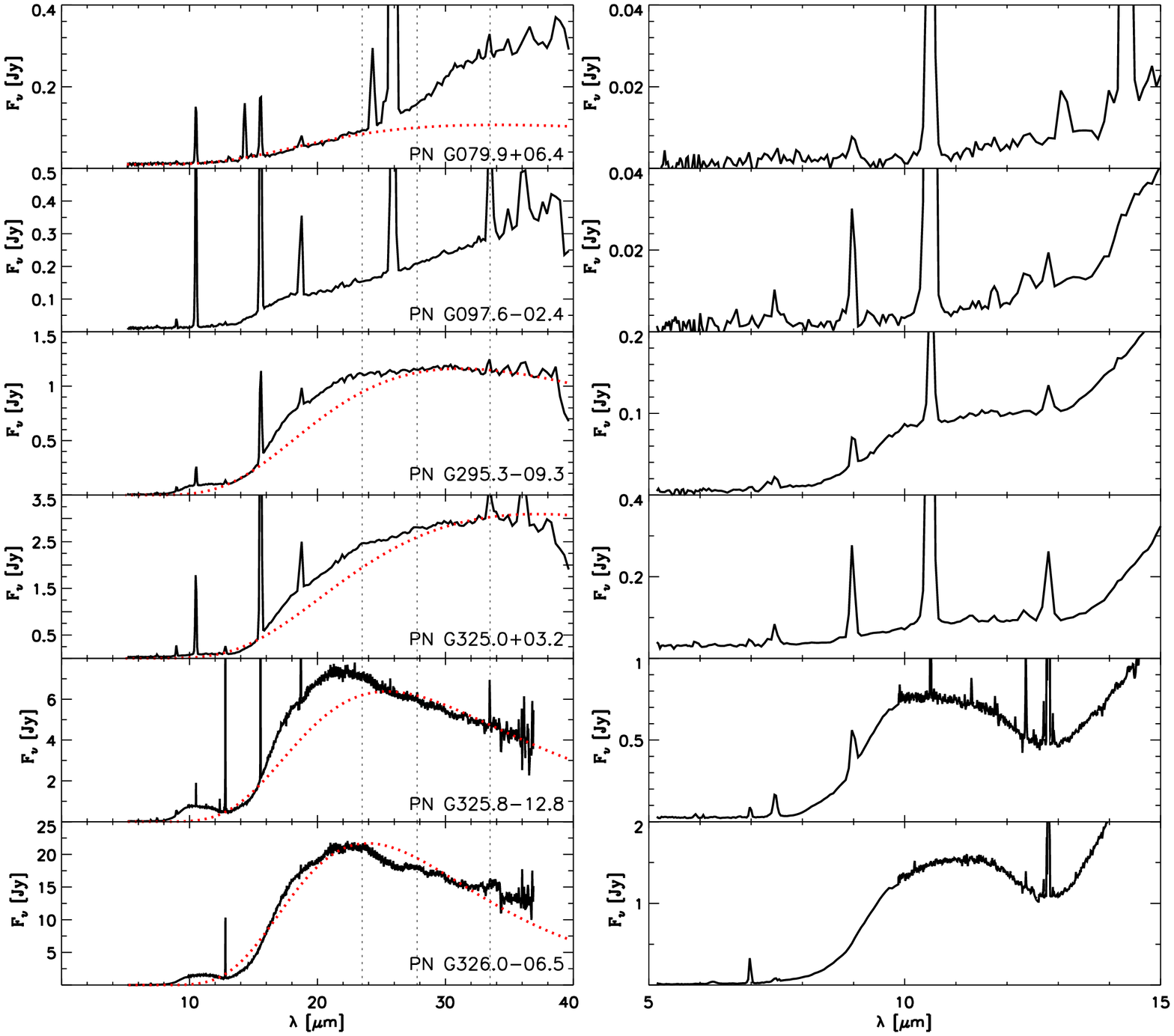}
\caption{ORD spectra, cont.}
\label{}
\end{figure}

\begin{figure}
\plotone{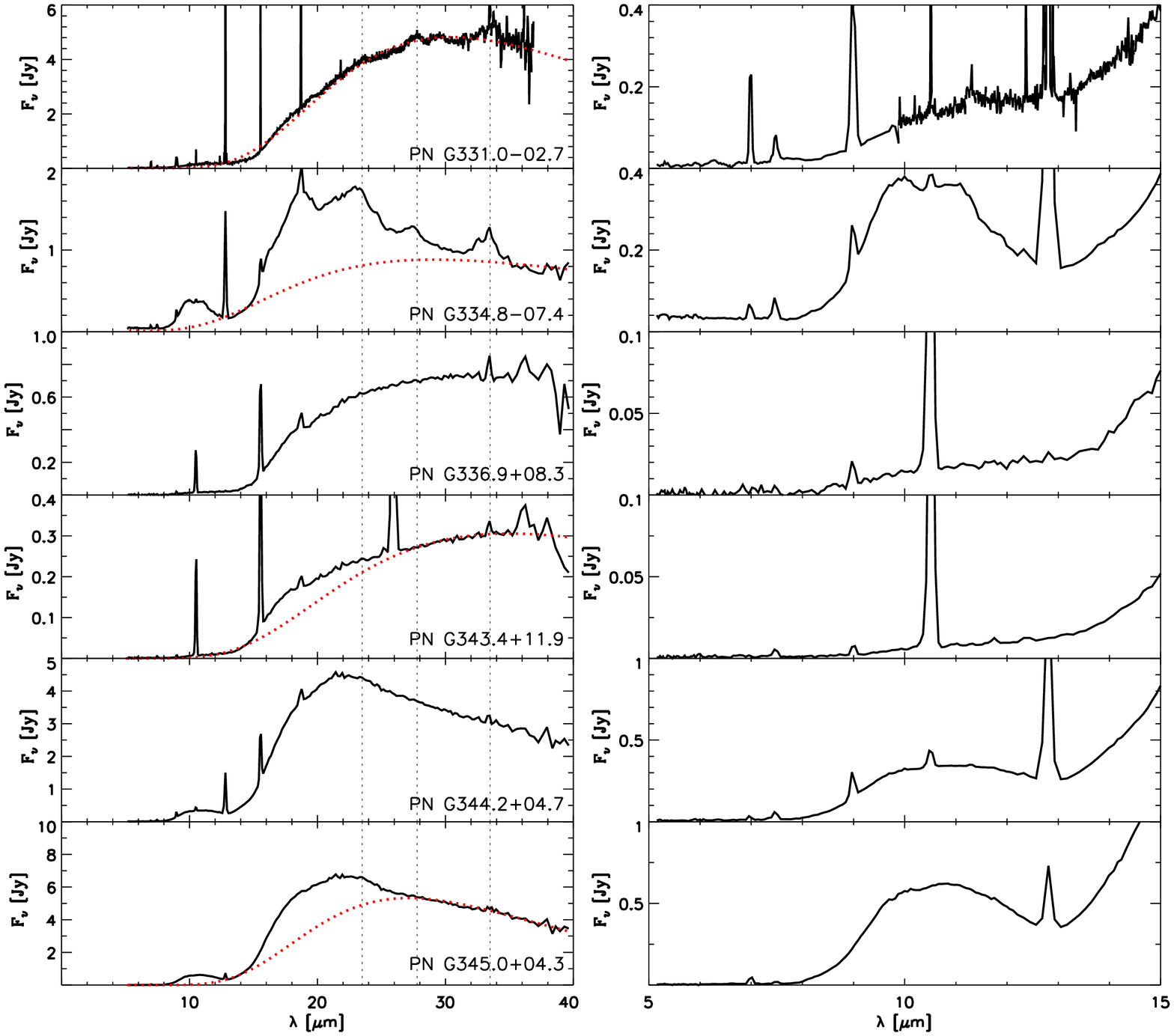}
\caption{ORD spectra, cont.}
\label{}
\end{figure}

\begin{figure}
\plotone{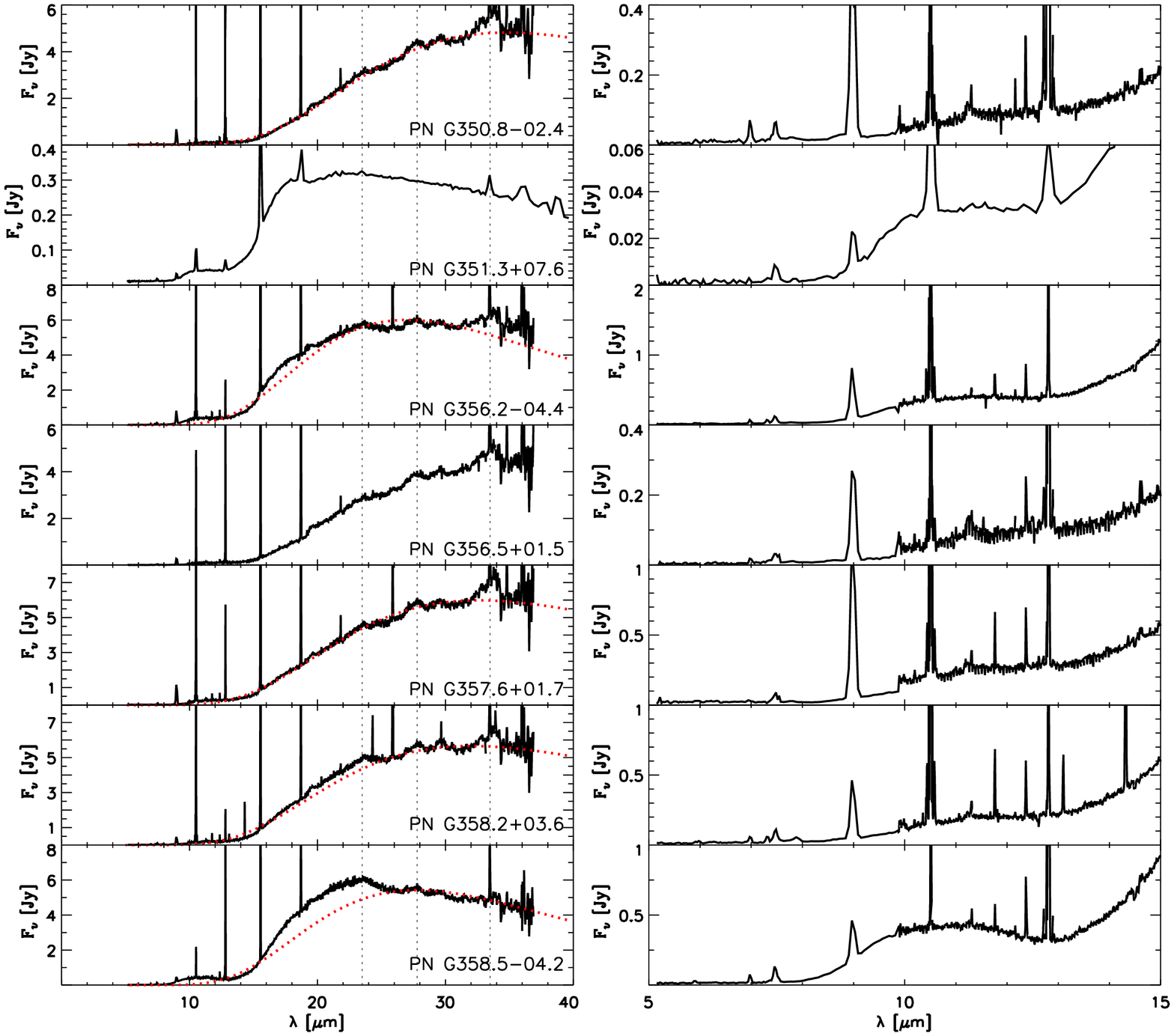}
\caption{ORD spectra, cont.}
\label{}
\end{figure}

\subsection{Segregation and evolution of dust-types}

In Figure 27  we show the distribution of dust types as projected on the Galactic plane, where X$_{\rm G}$=D cos(b) cos(l),
Y$_{\rm G}$=D cos(b) sin(l), D is the heliocentric distance, l and b are respectively the Galactic longitude and latitude.
Diamonds represent the CRD, squares the ORD, triangles the F, and circles the MCD PNe. It appears
that there is mild segregation of dust type within the Galactic distribution, in particular the MCD PNe tend
to be concentrated within the plot. To explore this segregation further in Figure 28 we show the histograms
of dust types as they are distributed in relation of their distance from the Galactic center, and we recover
the accumulation of MCD PNe with R$_{\rm G} <$ 9 kpc, unlike the other dust types. CRD and ORD PNe seem to
be rather uniformly distributed.  

Of the three (possible) halo PNe in our sample, two belong to the CRD class, both with aliphatic features, while one shows ORD features.  We do not observe a spatial segregation of the CRD the dust subtypes. On the other hand, by examining the latitude distribution of the ORD subtypes we find that the crystalline ORD PNe have typically lower Galactic latitude ($<|b|>=2.95\pm1.76$) than the amorphous ORD PNe ($<|b|>=6.33\pm3.51$). This is consistent with the former type of ORD PNe being the progeny of high mass AGB stars, while the latter are possibly the remnants of the lowest mass progenitors.

It is
worth noting that PN samples in the Galactic bulge, such as those studied by Gutenkunst et al. (2008) and
Perea-Calderon et al. (2009) show a majority of dual chemistry PNe, unlike other known samples. 
Our Galactic disk sample broadly selects against bulge PNe, but the mere definition of Galactic Bulge PN (e.g.,
Stanghellini \& Haywood 2010) is affected by the PN apparent angular radii, which can be very uncertain
for compact PNe. By using the definition of bulge PN by Stanghellini \& Haywood (2010) we determine that 26
 of our 150 PNe might belong to the bulge. We will discuss in more detail the dust properties of bulge, disk, and Magellanic Cloud 
 PNe. In the following analysis, and in all subsequent plots, we exclude potential bulge PNe from the
 Galactic sample. 

In Figure 29 we show the dust temperature plotted against the linear size of the PNe in our sample whose dust temperature
and distances are available. Different type PNe are plotted with different symbols (see Fig. 27 legend), and temperature error bars are also plotted,
while a typical error bar for the radii is given in the plot.
At zeroth order, if all PN progenitors were of identical mass and metallicity, and if the PNe
expand at constant expansion velocity, one would expect log T$_{\rm dust} \propto$ -0.4 log
R$_{\rm PN}$ if the dust grains do not evolve in size (see e.~g. Lenzuni et al. 1989). In the snapshot of evolutionary stages provided by
a sizable observational sample we see the scatter due to the different dust types, masses, metallicity, 
shell acceleration, and alternative evolutionary paths.  It appears that the
CRD PNe are concentrated toward the high dust tempeatures across a wide domain of radii. Furthermore, there is a rather tight correlation between dust
temperatures and physical radii of CRD PNe, with correlation coefficient of -0.99, which is suggestive of a 
rather narrow initial mass and metallicity range. The slope of the CRD PNe on this plane is shallower 
than that found for non-evolving grains (Lenzuni et al. 1989), suggesting  there is some type of evolution of the dust grains in CRD PNe. 

The solid line in Fig. 29 represents the least squares fit for the CRD PNe, log T${\rm _{dust}}$=-0.250 log
R$_{\rm PN}$+6.41, where the temperatures are measured in K and the nebular sizes in cm. The ORD PNe do not follow a narrow sequence in this plane. 
A possible explanation for the markedly different domains of dust temperatures in CRD and the bulk of ORD and MCD PNe 
could reside in the different heat capacity
of  dust grains as function of grain size and composition. Li \& Draine (2001) have shown that equilibrium temperatures of small carbonaceous grains are generally higher than for silicate grains in the ISM. PN dust should behave similarly to the ISM one, as the ISM is mostly composed by recycled dust from the evolution of LIMS, thus this could explain the observed distribution of temperatures of the CRD vs. the other dust type PNe. 

\begin{figure}
\plotone{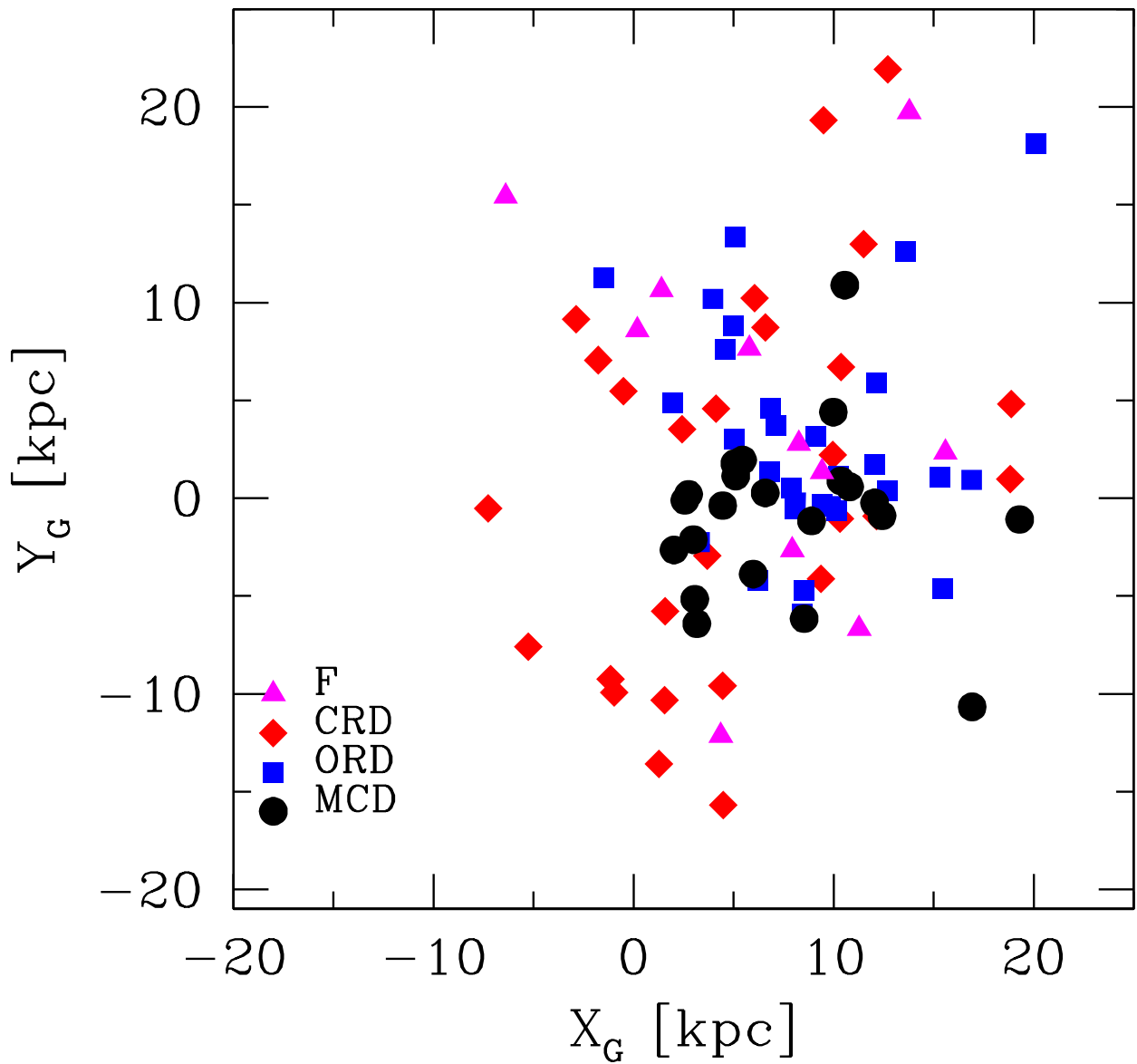}
\caption{Galactic X and Y location for F (triangles), CRD (diamonds), ORD (squares), and MCD (circles) PNe. It appears that most MCD PNe are more centrally concentrated than other types}

\label{}
\end{figure}

\begin{figure}
\plotone{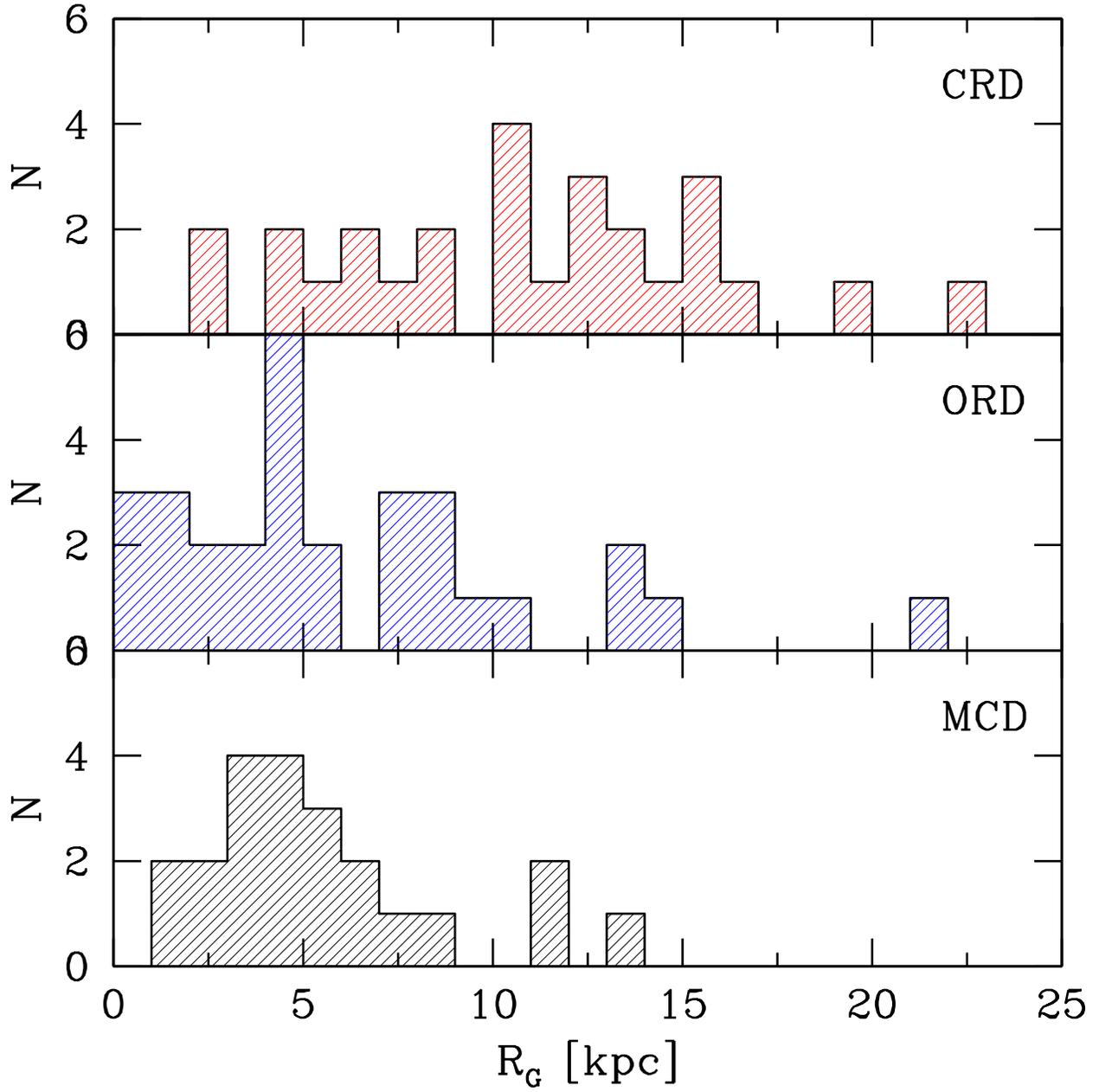}
\caption{Distribution of the distances to the galactic center for CRD (top), ORD (middle), and MCD (bottom) dust PNe.}
\label{}
\end{figure}

\begin{figure}
\plotone{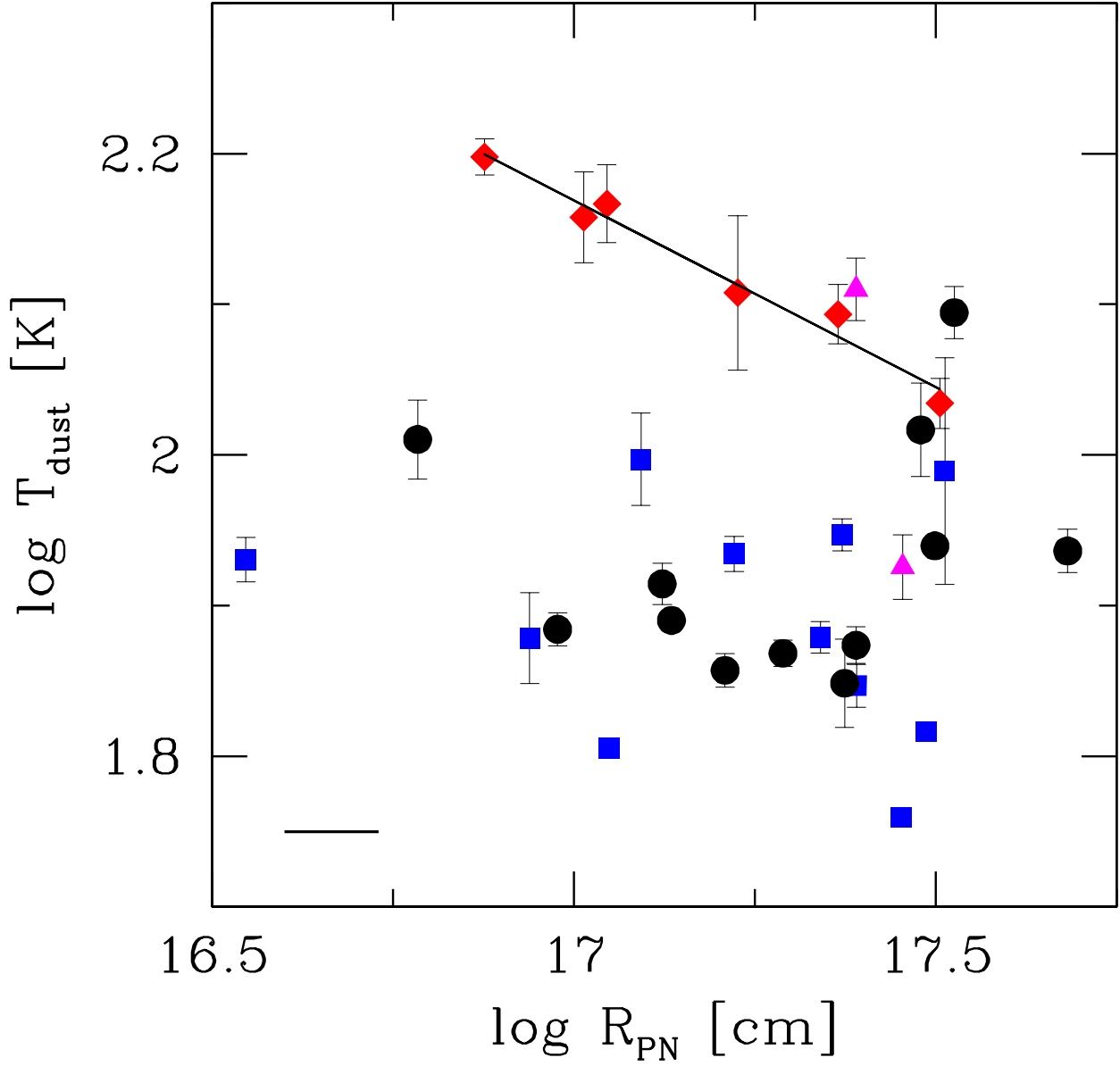}
\caption{Dust temperature plotted against linear nebular radii (in cm); Galactic disk sample. Symbols are like in Fig. 27. The solid line corresponds to the best fit to the CRD PNe. The error bars for T$-{\rm dust}$ have been derived from the continuum fits. The error bar on the left corner indicates a reasonable guess derived from the statistical distance uncertainty, corresponding to $\Delta$d=0.26.}
\label{}
\end{figure}

\begin{figure}
\plotone{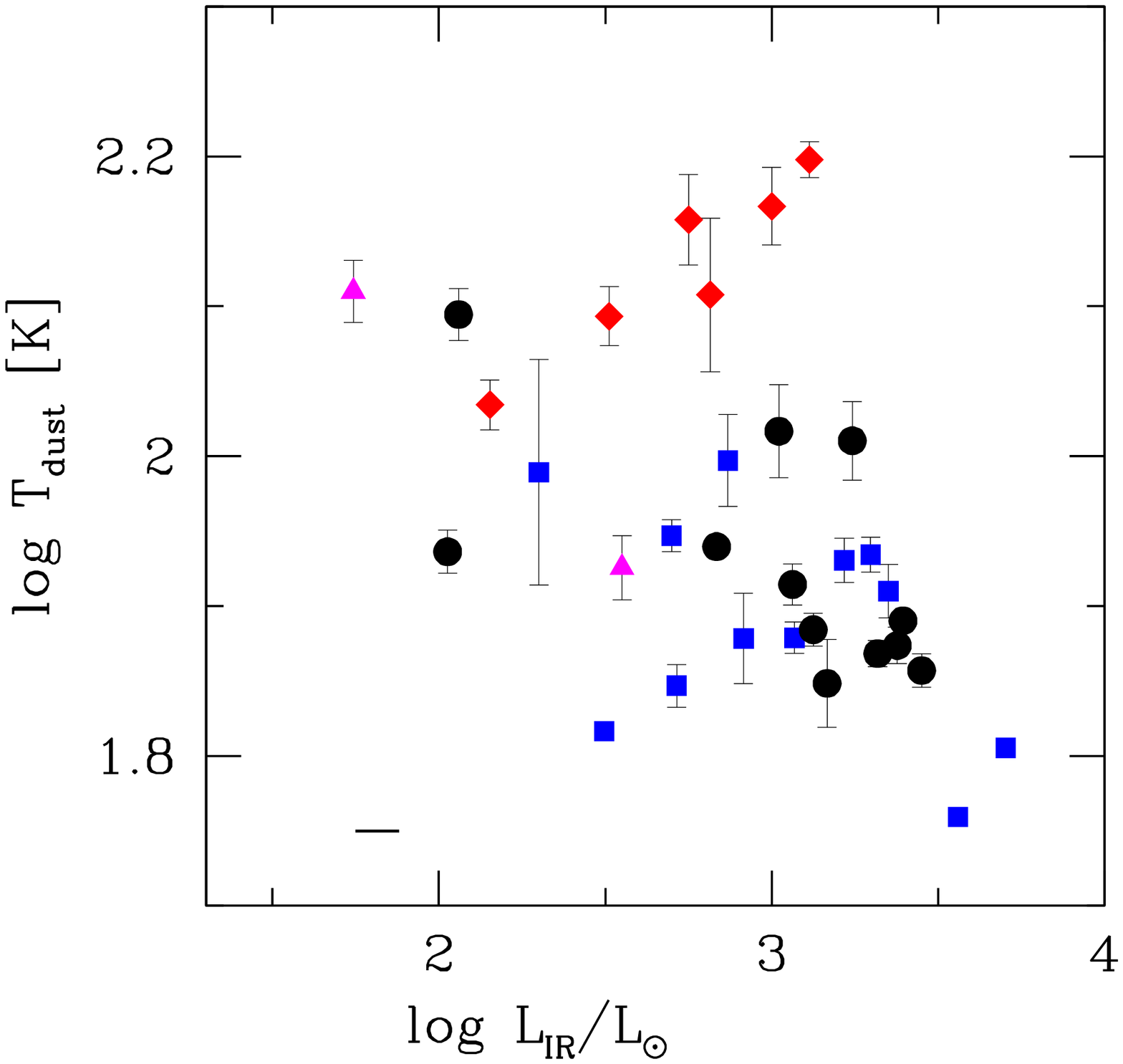}
\caption{Dust temperature vs. the integrated IR luminosity. Symbols are as in Figure 27. The error bars have the same meanings as in Fig. 29.}
\label{}
\end{figure}

In Figure 30 we show the dust temperature versus the IR luminosity (in solar units) as determined
from the black body fits of the continua. In this plot  the CRD
PNe are found in a  sequence, where PNe characterized by hot dust are more luminous in the IR. On the other
hand, ORD and MCD PNe are found scattered on this plane. If for example we look at PNe with log L$_{\rm
IR}$/L$_{\odot} >$3 we find several very hot CRD PNe, while both ORD and MCD PNe have low/intermediate dust
temperatures. This effect could be dominated by the different heat capacity of the carbon-rich and oxygen-rich 
dust grains. There could also be an effect of ORD PNe originating from a range of progenitors, including very
massive ones. One would expect to find  PNe with low- to intermediate-mass progenitors in the upper right portion 
of the plot, where typically we find CRD PNe.

In Figure 31 we plot the dust temperature against the nebular electron density.  There is no
particular segregation among the CRD and ORD or MCD PNe, apart from the high dust temperatures of CRD PNe
as noted above. 

In Figure 32 we show the IR luminosity of the PNe vs. the nebular radius. If we could assume that the expansion velocity is
uniform for all PNe, this plot would represent the evolution of IR luminosity in PNe. The expected decline of
luminosity with radius  is found as expected from a large PN population. Qualitatively, if the luminosity
of the CS is mostly radiated through the dust continuum, and if the expansion is uniform and non
accelerated, we can compare this plot with the log L/L$_{\odot}$ vs. t [yr] plot in Stanghellini and Renzini
(2000, Fig. 2 therein) and we see a  correspondence. 

It is well known that nebulae and CS should be studied together to gain insight on their
evolution and progenitor types. In the Galactic sample presented here the information about
the CS is still very scarce. Given the compactness of the PNe in our sample, their CS magnitudes are typically
unreachable from the ground, and the standard methods of temperature determinations from the Zanstra
analysis are thus not applicable. Spectral types, on the other hand, are available for only very few CS of
each dust class, thus correlations are not statistically sound. By correlating the CS spectral types
(Weidmann \& Gamen 2011) with the dust classes in this paper we noted that CS of the CRD PNe are of {\it
wels}  type in 6 of the 10 available spectra for this dust class, while most of the ORD PNe have Of
or O(H) CS (8 of the 12 available spectra). While there is certainly a mild correlation
between CS and dust class, the samples are too small to draw robust evolutionary scenarios.

For a sizable part of the sample there will be high quality {\it HST} magnitudes available in the near
future from the WFC3 images collected by us (Shaw et al., in preparation). For high excitation PNe, those
whose CS are hot enough to doubly-ionize nebular helium,  we can use the He II $\lambda$4686 flux as a
probe of stellar temperature. In Table 4, column (5) we list the $\lambda$4686 intensities found in the
literature (see selection in Stanghellini \& Haywood 2010), but the sample that includes reliable fluxes and dust
parameters (temperature, IR luminosity) is too small for meaningful comparative analysis.

We can use other criteria to characterize the hardness of the UV radiation of the CS both with IR
and optical nebular emission lines. For example, Morgan (1984) showed that a good measure of the hardness
of the ionizing stellar flux in a PN was the excitation class, or EC, derived as EC=0.45 I$_{\lambda
5007}$/I$_{\beta}$ for those PNe whose CS are not sufficiently hot to produce He II nebular
emission, and EC=5.54 (I$_{\lambda 4686}$/I$_{\beta}$ +0.78) for the high excitation PNe.  An estimate of
the hardness of the UV field could be also obtained directly from the IRS spectra, for example by using 
the [Ne III] (15.56 $\mu$m) to [Ne II] (12.81 $\mu$m) line fluxes ratio (Bernard-Salas et al. 2009).  We
list both the neon ratio described above, and calculated using our analysis of the IRS emission line spectra
(Stanghellini et al., in prep.), and the excitation constant, calculated using the parameters in
Stanghellini \& Haywood (2010), in Table 4. Naturally neither of these criteria are ideal substitutes for a
direct measurement of the stellar temperature, or even for an indirect measurement such as the stellar
temperatures derived via the Zanstra method. For example, the neon flux ratio from IR emission lines is a
reasonable stellar flux tracer only for intermediate excitation nebulae. Furthermore, the neon intensity
ratio method fails if the PN is lumpy,  non-homogeneous, or weather shocks dominate the emission.

\begin{figure}
\plotone{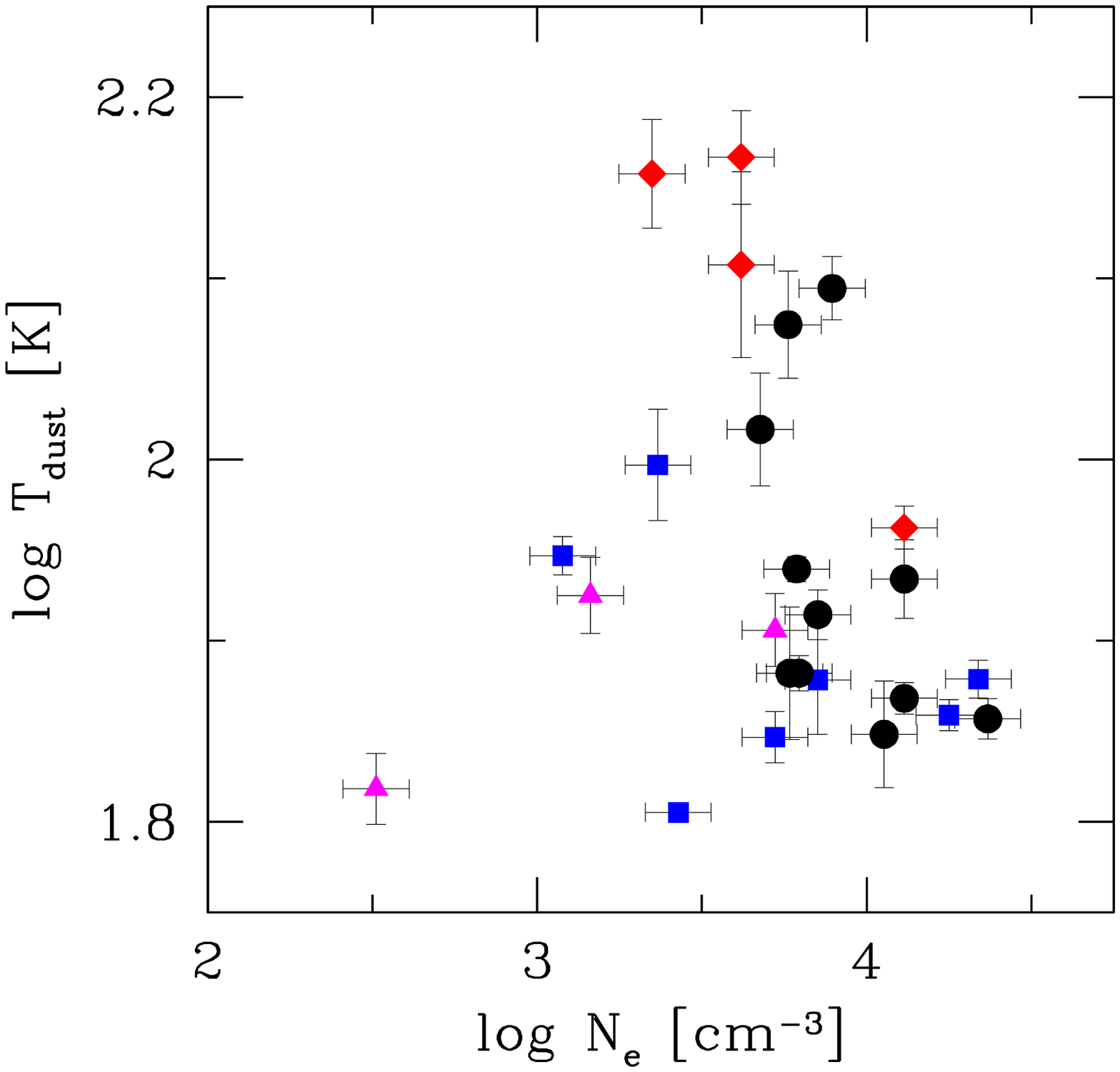}
\caption{T$_{\rm dust}$ vs. Electron density, calculated from [S~II] lines. Symbols are as in Figure 27. Temperature errorbars are as in Fig. 29, and for the electron densities it is assumed a 0.1 dex error (see text). }
\label{}
\end{figure}

\begin{figure}
\plotone{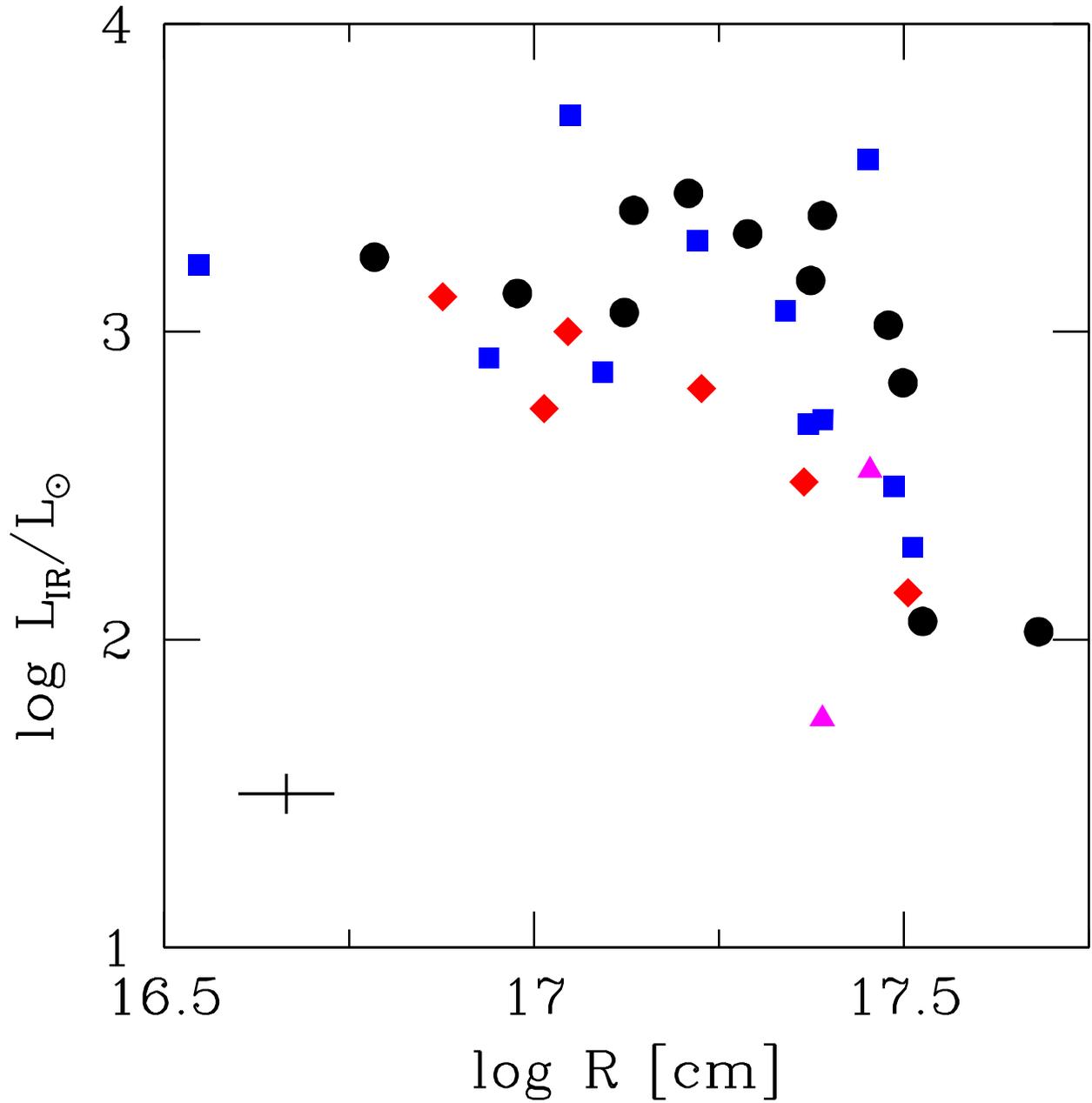}
\caption{IR luminosity against nebular radius, in pc. Symbols are as in Figure 27. The error bar on the left lower corner includes the assumption of a typical distance uncertainty from the statistical distance derivation. }
\label{}
\end{figure}

\begin{figure}
\plotone{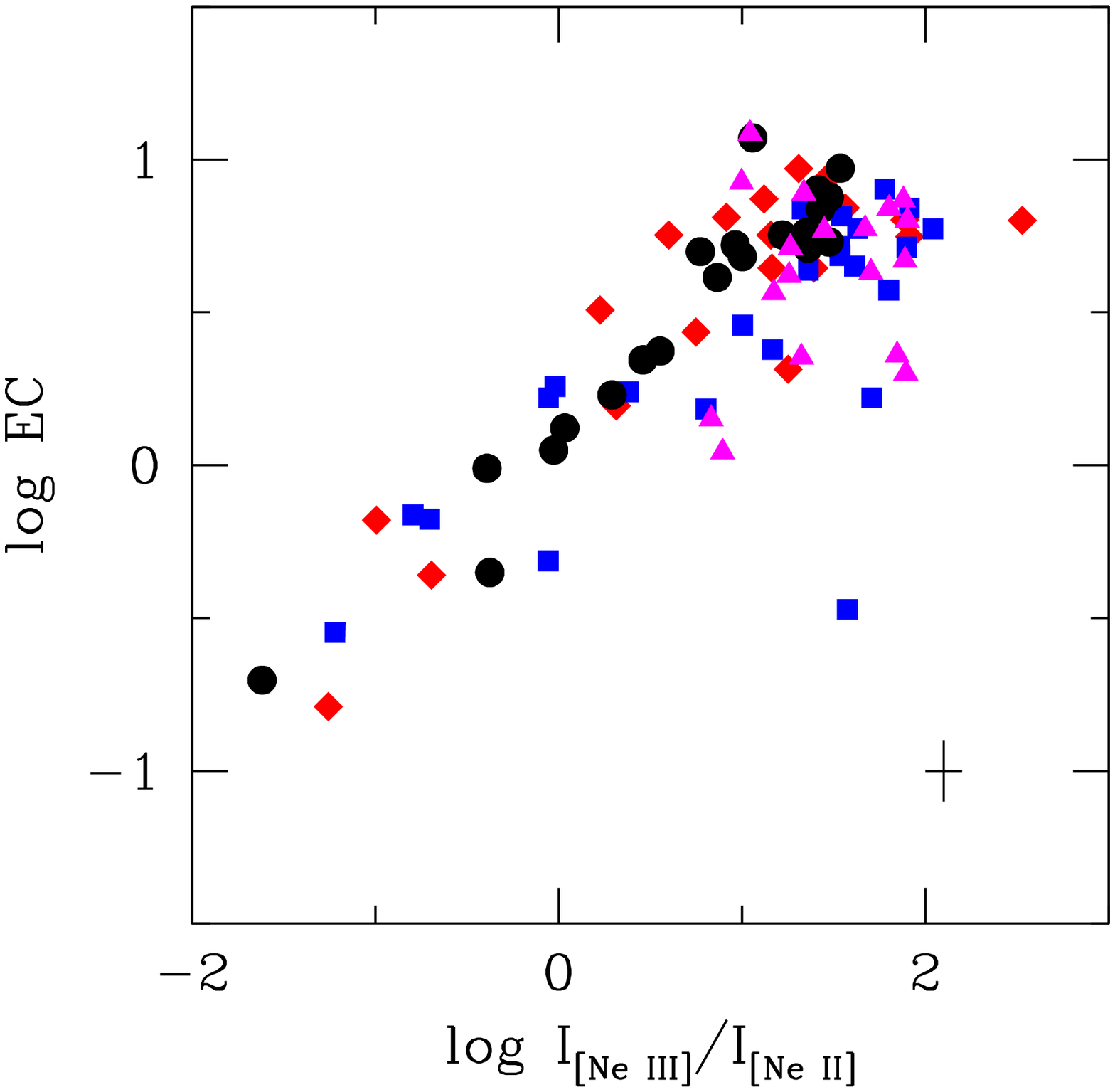}
\caption{The PN excitation class, from optical emission lines (see text),  versus the ratio of the IR line intensities of [Ne III] at $\lambda$15.56 $\mu$m and [Ne II] at $\lambda$12.81 $\mu$m . Symbols as in Figure 27. }
\label{}
\end{figure}

The [Ne III]/[Ne II] line ratio from the IRS spectra is a rough indication of the PN evolutionary stage, if the effects of shocks are
absent or moderate. In
fact, depending on where the line originate within the nebulae this ratio is tightly correlated to the
excitation class or not, and the difference is expected to be larger in PNe with complicated shapes. In
Figure 33 we show how the excitation class derived from the optical emission lines correlates quite well
with the UV flux- sensitive neon intensity ratio. We observe in the Figure that the neon ratio is a better
estimate of the PN excitation class in CRD and MCD PNe, where it correlates with the excitation constant
with correlation coefficient (between the two logarithmic values) of  0.88 and 0.94 for the CRD
and the MCD PNe respectively. On the other hand, the correlations is more scattered for ORD PNe, where the overall
correlation coefficient between the two logarithmic values is 0.75, and at high excitation the correlation
is very scattered. This finding complies with the hypothesis that in high excitation ORD PNe the neon ratio
is not a good indicator of the hardness of the stellar flux. The explanation could lie in the asymmetric
morphological type of a good fraction of ORD PNe, which was hinted in the Magellanic Cloud sample. 

The morphologies of our sample of Galactic PNe according to their dust type is illustrated in Table 5, where
we list, for each dust class, the number of PNe in each of the major morphological classes, following the
classification scheme by  Stanghellini et al. (2002). Sequentially, we give in the Table  the percentage of
PNe in the separate categories of round (R), elliptical (E), bipolar core (BC), bipolar (B) and
point-symmetric (P) PNe, as well as the cumulative classes of {\it symmetric} (round/elliptical) and {\it
asymmetric} (bipolar/bipolar core/point-symmetric) PNe. It is clear that the few F PNe in this Table do not
constitute  a statistically significative sample. Among the other dust types there are some notable
differences in the morphological distribution: CRD PNe are for the major part morphologically symmetric,
with just a quarter of the sample displaying asymmetric features; on the other hand, both ORD and MCD PNe
are almost equally distributed among the symmetric and asymmetric classes. It is also worth noting that
the peculiar class of point-symmetric PNe are well populated by MCD targets, a very interesting coincidence
of two peculiarities which is worth further study.

\begin{figure}
\plotone{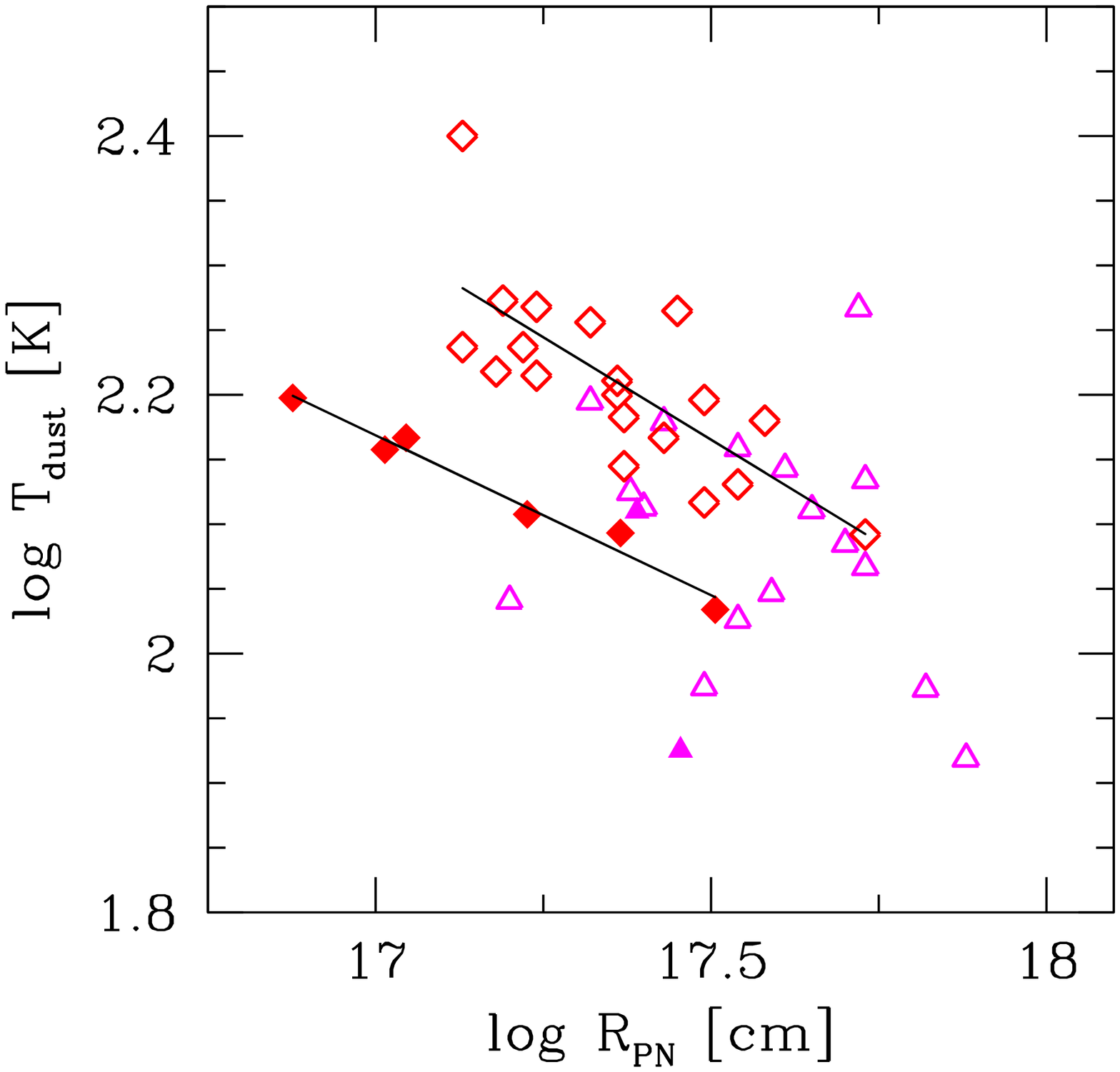}
\caption{Dust temperature plotted against linear nebular dimensions (in cm). Triangles: F, diamonds: CRD. Filled symbols, like in Figures 27 through 35, represent Galactic PNe. Open symbols represent the Magellanic Cloud sample from S07. Solid lines: best fit to the CRD PN loci in the two samples. }
\label{}
\end{figure}

\begin{figure}
\plotone{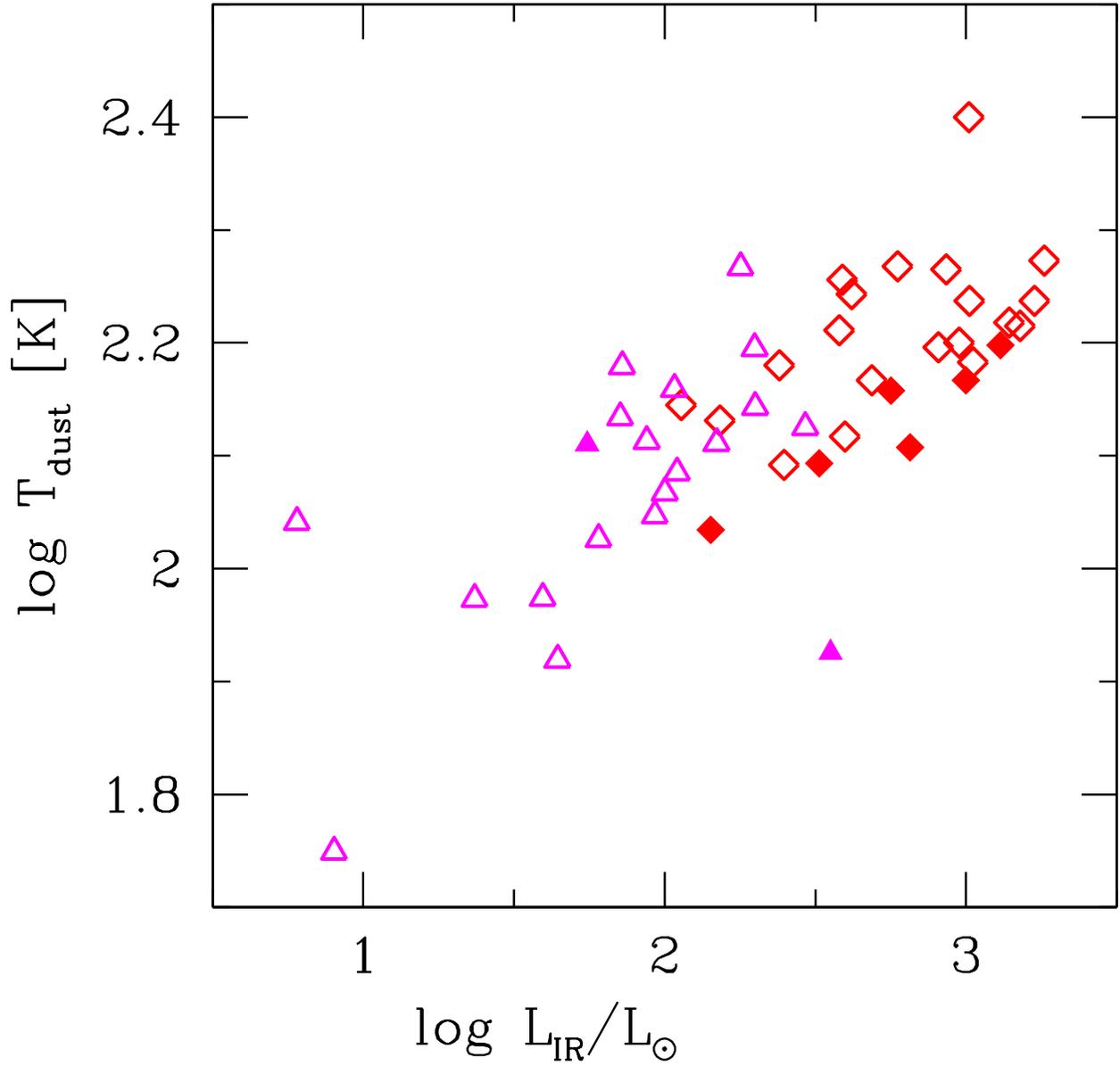}
\caption{Dust temperature vs. the integrated IR luminosity for Galactic and Magellanic PNe. Symbols are as in Figure 36. }
\label{}
\end{figure}

\begin{figure}
\plotone{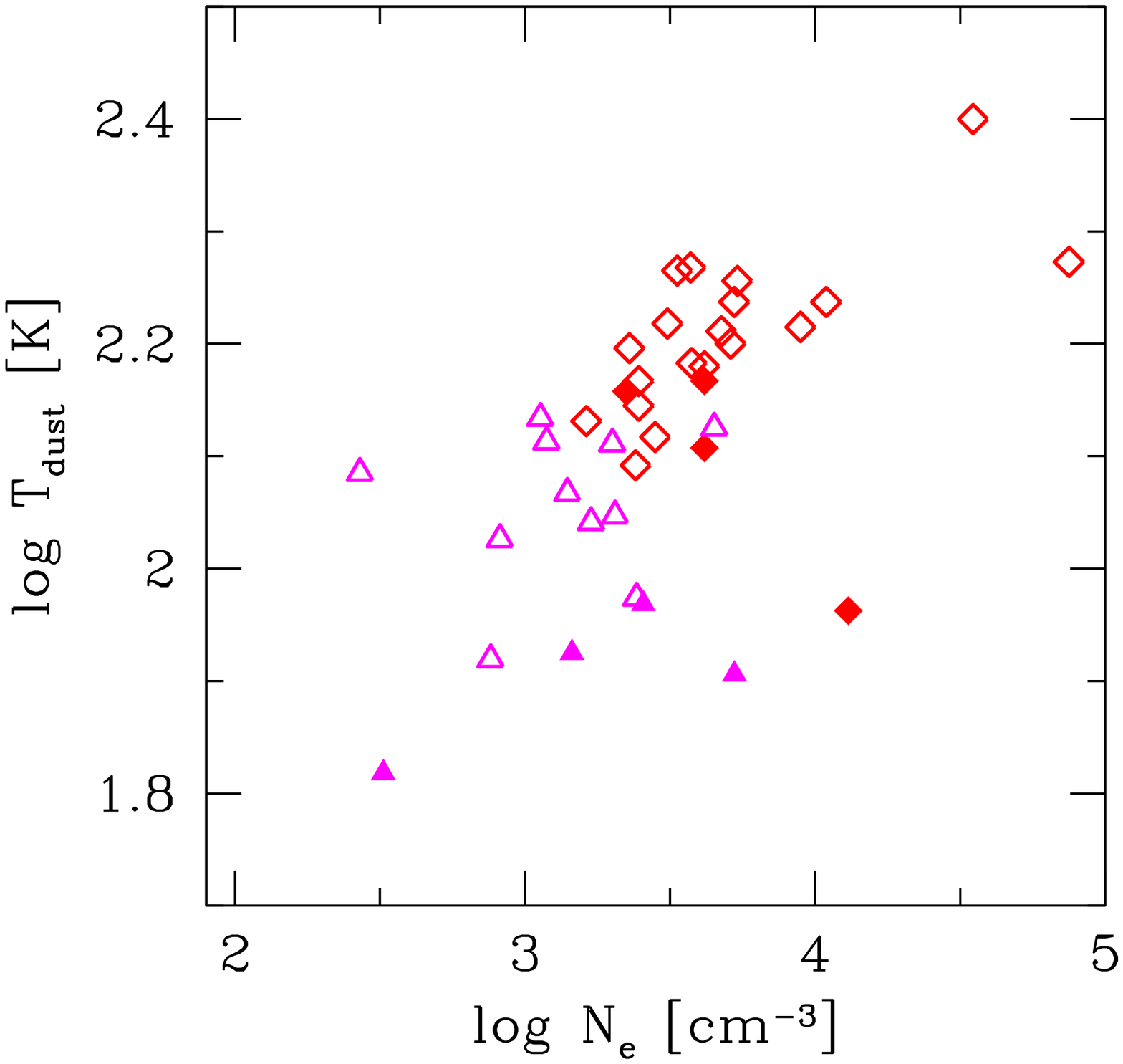}
\caption{T$_{\rm dust}$ vs.Electron density, calculated from [S~II] lines, for Galactic and Magellanic PNe. Symbols are as in Figure 36.}
\label{}
\end{figure}

\section{Comparison of the Galactic and the Magellanic Cloud PN samples}

In the previous sections we have shown the atlas of the spectra of $\sim$150 compact Galactic PNe
that we plan to study in much more detail. In this data paper we aim to show the general
characteristics of this sample, and compare it to the Magellanic Cloud PN IRS sample. 
Group and subgroup properties have been illustrated above;  here we
discuss what could be the leading evolutionary paths and the progenitor characteristics that might lead to
the observed configuration. A thorough analysis of these issues is planned for a future paper. 

The first notable comparison between the Galactic and Magellanic Cloud samples is the statistical
distribution of dust types within the different environments. In Table 6 we give, for each main dust class,
the PN frequency in the Galaxy (this paper, consisting in 150 PNe, 26 of which might be bulge PNe) and in the Magellanic Clouds,
where both the S07 and the Bernard-Salas et al. (2009) samples have been included, for a total of 66 PNe in
both Clouds. The major selection difference from the targets in the samples is the compactness of
Galactic PNe, while the Magellanic Cloud PN have been selected on PN optical brightness. Given the tight
correlation between physical size and brightness, we assume the selection criteria will not affect 
the following discussion. Of the 150 Galactic PNe only 25 do not show notable molecular/dust features
in their spectra, apart from a weak dust continuum, while in the Magellanic Cloud F PNe are  $\sim$41 percent of
the sample. Sloan et al. 's (2008) showed that the number of evolved, 
dust-free stars increases with decreasing metallicity, broadly in agreement with our findings.

Among the spectra with notable grain features, Galactic PNe are distributed almost uniformly among the
CRD, ORD, and MCD dust classes. In the Magellanic Clouds, on the other hand, most PNe are CRD, only a few
are ORD, and none have mixed-chemistry dust. In S07 we showed that the CRD PNe are also
enriched in gas-phase carbon, and they represent the progeny of the intermediate mass in the PN progenitor range. The
results of Table 6 indicate that PNe evolving from Magellanic Cloud progenitors are more likely to have
carbon-rich dust than Galactic PNe, probably indicating that the Galactic disk PN progenitors
are on average more massive than their homologous in the Clouds. Interestingly, none of the IRS observations
of Magellanic Cloud PNe (including those by different Authors) show any indication of MCD class PNe, while
this type is notable in the Galaxy. Whatever the mechanism to produce MCD PNe, it does not appear to be efficient 
at the Magellanic Cloud metallicity.

The PNe in our Galactic sample that might belong to the bulge have characteristics that are even more extremely different from the 
Magellanic Clouds than those of Galactic disk PNe. In fact, nearly all bulge PNe have dust features, and only 11$\%$ of them
have CRD types, while ORD (38$\%$), plus especially MCD (46$\%$) PNe are the majority. There seems to be continuity of dust class distribution from the 
bulge, to Galactic disk, to Magellanic Cloud PNe, where all the class frequencies change monotonically from one population to the next. 
It seems that metallicity plays a fundamental role in this picture: going from high (the bulge is $\alpha$-element enriched with respect to the Galactic disk, Zoccali et al. 2006) to low (the Magellanic Clouds are half or quarter solar in $\alpha$-element abundances) metallicity, the number of
dust-rich PNe decreases, as does the fraction of PNe with ORD and MCD, while the number of CRD PNe increases.

In Figure 34 we show the dust temperature vs. physical radii of the Galactic disk (filled symbols) and the
Magellanic Cloud (open symbols) PNe, respectively from this paper (Galactic disk only) and S07. Given that we did not find MCD
PNe in the Magellanic Clouds, and that the ORD PNe were a striking minority, here we only plot the F and CRD
PNe for a meaningful comparison of the two samples.  Distances, and consequently linear radii, are very
sound for Magellanic Cloud PNe, being based on the independently determined distances to the Clouds, while in the Galactic case
we use the statistical distance scale, which carries much larger, if reasonable, uncertainties. Nonetheless, the
qualitative behavior or the CRD PNe in the two samples is similar in this plot, indicating a possible 
similarity of dust evolution in very different metallicity environments. By comparing the linear radii
we see that we are sampling comparable expansion phases (assuming homogeneous expansion velocities), with the notable difference of  few Galactic PNe
of very small radii; this is sensible if we recall that the Galactic PN of the present sample were selected to
be compact, unevolved ones. 
In spite of the uncertainties, and the caveats associated with adopting a simple, black-body temperature for the dust continuum, we 
note that in the correlations for CRD PNe are very good if considered separately for each host galaxy, 
with the temperatures in the Magellanic Cloud sample being higher than in the Galactic sample at the same radii. 
The slope of the correlation is  slightly higher in the Magellanic Cloud
(T$_{\rm dust}\propto{\rm R_{PN}}^{-0.32}$) than in the Galactic  (T$_{\rm dust}\propto{\rm
R_{PN}}^{-0.25}$) case, in both cases the correlation is tight, with correlation coefficient of -0.99 and -0.75 for the Galactic disk and
the Magellanic Cloud PNe respectively. If we assume that the expansion rates are uniform in two samples
then we conclude that, at similar evolutionary stages,  Magellanic PNe have higher dust temperatures than
Galactic counterparts. If we assume that most CRD PNe have similar progenitor mass and metallicity (in each of the two
samples shown here) we can conclude that the cooling of PN dust depends on the metallicity of the
population.  In particular, if the PN dust grains in the Magellanic Clouds are smaller than those in Galactic PNe due to metallicity,
as seen in the ISM (e.~g., Sandstrom et al. 2011), they are expected to retain higher dust temperatures (Li \& Draine 2001), which is 
compatible with our analysis.

It is worth mentioning that the models of stellar evolution (Marigo et al. 1999; Karakas et al. 2003) predict that stars will experience the third dredge-up and the hot-bottom burning
at lower masses for lower metallicities. There will therefore be a difference in the carbon-rich star progenitor masses for Galactic and Magellanic Cloud PNe. If the correlation between carbon abundances and carbon dust, demonstrated in Magellanic Cloud PNe by S07, will hold for the Galactic PNe as well, then the sequences
of Galactic and Magellanic Cloud CRD PNe shown in Figure 34 could represent slightly different initial mass domains, which should be then taken into account in the comparison. To date, the
sample of CRD Galactic PNe with a carbon abundance determination is way too small for such an analysis to be meaningful.

In Figure 35 we show the T$_{\rm dust}$ vs/ IR luminosity for the Galactic (filled symbols) and Magellanic
Cloud (open symbols) PNe, for the F and CRD PNe. 
Both Galactic and extragalactic CRD PNe seem to populate a sequence where high luminosity correspond to high temperatures and vice versa. 
Galactic PNe have lower T$_{\rm dust}$ for similar luminosity, which we think is due to  metallicity effect on dust grain behavior, as illustrated above, 
implying the existence of smaller carbonaceous grains at low metallicity. 
Figure 36 shows the relation of dust temperature with electron density, equally dense CRD PNe seem to have hotter dust in the Magellanic Cloud sample than the Galactic sample. 

\section{Conclusions}

A large number of PNe with  Spitzer/IRS spectra are presented in this paper. With the present sample, the early properties of
ejected dust in Galactic planetary nebulae could be studied with sound statistical significance. The target
sample includes  all Galactic disk PNe with angular size smaller than $\sim$4 arcsec. A few targets turned out not to be bona fide 
PNe, leaving us with a pure sample of 150 Galactic PNe to study.
All PN spectra show dust continua and nebular emission lines. Solid state emission superimposed on
the continuum displays a variety of characteristics. The dust classification, performed similarly to what we
did for the Magellanic Cloud Spitzer/IRS spectra, consists of four major classes, determined by the type of
dust that is prevalent in the spectra, including CRD, ORD, MCD, and featureless spectra PNe. Molecular/dust emission is much more
apparent in this sample than in the Magellanic Cloud PNe, where featureless spectra with little or no continuum comprised
a large fraction of the sample. 

The most populated dust class in this Galactic sample is the ORD (or
oxygen-rich dust) class, including PNe featuring crystalline and amorphous silicates, and a few which show
both types of grains. Second most populated class is the mixed-chemistry PNe, where both carbon-rich and
oxygen-rich dust features are evident above the dust continuum. In contrast, in the Magellanic Clouds the
ORD PNe were a tiny minority, and MCD PNe were absent. 

By separating the pure disk population and the possible bulge PNe within the present sample we see that the 
bulge PNe are mostly of dual chemistry, as previously observed. Bulge, disk, and Magellanic Cloud PNe form a sequence where 
decreasing metallicity determines a decreasing fraction of ORD and MCD PNe, and an increasing fraction of CRD PNe.

Our analysis shows that CRD PNe define a rather narrow sequence when IR luminosity
and physical radii are plotted against the dust temperature, which we interpret as an 
evolutionary sequence. The consequence of these findings is that CRD PNe must originate from progenitors
with narrowly distributed mass, probably below $\sim$3 M$_{\odot}$, or, non-type I PNe (Peimbert 1978). ORD and MCD PNe might not display a similar  sequence 
because oxygen-rich grains
radiate more efficiently than their CRD counterparts and their
dust temperatures are lower than those of CRD PNe.  

When comparing the Galactic and Magellanic Cloud CRD PN, in particular their T$_{\rm dust}$ evolution and sequences, we find that dust temperatures 
are somewhat higher in the extragalactic than Galactic PNe for similar radii, which is in broad agreement with smaller carbonaceous dust
 grains observed at low metallicity in the ISM, and consequently possible smaller radiation efficiency in the Magellanic Cloud PNe than
  in their Galactic counterparts.  Another likely explanation for the observed correlation is the effect of line blanketing at lower metallicity.

The data presented here have been already analyzed for the presence of some complex molecules, such as 
C$_{60}$ (fullerene, Garc{\'{\i}}a-Hern{\'a}ndez et al. 2010). More molecular analysis is performed at
this time, as this represents the largest and most complete and homogeneous Spitzer/IRS Galactic PN data set
available. We foresee that our group and others will take full advantage of the reduced data presented here for detailed  future
analysis of dust features. 

\acknowledgements
Support for this work was provided by NASA through a grant
issued by JPL/Caltech for Spitzer Program GO 50261. We acknowledge support from the Faculty of the European Space Astronomy Centre (ESAC). 
D.A.G.H. and A.M. also acknowledge support provided by the Spanish Ministry of Science and Innovation (MICINN) under a 2008 JdC grant and under grant AYA-2007-64748. We thank an anonymous Referee for helping us improve a previous version of this paper.
E. V. acknowledges support provided by the Spanish Ministry of Science and Innovation (MICINN) under grant AYA2010-20630 and to the Marie Curie FP7-People-RG268111.

\clearpage

\begin{deluxetable}{lcccccr}

\tablecaption{Observing log}

\tablehead{
\colhead{Name}& \colhead{alias}& \colhead{$\alpha_{2000}$}& \colhead{$\delta_{2000}$}&  \colhead{IRS campaign}& \colhead{mode}& 
\colhead{t$_{\rm exp}$[s]}\\
(1)&(2)&(3)&(4)&(5)&(6)&(7)\\}

\startdata
   PN~G000.3-02.8     &      M 3-47  &   17h57m52.0s &    -30d08m15s &     010600&  0&  44.04 \\
   PN~G 000.6-01.3     &     Bl 3-15  &   17h52m44.1s &    -29d12m26s &     010600&  0&  44.04 \\
    PN~G000.6-02.3     &      H 2-32  &   17h56m32.8s &    -29d43m52s &     010600&  0&  44.04 \\
    PN~G000.8-01.5     &        Bl O  &   17h53m49.6s &    -28d59m10s &     010600&  1&  12.58 \\
    PN~G000.8-07.6     &      H 2-46  &   18h18m47.9s &    -32d00m23s &     010600&  0&  12.58 \\
    PN~G001.7-01.6     &      H 2-31  &   17h56m10.5s &    -28d19m58s &     010600&  0&  44.04 \\
    PN~G001.7-04.6     &      H 1-56  &   18h08m03.0s &    -29d50m17s &     010600&  0&  12.58 \\
   PN~G002.4-03.7     &      M 1-38  &   18h06m05.7s &    -28d40m28s &     010600&  1&  12.58 \\
    PN~G002.6-03.4     &      M 1-37  &   18h05m25.7s &    -28d22m03s &     010600&  1&  12.58 \\
    PN~G002.9-03.9     &      H 2-39  &   18h08m14.6s &    -28d31m54s &     010600&  0&  44.04 \\
    PN~G003.0-02.6     &       KFL 4  &   18h02m59.9s &    -27d46m46s &     010600&  0&  44.04 \\
  PN~G003.1+03.4     &      H 2-17  &   17h40m07.3s &    -24d25m41s &     010600&  1&  12.58 \\
    PN~G003.2-04.4     &      KFL 12  &   18h10m39.7s &    -28d25m06s &     010600&  0&  44.04 \\
    PN~G003.9-14.9     &       Hb  7  &   18h55m48.8s &    -32d21m23s &     010700&  0&  12.58 \\
    PN~G004.1-03.8     &      KFL 11  &   18h10m20.9s &    -27d22m19s &     010600&  0&  44.04 \\
   PN~G004.3-02.6     &      H 1-53  &   18h06m05.1s &    -26d35m30s &     010700&  0&  44.04 \\
    PN~G004.9-04.9     &      M 1-44  &   18h16m17.3s &    -27d04m31s &     010600&  1&  12.58 \\
    PN~G005.5+02.7     &      H 1-34  &   17h48m07.5s &    -22d46m46s &     010600&  1&  12.58 \\
    PN~G006.0+02.8     &     Th 4- 3  &   17h48m37.3s &    -22d16m48s &     010600&  1&  12.58 \\
   PN~G006.3+04.4     &      H 2-18  &   17h43m34.9s &    -21d15m44s &     010600&  0&  44.04 \\
   PN~G006.8-19.8     & Wray 16-423  &   19h22m01.0s &    -31d24m57s &     011600&  0&  44.04 \\
   PN~G006.8+02.0     &     Pe 2-10  &   17h53m43.9s &    -22d04m33s &     010600&  0&  44.04 \\
   PN~G007.5+04.3     &     Th 4- 1  &   17h46m26.9s &    -20d19m41s &     010600&  0&  44.04 \\
   PN~G008.1-04.7     &      M 2-39  &   18h22m09.6s &    -24d16m24s &     010600&  0&  12.58 \\
    PN~G008.2-04.8     &      M 2-42  &   18h22m40.5s &    -24d15m13s &     010600&  0&  12.58 \\
   PN~G008.6-02.6     &    MaC 1-11  &   18h14m58.6s &    -22d49m42s &     010600&  0&  44.04 \\
   PN~G008.6-07.0     &    He 2-406  &   18h32m01.9s &    -24d51m58s &     010600&  0&  12.58 \\
    PN~G009.3+04.1     &     Th 4- 6  &   17h50m57.1s &    -18d46m47s &     010600&  1&  12.58 \\
   PN~G010.6+03.2     &     Th 4-10  &   17h57m06.5s &    -18d06m42s &     010600&  1&  12.58 \\
   PN~G011.1+07.0     &    Sa 2-237  &   17h44m48.2s &    -15d51m04s &     010600&  0&  44.04 \\
    PN~G011.3-09.4     &      H 2-48  &   18h46m35.0s &    -23d26m47s &     010700&  1&  12.58 \\
    PN~G011.7-06.6     &      M 1-55  &   18h36m42.5s &    -21d48m58s &     010700&  1&  12.58 \\
    PN~G012.5-09.8     &      M 1-62  &   18h50m35.5s &    -22d40m01s &     010700&  0&  12.58 \\
   PN~G 012.6-02.7     &      M 1-45  &   18h23m15.2s &    -19d22m54s &     010600&  0&  12.58 \\
    PN~G014.0-05.5     &     V-V 3-5  &   18h36m40.0s &    -19d25m14s &     010600&  0&  12.58 \\
   PN~G 014.3-05.5     &     V-V 3-6  &   18h37m18.8s &    -19d08m08s &     010600&  0&  12.58 \\
    PN~G016.9-02.0     &    Sa 3-134  &   18h29m19.7s &    -15d07m39s &     010600&  1&  12.58 \\
    PN~G018.6-02.2     &      M 3-54  &   18h33m09.9s &    -13d50m12s &     010600&  0&  12.58 \\
    PN~G019.2-02.2     &      M 4-10  &   18h34m19.9s &    -13d18m17s &     010600&  0&  12.58 \\
   PN~G 019.4-05.3     &      M 1-61  &   18h45m55.0s &    -14d27m37s &     010700&  1&  12.58 \\
    PN~G019.7+03.2     &      M 3-25  &   18h15m16.9s &    -10d10m08s &     010600&  1&  12.58 \\
    PN~G019.8+05.6     &       CTS 1  &   18h07m04.5s &    -09d01m29s &     010600&  0&  44.04 \\
    PN~G023.8-01.7     &      K 3-11  &   18h40m58.5s &    -08d50m20s &     010300&  0&  12.58 \\
    PN~G023.9+01.2     &       MA 13  &   18h30m18.5s &    -07d22m22s &     010400&  0&  44.04 \\
   PN~G025.3-04.6     &      K 4- 8  &   18h54m29.2s &    -08d53m10s &     010700&  0&  12.58 \\
    PN~G026.0-01.8     &     Pe 2-15  &   18h45m36.2s &    -07d02m37s &     010700&  0&  44.04 \\
    PN~G027.6-09.6     &     IC 4846  &   19h16m16.1s &    -08d57m21s &     010400&  0&  12.58 \\
    PN~G031.0+04.1     &      K 3- 6  &   18h33m17.6s &    +00d11m47s &     010400&  1&  12.58 \\
    PN~G032.5-03.2     &      K 3-20  &   19h01m56.8s &    -01d43m43s &     011600&  0&  12.58 \\
    PN~G032.9-02.8     &      K 3-19  &   19h01m23.3s &    -01d14m04s &     010400&  0&  12.58 \\
    PN~G034.0+02.2     &      K 3-13  &   18h45m24.7s &    +02d01m23s &     010400&  1&  12.58 \\
    PN~G038.4-03.3     &      K 4-19  &   19h13m31.4s &    +03d19m22s &     010700&  0&  12.58 \\
   PN~G038.7-03.3     &      M 1-69  &   19h14m02.7s &    +03d32m03s &     010700&  0&  12.58 \\
    PN~G041.8+04.4     &      K 3-15  &   18h51m41.6s &    +09d54m53s &     011600&  1&  12.58 \\
    PN~G042.0+05.4     &      K 3-14  &   18h48m42.1s &    +10d30m14s &     010700&  0&  12.58 \\
    PN~G042.9-06.9     &    NC 6807  &   19h34m33.4s &    +05d41m04s &     010700&  1&  12.58 \\
    PN~G044.1+05.8     &      CTSS 2  &   18h50m55.8s &    +12d31m52s &     010700&  0&  44.04 \\
    PN~G045.9-01.9     &      K 3-33  &   19h22m26.8s &    +10d41m21s &     010400&  1&  12.58 \\
    PN~G048.1+01.1     &      K 3-29  &   19h15m30.7s &    +14d03m49s &     010400&  1&  12.58 \\
    PN~G048.5+04.2     &      K 4-16  &   19h04m35.3s &    +15d52m16s &     011600&  0&  12.58 \\
    PN~G051.0+02.8     &      WhMe 1  &   19h14m59.9s &    +17d22m45s &     010400&  1&  12.58 \\
    PN~G052.9-02.7     &      K 3-41  &   19h39m01.1s &    +16d25m43s &     010400&  0&  12.58 \\
   PN~G 052.9+02.7     &      K 3-31  &   19h19m02.8s &    +19d02m20s &     010400&  1&  12.58 \\
    PN~G055.1-01.8     &      K 3-43  &   19h40m10.8s &    +18d54m07s &     010400&  0&  12.58 \\
    PN~G055.5-00.5     &      M 1-71  &   19h36m27.1s &    +19d42m23s &     010400&  1&  12.58 \\
    PN~G058.9+01.3     &      K 3-40  &   19h36m09.3s &    +23d45m08s &     010400&  0&  12.58 \\
    PN~G059.4+02.3     &      K 3-37  &   19h33m34.0s &    +24d37m46s &     010400&  0&  12.58 \\
    PN~G059.9+02.0     &      K 3-39  &   19h35m54.6s &    +24d54m48s &     010400&  1&  12.58 \\
   PN~G 060.5+01.8     &    He 2-440  &   19h38m08.5s &    +25d15m40s &     010400&  1&  12.58 \\
    PN~G063.8-03.3     &      K 3-54  &   20h04m43.6s &    +25d31m38s &     010400&  0&  12.58 \\
    PN~G067.9-00.2     &      K 3-52  &   20h03m11.6s &    +30d32m34s &     010400&  1&  12.58 \\
    PN~G068.7+01.9     &      K 4-41  &   19h56m34.2s &    +32d22m12s &     010400&  1&  12.58 \\
    PN~G068.7+14.8     &      Sp 4-1  &   19h00m09.2s &    +38d26m07s &     010400&  0&  44.04 \\
    PN~G069.2+02.8     &      K 3-49  &   19h54m00.8s &    +33d22m12s &     010400&  1&  12.58 \\
    PN~G077.7+03.1     &      Kj 2  &   20h15m22.4s &    +40d34m46s &     009400&  1&  12.58 \\
   PN~G 079.9+06.4     &      K 3-56  &   20h06m21.7s &    +44d14m40s &     009400&  0&  12.58 \\
    PN~G082.5+11.3     &    NC 6833  &   19h49m12.5s &    +48d59m60s &     011400&  0&  44.04 \\
    PN~G088.7+04.6     &      K 3-78  &   20h44m45.5s &    +50d23m52s &     009400&  0&  12.58 \\
    PN~G095.2+00.7     &      K 3-62  &   21h31m50.4s &    +52d33m52s &     009400&  1&  12.58 \\
    PN~G097.6-02.4     &      M 2-50  &   21h57m06.9s &    +51d44m22s &     009400&  0&  12.58 \\
    PN~G104.1+01.0     &     Bl 2- 1  &   22h20m16.7s &    +58d14m18s &     010600&  1&  12.58 \\
    PN~G107.4-00.6     &      K 4-57  &   22h48m34.5s &    +58d29m10s &     010600&  1&  12.58 \\
    PN~G107.4-02.6     &      K 3-87  &   22h54m23.6s &    +56d43m38s &     009500&  0&  12.58 \\
    PN~G112.5-00.1     &      Kj 8  &   23h23m22.8s &    +60d59m18s &     009500&  0&  44.04 \\
    PN~G184.0-02.1     &      M 1- 5  &   05h46m50.1s &    +24d22m01s &     010700&  1&  12.58 \\
    PN~G205.8-26.7     &    MaC 2- 1  &   05h03m39.7s &    -06d04m01s &     010700&  0&  44.04 \\
    PN~G235.3-03.9     &      M 1-12  &   07h19m21.6s &    -21d43m57s &     010800&  1&  12.58 \\
    PN~G263.0-05.5     &       PB  2  &   08h20m52.6s &    -46d17m19s &     010900&  0&  12.58  \\
    PN~G264.4-12.7     &    He 2-  5  &   07h47m21.5s &    -51d09m01s &     010800&  0&  12.58 \\
    PN~G274.1+02.5     &    He 2- 34  &   09h41m13.9s &    -49d22m46s &     011400&  1&  12.58 \\
    PN~G275.3-04.7     &    He 2- 21  &   09h14m04.7s &    -55d22m28s &     011800&  0&  44.04 \\
    PN~G278.6-06.7     &    He 2- 26  &   09h19m44.4s &    -59d06m22s &     011900&  0&  12.58 \\
    PN~G285.4+01.5     &     Pe 1- 1  &   10h38m27.6s &    -56d47m05s &     009400&  1&  12.58 \\
    PN~G285.4+02.2     &     Pe 2- 7  &   10h40m56.6s &    -56d14m24s &     009400&  0&  12.58 \\
    PN~G286.0-06.5     &    He 2- 41  &   10h07m58.9s &    -63d49m52s &     011000&  0&  12.58 \\
    PN~G289.8+07.7     &    He 2- 63  &   11h24m17.1s &    -52d45m47s &     011000&  0&  44.04 \\
    PN~G294.9-04.3     &    He 2- 68  &   11h31m45.9s &    -65d58m15s &     011100&  3&  12.58 \\
    PN~G295.3-09.3     &    He 2- 62  &   11h17m14.2s &    -70d55m06s &     009400&  0&  12.58 \\
    PN~G296.3-03.0     &    He 2- 73  &   11h48m38.7s &    -65d08m39s &     011100&  3&  12.58 \\
    PN~G297.4+03.7     &    He 2- 78  &   12h09m24.5s &    -58d48m23s &     011400&  0&  12.58 \\
    PN~G300.7-02.0     &    He 2- 86  &   12h30m30.5s &    -64d52m07s &     012000&  1&  12.58 \\
    PN~G307.5-04.9     &     MyCn 18  &   13h39m34.9s &    -67d22m50s &     009400&  1&  12.58 \\
    PN~G309.0+00.8     &    He 2- 96  &   13h42m36.0s &    -61d22m28s &     009400&  1&  12.58 \\
    PN~G309.5-02.9     &    MaC 1- 2  &   13h54m37.5s &    -65d05m33s &     009400&  0&  44.04 \\
    PN~G311.1+03.4     &    He 2-101  &   13h54m55.6s &    -58d27m15s &     009400&  1&  12.58 \\
   PN~G321.3+02.8     &    He 2-115  &   15h05m16.7s &    -55d11m09s &     010600&  1&  12.58 \\
    PN~G324.2+02.5     &    He 2-125  &   15h23m33.8s &    -53d57m30s &     010600&  0&  12.58 \\
    PN~G324.8-01.1     &    He 2-133  &   15h41m58.7s &    -56d36m24s &     010600&  1&  12.58 \\
    PN~G325.0+03.2     &    He 2-129  &   15h25m41.8s &    -52d56m32s &     009500&  0&  12.58 \\
    PN~G325.8-12.8     &    He 2-182  &   16h54m35.0s &    -64d14m27s &     010600&  1&  12.58 \\
    PN~G326.0-06.5     &    He 2-151  &   16h15m42.1s &    -59d53m60s &     010600&  1&  12.58 \\
   PN~G327.1-01.8     &    He 2-140  &   15h58m08.0s &    -55d41m49s &     010600&  1&  12.58 \\
    PN~G327.8-06.1     &    He 2-158  &   16h23m36.1s &    -58d25m23s &     010600&  0&  12.58 \\
    PN~G327.9-04.3     &    He 2-147  &   16h14m00.9s &    -56d59m26s &     010600&  1&  12.58 \\
    PN~G329.4-02.7     &    He 2-149  &   16h14m27.7s &    -54d53m41s &     010600&  0&  12.58 \\
   PN~G331.0-02.7     &    He 2-157  &   16h22m14.1s &    -53d40m53s &     010600&  1&  12.58 \\
    PN~G334.8-07.4     &  SaSt 2- 12  &   17h03m13.2s &    -54d01m46s &     010600&  0&  12.58 \\
    PN~G336.3-05.6     &    He 2-186  &   16h59m45.2s &    -51d47m59s &     010600&  0&  12.58 \\
    PN~G336.9+08.3     &   StWr 4-10  &   16h02m14.5s &    -41d39m39s &     010600&  0&  12.58 \\
    PN~G340.9-04.6     &      Sa 1-5  &   17h11m36.7s &    -47d30m52s &     010600&  0&  12.58 \\
    PN~G341.5-09.1     &    He 2-248  &   17h36m16.9s &    -49d31m35s &     010600&  0&  44.04 \\
    PN~G343.4+11.9     &      H 1- 1  &   16h13m30.6s &    -34d41m41s &     010600&  0&  44.04 \\
    PN~G344.2+04.7     &      Vd 1-1  &   16h42m38.4s &    -39d00m30s &     010600&  0&  12.58 \\
    PN~G344.4-06.1     & Wray 16-278  &   17h30m15.0s &    -45d28m34s &     010600&  0&  44.04 \\
    PN~G344.4+02.8     &      Vd 1-5  &   16h51m37.8s &    -40d08m56s &     010600&  0&  44.04 \\
    PN~G344.8+03.4     &      Vd 1-3  &   16h49m41.9s &    -39d26m57s &     009500&  0&  44.04 \\
    PN~G345.0+04.3     &      Vd 1-2  &   16h46m50.6s &    -38d42m55s &     010600&  0&  12.58 \\
    PN~G348.4-04.1     &      H 1-21  &   17h32m57.7s &    -41d04m14s &     010600&  0&  12.58 \\
    PN~G348.8-09.0     &    He 2-306  &   17h56m46.2s &    -43d08m55s &     010600&  0&  12.58 \\
    PN~G350.8-02.4     &      H 1-22  &   17h32m22.0s &    -37d57m23s &     010600&  1&  12.58 \\
   PN~G351.3+07.6     &      H 1- 4  &   16h53m41.4s &    -31d46m32s &     010600&  0&  44.04 \\
    PN~G351.9-01.9     & Wray 16-286  &   17h33m00.6s &    -36d43m51s &     010600&  1&  12.58 \\
    PN~G352.6+03.0     &      H 1- 8  &   17h14m48.7s &    -33d30m43s &     010600&  0&  44.04 \\
    PN~G354.2+04.3     &      M 2-10  &   17h14m13.4s &    -31d25m36s &     010600&  0&  12.58 \\
    PN~G354.9+03.5     &     Th 3- 6  &   17h19m20.1s &    -31d12m39s &     010600&  1&  12.58 \\
    PN~G355.2-02.5     &      H 1-29  &   17h44m13.7s &    -34d17m32s &     010600&  1&  12.58 \\
    PN~G355.7-03.0     &      H 1-33  &   17h47m58.4s &    -34d13m51s &     010600&  0&  44.04 \\
   PN~G355.9+02.7     &     Th 3-10  &   17h24m40.8s &    -30d51m58s &     010600&  1&  12.58 \\
    PN~G356.2-04.4     &      Cn 2-1  &   17h54m32.9s &    -34d22m20s &     010600&  1&  12.58 \\
    PN~G356.5-03.6     &      H 2-27  &   17h51m50.5s &    -33d47m35s &     010600&  1&  12.58 \\
    PN~G356.5+01.5     &     Th 3-55  &   17h30m58.7s &    -31d01m05s &     010600&  1&  12.58 \\
    PN~G356.8+03.3     &     Th 3-12  &   17h25m06.0s &    -29d45m16s &     010600&  1&  12.58 \\
    PN~G357.1-06.1     &      M 3-50  &   18h04m15.4s &    -34d34m18s &     010600&  0&  44.04 \\
    PN~G357.1+01.9     &     Th 3-24  &   17h30m57.9s &    -30d23m05s &     010600&  0&  44.04 \\
    PN~G357.2+02.0     &      H 2-13  &   17h31m14.6s &    -30d16m21s &     010600&  0&  44.04 \\
    PN~G357.6+01.7     &      H 1-23  &   17h32m47.0s &    -30d00m16s &     010600&  1&  12.58 \\
    PN~G357.6+02.6     &      H 1-18  &   17h29m42.7s &    -29d32m49s &     010600&  1&  12.58 \\
    PN~G358.2+03.6     &      M 3-10  &   17h27m20.1s &    -28d27m50s &     010600&  1&  12.58 \\
    PN~G358.3+01.2     &        Bl B  &   17h37m06.7s &    -29d46m01s &     010600&  0&  44.04 \\
    PN~G358.5-04.2     &      H 1-46  &   17h59m02.4s &    -32d21m42s &     010600&  1&  12.58 \\
   PN~G358.5+02.9     &      Al 2-F  &   17h30m36.8s &    -28d41m48s &     010600&  0&  44.04 \\
   PN~G358.6+01.8     &      M 4- 6  &   17h35m13.9s &    -29d03m09s &     010600&  1&  12.58 \\
    PN~G358.7-02.7     &      Al 2-R  &   17h53m45.1s &    -31d31m11s &     010600&  0&  44.04 \\
    PN~G358.7+05.2     &      M 3-40  &   17h22m28.2s &    -27d08m41s &     010600&  1&  12.58 \\
    PN~G358.9-03.7     &      H 1-44  &   17h58m10.5s &    -31d42m55s &     010600&  1&  12.58 \\
    PN~G358.9+03.4     &      H 1-19  &   17h30m02.5s &    -27d59m16s &     010600&  1&  12.58 \\
    PN~G359.3+03.6     &      Al 2-E  &   17h30m14.3s &    -27d30m18s &     010600&  1&  12.58 \\
    PN~G359.4+02.3     &     Th 3-32  &   17h35m22.1s &    -28d12m54s &     010600&  0&  44.04 \\

\enddata
\end{deluxetable}

\clearpage

\begin{deluxetable}{lllr}

\tablecaption{Classification scheme}

\tablehead{ 
\colhead{class} &  \colhead{description}& \colhead{subclass}& \colhead{N$_{\rm PN}$} \\
(1)&(2)&(3)&(4)\\}

\startdata

{\bf F}&	dust continuum & & 25 (17$\%$) \\
\hline \\
{\bf CRD}& carbon-rich dust & &38 (25$\%$)\\

         &		                                                                    & aromatic& 13\\
         &                                                                                    & aliphatic& 22 \\
         &									&aromatic/aliphatic& 3\\
\hline\\
{\bf ORD}& oxygen-rich dust & &45 (30$\%$)\\
         &                                                                                        &crystalline & 16\\
       &                                                                                     &amorphous& 24   \\
           &									&crystalline/amorphous& 1\\
\hline\\
{\bf MCD} & mixed-chemistry dust & &  42 (28$\%$)\\
 
        \enddata
\end{deluxetable}

 \clearpage

\begin{deluxetable}{lcrrrrrrr}
\tablecaption{Dust classification and dust parameters}

\tablehead{
\colhead{Name}& 
\colhead{class}& 
\colhead{sub.}& 
\colhead{fit}& 
\colhead{$\alpha$}&
\colhead{F$_{\rm ref.}$}&
\colhead{T$_{\rm dust}$}&  
\colhead{log L$_{\rm IR}$}& 
 \colhead{IRE}\\
 
 \colhead{}& 
\colhead{}& 
\colhead{$^a$}& 
\colhead{$^c$}& 
\colhead{}&
\colhead{[Jy]}&
\colhead{[K]}&  
\colhead{[L$_{\odot}]$}& 
 \colhead{}\\

\colhead{(1)}&
\colhead{(2)}&
\colhead{(3)}&
\colhead{(4)}&
\colhead{(5)}&
\colhead{(6)}&
\colhead{(7)}&
\colhead{(8)}&
\colhead{(9)}\\
}

\startdata
PN~G000.3-02.8 &   F & 0 & B & 1.02 & $\dots$&  65.81 $\pm$ 2.58 & $\dots$ & $\dots$ \\
PN~G000.6-01.3 & CRD & 3 & N & $\dots$ &  $\dots$ & $\dots$& $\dots$ & $\dots$ \\
PN~G000.6-02.3 & MCD & 7 & N & $\dots$ &  $\dots$ & $\dots$& $\dots$ & $\dots$ \\
PN~G000.8-01.5 & MCD & 7 & A & 2.20 & 4.77$^e$   & 79.38 $\pm$ 1.80 & $\dots$ & $\dots$ \\
PN~G000.8-07.6 & CRD & 1$^b$ & N & $\dots$ &  $\dots$ & $\dots$& $\dots$ & $\dots$ \\
PN~G001.7-01.6 & MCD & 7 & B & 2.58 & $\dots$&  76.23 $\pm$ 5.56 & $\dots$ & $\dots$ \\
PN~G001.7-04.6 & ORD & 4 & A & 2.22 & 1.41$^d$   &  82.31 $\pm$ 1.76 & 2.920 & 2.920 \\
PN~G002.4-03.7 & MCD & 7 & A & 2.79 & 5.21$^e$   &  77.05 $\pm$ 0.90 & 2.837 & 2.837 \\
PN~G002.6-03.4 & MCD & 7 & A & 2.68 & 9.19$^e$   & 75.74 $\pm$ 0.88 & $\dots$ & $\dots$ \\
PN~G002.9-03.9 & ORD & 5$^b$ & N & $\dots$ & $\dots$&  $\dots$& $\dots$ & $\dots$ \\
PN~G003.0-02.6 & CRD & 2$^b$ & N & $\dots$ & $\dots$&  $\dots$& $\dots$ & $\dots$ \\
PN~G003.1+03.4 & MCD & 7 & A & 1.95$^e$ & 3.92 $\pm$ 0.49$^d$ & 79.12 $\pm$ 1.83 & 3.064 & 3.064 \\
PN~G003.2-04.4 & ORD & 5$^b$ & B & 2.39 & $\dots$&  77.88 $\pm$ 3.31 & 2.346 & 2.346 \\
PN~G003.9-14.9 & ORD & 5$^b$ & A & 5.52 & 1.93   &  70.27 $\pm$ 2.00 & 2.714 & 2.714 \\
PN~G004.1-03.8 & ORD & 5$^b$ & B & 0.00 & $\dots$&  126.00 $\pm$ 4.55 & 1.768 & 1.768 \\
PN~G004.3-02.6 & MCD & 7 & N & $\dots$ & $\dots$&  $\dots$& $\dots$ & $\dots$ \\
PN~G004.9-04.9 & MCD & 7 & C & 0.16 & 5.19   &  98.87 $\pm$ 3.18 & 2.669 & 2.669 \\
PN~G005.5+0.27 & MCD & 7 & C & 6.63 & 13.30$^e$   &  57.07 $\pm$ 1.13 & $\dots$ & $\dots$ \\
PN~G006.0+02.8 & MCD & 7 & A & 6.38 & 1.11$^d$   & 58.13 $\pm$ 0.74 & $\dots$ & $\dots$ \\
PN~G006.3+04.4 & ORD & 5$^b$ & N & $\dots$ &  $\dots$ & $\dots$& $\dots$ & $\dots$ \\
PN~G006.8+02.0 & MCD & 7 & N & $\dots$ & $\dots$& $\dots$& $\dots$ & $\dots$ \\
PN~G006.8-19.8 & CRD & 3 & N & $\dots$ & $\dots$&  $\dots$& $\dots$ & $\dots$ \\
PN~G007.5+04.3 & ORD & 4 & N & $\dots$ & $\dots$&  $\dots$& $\dots$ & $\dots$ \\
PN~G008.1-04.7 & ORD & 4 & N & $\dots$ & $\dots$&  $\dots$& $\dots$ & $\dots$ \\
PN~G008.2-04.8 &   F & 0 & N & $\dots$ & $\dots$&  $\dots$& $\dots$ & $\dots$ \\
PN~G008.6-02.6 & ORD & 5 & N & $\dots$ & $\dots$&  $\dots$& $\dots$ & $\dots$ \\
PN~G008.6-07.0 &   F & 0 & N & $\dots$ & $\dots$&  $\dots$& $\dots$ & $\dots$ \\
PN~G009.3+04.1 & ORD & 6$^b$ & C & 2.19 & 0.85$^e$   &  87.89 $\pm$ 1.40 & $\dots$ & $\dots$ \\
PN~G010.6+03.2 & CRD & 1 & A & 1.58 & 1.82$^d$   &  91.72 $\pm$ 2.19 & $\dots$ & $\dots$ \\
PN~G011.1+07.0 & ORD & 5 & N & $\dots$ & $\dots$&  $\dots$& $\dots$ & $\dots$ \\
PN~G011.3-09.4 & ORD & 5 & N & $\dots$ & $\dots$&  $\dots$& $\dots$ & $\dots$ \\
PN~G011.7-06.6 &  NA & 8 & N & $\dots$ & $\dots$&  $\dots$& $\dots$ & $\dots$ \\
PN~G012.5-09.8 & CRD & 2$^b$ & N & $\dots$ &  $\dots$ & $\dots$& $\dots$ & $\dots$ \\
PN~G012.6-02.7 & MCD & 7 & N & $\dots$ & $\dots$&  $\dots$& $\dots$ & $\dots$ \\
PN~G014.0-05.5 &   F & 0 & C & 1.34 & 1.91 $\pm$ 0.58$^d$ &  77.46 $\pm$ 3.42 & $\dots$ & $\dots$ \\
PN~G014.3-05.5 & CRD & 1 & N & $\dots$ & $\dots$&  $\dots$& $\dots$ & $\dots$ \\
PN~G016.9-02.0 & MCD & 7 & N & $\dots$ & $\dots$&  $\dots$& $\dots$ & $\dots$ \\
PN~G018.6-02.2 &   F & 0 & N & $\dots$ & $\dots$&  $\dots$& $\dots$ & $\dots$ \\
PN~G019.2-02.2 & ORD & 6 & B & 3.28 & $\dots$&  75.58 $\pm$ 4.54 & 2.915 & 2.915 \\
PN~G019.4-05.3 & MCD & 7 & N & $\dots$ & $\dots$& $\dots$& $\dots$ & $\dots$ \\
PN~G019.7+03.2 & MCD & 7 & A & 1.00 & 10.37 $\pm$ 2.18$^d$ & 102.40 $\pm$ 5.36 & 3.242 & 3.242 \\
PN~G019.8+05.6 &   F & 0 & N & $\dots$ & $\dots$& $\dots$ & $\dots$ & $\dots$ \\
PN~G023.8-01.7 & MCD & 7 & A & 3.36 & 8.77 $\pm$ 2.11$^d$ &  74.76 $\pm$ 1.83 & 3.377 & 3.377 \\
PN~G023.9+01.2 & MCD & 7 & A & 2.85 & 10.93 $\pm$ 1.70$^d$ &  76.23 $\pm$ 1.49 & $\dots$ & $\dots$ \\
PN~G025.3-04.6 & ORD & 5 & N & $\dots$ & $\dots$&  $\dots$& $\dots$ & $\dots$ \\
PN~G026.0-01.8 & ORD & 5$^b$ & N & $\dots$ &  $\dots$ & $\dots$& $\dots$ & $\dots$ \\
PN~G027.6-09.6 & ORD & 5 & N & $\dots$ & $\dots$ & $\dots$& $\dots$ & $\dots$ \\
PN~G031.0+04.1 & ORD & 4 & A & 4.12 & 5.50 $\pm$ 0.54$^d$ &  85.20 $\pm$ 2.53 & 3.218 & 3.218 \\
PN~G032.5-03.2 & MCD & 7 & A & 0.88 & 1.36$^d$   &  104.20 $\pm$ 6.46 & $\dots$ & $\dots$ \\
PN~G032.9-02.8 & CRD & 2 & A & 0.00 & 2.56$^e$   &  146.80 $\pm$ 7.62 & 3.000 & 3.000 \\
PN~G034.0+02.2 & ORD & 4 & A & 0.92 & 3.56$^e$   & 88.48 $\pm$ 1.87 & 2.699 & 2.699 \\
PN~G038.4-03.3 &   F & 0 & A & 1.62 & 2.14 $\pm$ 0.01$^d$ & 84.15 $\pm$ 3.54 & $\dots$ & $\dots$ \\
PN~G038.7-03.3 &   F & 0 & A & 2.11 & 3.23 $\pm$ 0.73$^d$ &  80.52 $\pm$ 3.25 & $\dots$ & $\dots$ \\
PN~G041.8+04.4 & CRD & 3 & A & 1.50 & 0.52$^d$   & 127.20 $\pm$ 5.24 & $\dots$ & $\dots$ \\
PN~G042.0+05.4 & ORD & 5 & N & $\dots$ & $\dots$&  $\dots$& $\dots$ & $\dots$ \\
PN~G042.9-06.9 & ORD & 6 & A & 7.16 & 0.65$^d$   & 63.87 $\pm$ 0.66 & 3.703 & 3.703 \\
PN~G044.1+05.8 &  NA & 8 & N & $\dots$ & $\dots$&  $\dots$& $\dots$ & $\dots$ \\
PN~G045.9-01.9 & MCD & 7 & A & 2.98 & 3.46$^e$   & 77.65 $\pm$ 0.79 & 3.394 & 3.394 \\
PN~G048.1+01.1 & CRD & 1 & N & $\dots$ & $\dots$&  $\dots$& $\dots$ & $\dots$ \\
PN~G048.5+04.2 & CRD & 2$^b$ & B & 0.00 & $\dots$&  108.20 $\pm$ 3.60 & 2.153 & 2.153 \\
PN~G051.0+02.8 & ORD & 4 & A & 0.00 & 6.39 $\pm$ 0.54$^d$ &  224.60 $\pm$ 12.59 & $\dots$ & $\dots$ \\
PN~G052.9+02.7 & CRD & 2 & N & $\dots$ & $\dots$&  $\dots$& $\dots$ & $\dots$ \\
PN~G052.9-02.7 &   F & 0 & N & $\dots$ & $\dots$&  $\dots$& $\dots$ & $\dots$ \\
PN~G055.1-01.8 &   F & 0 & N & $\dots$ & $\dots$& $\dots$& $\dots$ & $\dots$ \\
PN~G055.5-00.5 & CRD & 2 & N & $\dots$ & $\dots$&  $\dots$& $\dots$ & $\dots$ \\
PN~G058.9+01.3 & ORD & 4 & N & $\dots$ & $\dots$&  $\dots$& $\dots$ & $\dots$ \\
PN~G059.4+02.3 & CRD & 2 & A & 0.00 & 0.80$^d$   &  124.00 $\pm$ 4.89 & 2.512 & 2.512 \\
PN~G059.9+02.0 & CRD & 2 & N & $\dots$ & $\dots$& $\dots$& $\dots$ & $\dots$ \\
PN~G060.5+01.8 & ORD & 4 & A & 2.31 & 2.70 $\pm$ 0.16$^d$ &  85.94 $\pm$ 2.01 & 3.297 & 3.297 \\
PN~G063.8-03.3 & CRD & 2 & N & $\dots$ & $\dots$&  $\dots$& $\dots$ & $\dots$ \\
PN~G067.9-00.2 & ORD & 4 & C & 0.00 & 35.94 $\pm$ 2.70$^d$ &  124.50 $\pm$ 1.93 & 3.164 & 3.164 \\
PN~G068.7+01.9 & ORD & 4 & A & 1.46 & 2.44 $\pm$ 0.10$^d$ & 86.26 $\pm$ 1.71 & 2.805 & 2.805 \\
PN~G068.7+14.8 & CRD & 1 & N & $\dots$ & $\dots$& $\dots$ &  $\dots$ & $\dots$ \\
PN~G069.2+02.8 & ORD & 5 & A & 3.03 & 1.10 $\pm$ 0.36$^d$ &  81.28 $\pm$ 2.89 & 3.351 & 3.351 \\
PN~G077.7+03.1 &   F & 0 & N & $\dots$ & $\dots$& $\dots$ &  $\dots$ & $\dots$ \\
PN~G079.9+06.4 & ORD & 5$^b$ & B & 0.70 & $\dots$&  118.00 $\pm$ 4.39 & $\dots$ & $\dots$ \\
PN~G082.5+11.3 &   F & 0 & N & $\dots$ & $\dots$& $\dots$ & $\dots$&  $\dots$ \\
PN~G088.7+04.6 &   F & 0 & A & 1.59 & 1.99 $\pm$ 0.36$^d$ & 84.24 $\pm$ 3.63 & 2.550 & 2.550 \\
PN~G095.2+00.7 & CRD & 2 & N & $\dots$ & $\dots$& $\dots$ &  $\dots$ & $\dots$ \\
PN~G097.6-02.4 & ORD & 5$^b$ & N & $\dots$ & $\dots$&  $\dots$& $\dots$ & $\dots$ \\
PN~G104.1+01.0 & CRD & 1 & A & 0.00 & 2.66 $\pm$ 0.10$^d$ &  157.70 $\pm$ 3.83 & 3.114 & 3.114 \\
PN~G107.4-00.6 &  NA & 8 & N & $\dots$ & $\dots$&  $\dots$& $\dots$ & $\dots$ \\
PN~G107.4-02.6 & CRD & 2$^b$ & N & $\dots$ & $\dots$ & $\dots$& $\dots$ & $\dots$ \\
PN~G112.5-00.1 &   F & 0 & N & $\dots$ & $\dots$&  $\dots$& $\dots$ & $\dots$ \\
PN~G184.0-02.1 & CRD & 2 & N & $\dots$ & $\dots$& $\dots$& $\dots$ & $\dots$ \\
PN~G205.8-26.7 & CRD & 2 & B & 1.02 & $\dots$&  122.30 $\pm$ 13.89 & $\dots$ & $\dots$ \\
PN~G235.3-03.9 & CRD & 2 & N & $\dots$ & $\dots$ $\dots$& $\dots$ & $\dots$ \\
PN~G263.0-05.5 & CRD & 2 & N & $\dots$ & $\dots$& $\dots$& $\dots$ & $\dots$ \\
PN~G264.4-12.7 & CRD & 2 & N & $\dots$ & $\dots$& $\dots$& $\dots$ & $\dots$ \\
PN~G274.1+02.5 &  NA & 8 & N & $\dots$ & $\dots$& $\dots$& $\dots$ & $\dots$ \\
PN~G275.3-04.7 & CRD & 2 & N & $\dots$ & $\dots$&  $\dots$& $\dots$ & $\dots$ \\
PN~G278.6-06.7 & CRD & 2 & A & 0.53 & 1.52$^e$   &  128.10 $\pm$ 13.15 & 2.815 & 2.815 \\
PN~G285.4+01.5 & CRD & 1 & N & $\dots$ &  $\dots$ & $\dots$& $\dots$ & $\dots$ \\
PN~G285.4+02.2 &   F & 0 & N & $\dots$ & $\dots$ & $\dots$& $\dots$ & $\dots$ \\
PN~G286.0-06.5 & CRD & 2 & A & 0.00 & 0.57$^e$   &  143.80 $\pm$ 8.65 & 2.750 & 2.750 \\
PN~G289.8+07.7 &   F & 0 & B & 0.00 & $\dots$&  128.80 $\pm$ 5.31 & 1.743 & 1.743 \\
PN~G294.9-04.3 & CRD & 2 & N & $\dots$ &  $\dots$ & $\dots$& $\dots$ & $\dots$ \\
PN~G295.3-09.3 & ORD & 5 & A & 1.78 & 0.87$^e$   &  97.33 $\pm$ 7.08 & $\dots$ & $\dots$ \\
PN~G296.3-03.0 & MCD & 7 & A & 2.27 & 5.00 $\pm$ 0.43$^d$ &  82.11 $\pm$ 2.26 & 3.063 & 3.063 \\
PN~G297.4+03.7 & CRD & 1 & N & $\dots$ & $\dots$& $\dots$& $\dots$ & $\dots$ \\
PN~G300.7-02.0 & MCD & 7 & A & 3.65 & 18.48 $\pm$ 1.11$^d$ & 71.93 $\pm$ 1.62 & 3.450 & 3.450 \\
PN~G307.5-04.9 & MCD & 7 & A & 2.42 & 17.17 $\pm$ 0.78$^d$ & 87.02 $\pm$ 1.20 & 2.834 & 2.834 \\
PN~G309.0+00.8 & MCD & 7 & C & 2.07 & 18.27 $\pm$ 1.43$^d$ & 90.02 $\pm$ 2.37 & $\dots$ & $\dots$ \\
PN~G309.5-02.9 & CRD & 1$^b$ & N & $\dots$ & $\dots$&  $\dots$& $\dots$ & $\dots$ \\
PN~G311.1+03.4 &  NA & 8 & N & $\dots$ & $\dots$& $\dots$ &  $\dots$ & $\dots$ \\
PN~G321.3+02.8 & CRD & 2 & N & $\dots$ & $\dots$& $\dots$ & $\dots$ & $\dots$ \\
PN~G324.2+02.5 & MCD & 7 & A & 3.26 & 2.86 $\pm$ 0.56$^d$ &  70.53 $\pm$ 4.14 & 3.167 & 3.167 \\
PN~G324.8-01.1 & MCD & 7 & A & 3.12 & 19.45 $\pm$ 2.31 &  76.59 $\pm$ 1.67 & 3.125 & 3.125 \\
PN~G325.0+03.2 & ORD & 5 & A & 0.98 & 1.33$^d$   &  99.32 $\pm$ 6.09 & 2.868 & 2.868 \\
PN~G325.8-12.8 & ORD & 5 & A & 5.63 & 0.72$^d$   &  65.55 $\pm$ 0.59 & 2.496 & 2.496 \\
PN~G326.0-06.5 & ORD & 5 & A & 7.46 & 4.70 $\pm$ 0.93$^d$ &  57.45 $\pm$ 0.39 & 3.560 & 3.560 \\
PN~G327.1-01.8 & MCD & 7 & A & 3.12 & 11.77 $\pm$ 3.41$^d$ &  73.83 $\pm$ 1.28 & 3.318 & 3.318 \\
PN~G327.8-06.1 & MCD & 7 & A & 0.68 & 1.12$^d$   &  103.90 $\pm$ 6.45 & 3.022 & 3.022 \\
PN~G327.9-04.3 &  NA & 8 & N & $\dots$ & $\dots$&  $\dots$& $\dots$ & $\dots$ \\
PN~G329.4-02.7 &   F & 0 & N & $\dots$ & $\dots$&  $\dots$& $\dots$ & $\dots$ \\
PN~G331.0-02.7 & ORD & 4 & A & 3.26 & 2.61$^d$   & 75.65 $\pm$ 1.59 & 3.068 & 3.068 \\
PN~G334.8-07.4 & ORD & 4 & A & 0.81 & 0.41$^e$   &  133.10 $\pm$ 10.75 & $\dots$ & $\dots$ \\
PN~G336.3-05.6 & CRD & 1 & N & $\dots$ & $\dots$&  $\dots$& $\dots$ & $\dots$ \\
PN~G336.9+08.3 & ORD & 5 & N & $\dots$ & $\dots$&  $\dots$& $\dots$ & $\dots$ \\
PN~G340.9-04.6 &   F & 0 & N & $\dots$ & $\dots$&  $\dots$& $\dots$ & $\dots$ \\
PN~G341.5-09.1 &   F & 0 & N & $\dots$ & $\dots$&  $\dots$& $\dots$ & $\dots$ \\
PN~G343.4+11.9 & ORD & 5 & B & 1.28 & $\dots$&  97.56 $\pm$ 14.71 & 2.301 & 2.301 \\
PN~G344.2+04.7 & ORD & 5 & N & $\dots$ & $\dots$&  $\dots$& $\dots$ & $\dots$ \\
PN~G344.4+02.8 & CRD & 2$^b$ & N & $\dots$ & $\dots$ & $\dots$& $\dots$ & $\dots$ \\
PN~G344.8+03.4 & CRD & 1 & N & $\dots$ & $\dots$&  $\dots$& $\dots$ & $\dots$ \\
PN~G345.0+04.3 & ORD & 5 & A & 4.62 & 0.74$^d$   &  69.58 $\pm$ 3.29 & $\dots$ & $\dots$ \\
PN~G348.4-04.1 & MCD & 7 & N & $\dots$ & $\dots$&  $\dots$& $\dots$ & $\dots$ \\
PN~G348.8-09.0 &   F & 0 & A & 1.53 & 2.16$^d$   & 88.05 $\pm$ 2.17 & $\dots$ & $\dots$ \\
PN~G350.8-02.4 & ORD & 4 & A & 2.74 & 1.60$^d$   & 72.27 $\pm$ 1.24 & $\dots$ & $\dots$ \\
PN~G351.3+07.6 & ORD & 5 & N & $\dots$ & $\dots$&  $\dots$& $\dots$ & $\dots$ \\
PN~G351.9-01.9 & MCD & 7 & N & $\dots$ & $\dots$&  $\dots$& $\dots$ & $\dots$ \\
PN~G352.6+03.0 & MCD & 7 & B & 2.38 & $\dots$&  77.00 $\pm$ 5.25 & 2.988 & 2.988 \\
PN~G354.2+04.3 & CRD & 1 & N & $\dots$ & $\dots$&  $\dots$& $\dots$ & $\dots$ \\
PN~G354.9+03.5 & MCD & 7 & N & $\dots$ & $\dots$&  $\dots$& $\dots$ & $\dots$ \\
PN~G355.2-02.5 & MCD & 7 & A & 0.22 & 1.53$^e$   & 124.30 $\pm$ 4.33 & 2.059 & 2.059 \\
PN~G355.7-03.0 & CRD & 1 & B & 1.08 & $\dots$&  97.23 $\pm$ 6.07 & 2.770 & 2.770 \\
PN~G355.9+02.7 & MCD & 7 & C & 0.70 & 7.81 $\pm$ 2.33$^d$ &  96.69 $\pm$ 5.66 & 3.179 & 3.179 \\
PN~G356.2-04.4 & ORD & 6 & A & 4.17 & 2.98 $\pm$ 0.14$^d$ &  74.39 $\pm$ 1.81 & 3.062 & 3.062 \\
PN~G356.5+01.5 & ORD & 4 & C & 1.28 & 7.21$^e$   &  88.89 $\pm$ 1.78 & 2.446 & 2.446 \\
PN~G356.5-03.6 &   F & 0 & C & 0.46 & 2.03$^d$   &  92.95 $\pm$ 1.27 & $\dots$ & $\dots$ \\
PN~G356.8+03.3 & MCD & 7 & N & $\dots$ & $\dots$&  $\dots$& $\dots$ & $\dots$ \\
PN~G357.1+01.9 &   F & 0 & N & $\dots$ & $\dots$&  $\dots$& $\dots$ & $\dots$ \\
PN~G357.1-06.1 &   F & 0 & N & $\dots$ & $\dots$& $\dots$& $\dots$ & $\dots$ \\
PN~G357.2+02.0 & MCD & 7$^b$ & B & 0.19 & $\dots$& 118.70 $\pm$ 7.02 & $\dots$ & $\dots$ \\
PN~G357.6+01.7 & ORD & 4 & A & 2.29 & 5.35$^d$   &  83.75 $\pm$ 2.01 & 3.153 & 3.153 \\
PN~G357.6+02.6 & MCD & 7 & A & 1.86 & 6.59$^e$   &  86.34 $\pm$ 2.49 & 2.026 & 2.026 \\
PN~G358.2+03.6 & ORD & 4 & A & 2.08 & 2.65$^d$   & 89.06 $\pm$ 2.60 & 3.116 & 3.116 \\
PN~G358.3+01.2 & MCD & 7 & N & $\dots$ & $\dots$&  $\dots$& $\dots$ & $\dots$ \\
PN~G358.5+02.9 &   F & 0 & B & 0.97 & $\dots$&  110.80 $\pm$ 5.39 & $\dots$ & $\dots$ \\
PN~G358.5-04.2 & ORD & 6 & A & 3.86 & 2.17 $\pm$ 0.35$^d$ &  75.65 $\pm$ 1.61 & 1.616 & 1.616 \\
PN~G358.6+01.8 & MCD & 7 & A & 1.91$^e$ & 4.28   &  90.02 $\pm$ 2.42 & $\dots$ & $\dots$ \\
PN~G358.7+05.2 & MCD & 7 & C & $\dots$ & $\dots$   &73.37 $\pm$ 1.94 & $\dots$ & $\dots$ \\
PN~G358.7-02.7 &  NA & 8 & N & $\dots$ & $\dots$& $\dots$& $\dots$ & $\dots$ \\
PN~G358.9+03.4 & MCD & 7 & A & 2.52 & 5.17 $\pm$ 1.02$^d$ & 79.98 $\pm$ 1.93 & 2.761 & 2.761 \\
PN~G358.9-03.7 & MCD & 7 & N & $\dots$ & $\dots$& $\dots$ &  $\dots$ & $\dots$ \\
PN~G359.3+03.6 &   F & 0 & A & 0.99 & 1.23$^d$   &  86.01 $\pm$ 1.79 & $\dots$ & $\dots$ \\
PN~G359.4+02.3 & MCD & 7 & A & 2.14 & 3.98$^d$   & 85.92 $\pm$ 3.75 & $\dots$ & $\dots$ \\

\enddata
\\
$^a$ Subclasses: 0=featureless, 1=aromatic, 2=aliphatic, 3=aromatic/aliphatic, 4=crystalline, 5=amorphous, 6= crystalline/amorphous, 7=MCD, 8=other;
$^b$: Class or subclass uncertain.
$^c$ Black body fit type: NC= not converging (see text); NA: not available (not a PN); A=good fit, excellent F$_{\rm ref}$ (60 or 65 $\mu$m) constraints; B=good fit, no constraints available; C=good fit, poor constraints (not used in the plots). $^d$ F$_{\rm ref}$=F$_{65 \mu m}$ from Akari. $^e$ F$_{\rm ref}$=F$_{60 \mu m}$ from IRAS. $^f$ Flux from out grey-body fit, calculated either at 60 or 65 $\mu$m, depending on F$_{\rm ref}$.

\end{deluxetable}

\clearpage

\begin{deluxetable}{lrrrrrr}
\tablecaption{Other PN parameters}

\tablehead{
\colhead{Name}& \colhead{R$_{\rm G}$}& \colhead{log R$_{\rm PN}$}& \colhead{log N$_{\rm e}$}& 
\colhead{I$_{4686}$}&  
\colhead{Ne ratio} & \colhead{EC}\\
& [kpc]& [cm]& [cm$^{-3}$]& [I$\beta$=100]& & \\
(1)& (2)& (3)& (4)&  (5)& (6)& (7)\\}

\startdata
 PN~G000.3-02.8   &$\dots$   &$\dots$   & 2.511   &$\dots$   &  1.257   &  4.181  \\
 PN~G000.6-01.3   &$\dots$   &$\dots$   & 3.357   &$\dots$   & -1.204   &$\dots$  \\
 PN~G000.6-02.3   &$\dots$   &$\dots$   & 3.280   &$\dots$   &$\dots$   &$\dots$  \\
 PN~G000.8-01.5   &$\dots$   &$\dots$   & 4.166   &$\dots$   &$\dots$   &$\dots$  \\
 PN~G000.8-07.6   &$\dots$   &$\dots$   & 3.304   &$\dots$   &  1.162   &  4.415  \\
 PN~G001.7-01.6   &$\dots$   &$\dots$   & 3.767   &$\dots$   &$\dots$   &$\dots$  \\
 PN~G001.7-04.6   & 4.731   &17.46   & 3.787   &  2.5   &  1.133   &  2.839  \\
 PN~G002.4-03.7   & 1.439   &17.39   & 4.352   &$\dots$   &$\dots$   &$\dots$  \\
 PN~G002.6-03.4   &$\dots$   &$\dots$   & 3.722   &$\dots$   &$\dots$   &$\dots$  \\
 PN~G002.9-03.9   &$\dots$   &$\dots$   &$\dots$   & 35.0   &$\dots$   &  6.260  \\
 PN~G003.0-02.6   &10.880   &17.63   &$\dots$   & 34.0   &$\dots$   &  6.205  \\
 PN~G003.1+03.4   & 2.868   &17.47   & 3.215   &$\dots$   &$\dots$   &$\dots$  \\
 PN~G003.2-04.4   & 8.955   &17.55   &$\dots$   &$\dots$   &  1.516   &  4.127  \\
 PN~G003.9-14.9   & 0.547   &17.39   & 3.722   &$\dots$   &  1.801   &  3.740  \\
 PN~G004.1-03.8   & 7.424   &17.57   &$\dots$   &$\dots$   &  1.097   &  4.095  \\
 PN~G004.3-02.6   & 5.251   &17.71   &$\dots$   &$\dots$   &  0.460   &  2.209  \\
 PN~G004.9-04.9   & 2.521   &17.50   & 3.428   &$\dots$   & -0.548   &  0.112  \\
 PN~G005.5+0.27   &$\dots$   &$\dots$   &$\dots$   &$\dots$   &$\dots$   &  0.814  \\
 PN~G006.0+02.8   &$\dots$   &$\dots$   & 2.707   &$\dots$   &$\dots$   &$\dots$  \\
 PN~G006.3+04.4   & 2.515   &17.47   &$\dots$   &  4.5   &  1.829   &  6.422  \\
 PN~G006.8+02.0   &$\dots$   &$\dots$   &$\dots$   &$\dots$   &  0.292   &  1.696  \\
 PN~G006.8-19.8   &$\dots$   &$\dots$   & 3.206   & 13.4   &$\dots$   &  5.064  \\
 PN~G007.5+04.3   &$\dots$   &$\dots$   &$\dots$   &$\dots$   &$\dots$   &  1.255  \\
 PN~G008.1-04.7   & 4.401   &17.47   & 3.240   &$\dots$   &  0.733   &  2.416  \\
 PN~G008.2-04.8   & 1.967   &17.46   & 3.240   &$\dots$   &  1.516   &  4.181  \\
 PN~G008.6-02.6   &$\dots$   &$\dots$   &$\dots$   &$\dots$   &  1.614   &  4.477  \\
 PN~G008.6-07.0   & 7.938   &17.55   & 2.825   & 15.0   &  1.264   &  5.152  \\
 PN~G009.3+04.1   &$\dots$   &$\dots$   &$\dots$   &$\dots$   &  1.535   &  4.851  \\
 PN~G010.6+03.2   &$\dots$   &$\dots$   & 4.114   &$\dots$   &  0.316   &  1.561  \\
 PN~G011.1+07.0   &$\dots$   &$\dots$   &$\dots$   & 50.0   &$\dots$   &  7.091  \\
 PN~G011.3-09.4   & 1.822   &16.98   & 4.114   &$\dots$   & -1.220   &  0.283  \\
 PN~G011.7-06.6   & 4.816   &17.79   & 4.270   &$\dots$   &$\dots$   &$\dots$  \\
 PN~G012.5-09.8   & 2.959   &17.45   &$\dots$   &  7.8   &$\dots$   &  5.845  \\
 PN~G012.6-02.7   & 3.112   &17.59   & 4.058   &$\dots$   &$\dots$   &$\dots$  \\
 PN~G014.0-05.5   &$\dots$   &$\dots$   &$\dots$   & 29.0   &  1.671   &  5.928  \\
 PN~G014.3-05.5   &11.900   &17.12   & 3.722   &  5.9   &$\dots$   &  1.975  \\
 PN~G016.9-02.0   &$\dots$   &$\dots$   & 3.508   &$\dots$   &  1.477   &  7.537  \\
 PN~G018.6-02.2   & 2.796   &17.56   &$\dots$   & 47.0   &  1.802   &  6.925  \\
 PN~G019.2-02.2   & 3.370   &16.94   & 3.852   &  4.0   &  1.778   &  8.010  \\
 PN~G019.4-05.3   & 3.433   &16.86   & 4.114   &$\dots$   &  0.865   &  4.113  \\
 PN~G019.7+03.2   & 3.208   &16.78   &$\dots$   &$\dots$   &  1.414   &  7.956  \\
 PN~G019.8+05.6   &$\dots$   &$\dots$   & 3.576   & 28.0   &  1.449   &  5.872  \\
 PN~G023.8-01.7   & 4.833   &17.39   &$\dots$   &$\dots$   &$\dots$   &$\dots$  \\
 PN~G023.9+01.2   &$\dots$   &$\dots$   & 3.795   &$\dots$   &$\dots$   &$\dots$  \\
 PN~G025.3-04.6   &$\dots$   &$\dots$   &$\dots$   &$\dots$   &  1.630   &  5.944  \\
 PN~G026.0-01.8   & 7.225   &17.42   &$\dots$   &$\dots$   &  1.912   &  6.934  \\
 PN~G027.6-09.6   & 3.825   &17.13   &$\dots$   &  0.6   &  1.709   &  1.660  \\
 PN~G031.0+04.1   & 4.238   &16.55   &$\dots$   &$\dots$   &  1.575   &  0.337  \\
 PN~G032.5-03.2   &$\dots$   &$\dots$   &$\dots$   &$\dots$   &$\dots$   &  0.175  \\
 PN~G032.9-02.8   & 7.109   &17.05   & 3.619   &$\dots$   &  1.455   &  8.572  \\
 PN~G034.0+02.2   & 4.755   &17.37   & 3.077   &$\dots$   &  1.328   &  6.880  \\
 PN~G038.4-03.3   &$\dots$   &$\dots$   & 3.162   &$\dots$   &$\dots$   &$\dots$  \\
 PN~G038.7-03.3   &$\dots$   &$\dots$   & 3.722   & 11.4   &  1.894   &  1.993  \\
 PN~G041.8+04.4   &$\dots$   &$\dots$   &$\dots$   &$\dots$   &$\dots$   &  0.135  \\
 PN~G042.0+05.4   &21.790   &17.31   & 4.046   &$\dots$   &$\dots$   &$\dots$  \\
 PN~G042.9-06.9   &13.790   &17.05   & 3.428   &  0.3   &  1.531   &  5.184  \\
 PN~G044.1+05.8   &$\dots$   &$\dots$   &$\dots$   &$\dots$   &$\dots$   &  1.953  \\
 PN~G045.9-01.9   &11.190   &17.13   &$\dots$   &$\dots$   & -0.391   &  0.977  \\
 PN~G048.1+01.1   & 6.010   &16.66   & 3.593   &$\dots$   &  1.309   &  9.328  \\
 PN~G048.5+04.2   &13.460   &17.51   &$\dots$   & 66.0   &$\dots$   &  7.978  \\
 PN~G051.0+02.8   &$\dots$   &$\dots$   &$\dots$   &$\dots$   &$\dots$   &  6.412  \\
 PN~G052.9+02.7   & 8.846   &17.72   &$\dots$   &$\dots$   &  1.535   &  6.759  \\
 PN~G052.9-02.7   & 7.962   &17.06   &$\dots$   &$\dots$   &$\dots$   &  8.532  \\
 PN~G055.1-01.8   &20.580   &17.70   &$\dots$   &$\dots$   &  1.879   &  7.371  \\
 PN~G055.5-00.5   & 6.597   &17.09   & 4.114   &$\dots$   &  1.252   &  2.061  \\
 PN~G058.9+01.3   & 8.342   &17.43   &$\dots$   &$\dots$   &  1.167   &  2.389  \\
 PN~G059.4+02.3   &10.420   &17.37   &$\dots$   &$\dots$   &  2.527   &  6.327  \\
 PN~G059.9+02.0   &22.410   &17.11   &$\dots$   &$\dots$   &$\dots$   &$\dots$  \\
 PN~G060.5+01.8   & 9.297   &17.22   &$\dots$   &$\dots$   & -0.055   &  1.660  \\
 PN~G063.8-03.3   &19.380   &17.01   &$\dots$   &$\dots$   &$\dots$   &$\dots$  \\
 PN~G067.9-00.2   & 7.743   &16.50   & 3.787   &$\dots$   &  0.804   &  1.526  \\
 PN~G068.7+01.9   &10.950   &17.41   &$\dots$   &$\dots$   &  1.898   &  5.166  \\
 PN~G068.7+14.8   &$\dots$   &$\dots$   &$\dots$   &  2.4   &$\dots$   &  2.682  \\
 PN~G069.2+02.8   &13.670   &16.33   &$\dots$   &$\dots$   &$\dots$   &  0.153  \\
 PN~G077.7+03.1   &$\dots$   &$\dots$   & 3.473   &$\dots$   &  1.044   & 12.150  \\
 PN~G079.9+06.4   &$\dots$   &$\dots$   &$\dots$   & 77.0   &$\dots$   &  8.587  \\
 PN~G082.5+11.3   &12.530   &16.82   &$\dots$   &$\dots$   &  0.895   &  1.102  \\
 PN~G088.7+04.6   &11.620   &17.45   &$\dots$   &$\dots$   &  1.846   &  2.290  \\
 PN~G095.2+00.7   &10.110   &17.03   &$\dots$   &$\dots$   &  1.158   &  5.648  \\
 PN~G097.6-02.4   &14.730   &17.57   &$\dots$   &  7.0   &  2.041   &  5.935  \\
 PN~G104.1+01.0   &12.060   &16.88   &$\dots$   &$\dots$   &  0.913   &  6.476  \\
 PN~G107.4-00.6   &$\dots$   &$\dots$   &$\dots$   &$\dots$   &$\dots$   &  0.410  \\
 PN~G107.4-02.6   &14.200   &17.62   &$\dots$   & 92.0   &$\dots$   &  9.418  \\
 PN~G112.5-00.1   &21.100   &17.68   & 2.033   &$\dots$   &  0.833   &  1.418  \\
 PN~G184.0-02.1   &15.270   &17.09   &$\dots$   &$\dots$   &  0.012   &$\dots$  \\
 PN~G205.8-26.7   &$\dots$   &$\dots$   &$\dots$   &$\dots$   &$\dots$   &  4.045  \\
 PN~G235.3-03.9   &15.260   &17.10   & 4.689   &$\dots$   &$\dots$   &  0.072  \\
 PN~G263.0-05.5   &12.990   &17.31   & 3.762   &  9.2   &  1.915   &  5.598  \\
 PN~G264.4-12.7   &13.390   &17.33   & 3.949   &$\dots$   &  0.750   &  2.723  \\
 PN~G274.1+02.5   &$\dots$   &$\dots$   &$\dots$   &$\dots$   &$\dots$   &  1.620  \\
 PN~G275.3-04.7   &15.170   &17.33   &$\dots$   & 28.0   &$\dots$   &  5.872  \\
 PN~G278.6-06.7   &12.160   &17.23   & 3.619   &  6.0   &  1.887   &  6.354  \\
 PN~G285.4+01.5   & 8.625   &17.27   & 4.114   &$\dots$   &  0.601   &  5.656  \\
 PN~G285.4+02.2   &$\dots$   &$\dots$   & 2.909   & 78.0   &$\dots$   &  8.642  \\
 PN~G286.0-06.5   &16.060   &17.01   & 3.348   &$\dots$   &  1.391   &  4.419  \\
 PN~G289.8+07.7   &12.660   &17.39   &$\dots$   & 45.0   &$\dots$   &  6.814  \\
 PN~G294.9-04.3   &10.210   &17.11   &$\dots$   &$\dots$   & -0.994   &  0.661  \\
 PN~G295.3-09.3   &$\dots$   &$\dots$   &$\dots$   &$\dots$   &  1.362   &  4.338  \\
 PN~G296.3-03.0   & 8.029   &17.12   & 3.852   & 19.0   &  1.475   &  5.374  \\
 PN~G297.4+03.7   &$\dots$   &$\dots$   & 3.299   &$\dots$   & -1.256   &  0.162  \\
 PN~G300.7-02.0   & 7.140   &17.21   & 4.368   &$\dots$   &  1.002   &  4.815  \\
 PN~G307.5-04.9   & 6.532   &17.50   & 3.787   &  0.5   &  0.034   &  1.323  \\
 PN~G309.0+00.8   &$\dots$   &$\dots$   &$\dots$   &$\dots$   &  0.773   &  4.995  \\
 PN~G309.5-02.9   &$\dots$   &$\dots$   & 3.302   &$\dots$   &  1.563   &  6.934  \\
 PN~G311.1+03.4   &$\dots$   &$\dots$   &$\dots$   &$\dots$   &  0.289   &  0.608  \\
 PN~G321.3+02.8   & 5.232   &17.03   & 4.647   &$\dots$   &  0.226   &  3.213  \\
 PN~G324.2+02.5   & 6.178   &17.37   & 4.053   &$\dots$   &$\dots$   &$\dots$  \\
 PN~G324.8-01.1   & 5.434   &16.98   &$\dots$   &$\dots$   &  1.352   &  5.769  \\
 PN~G325.0+03.2   & 5.935   &17.09   & 3.366   &$\dots$   &  1.543   &  6.530  \\
 PN~G325.8-12.8   & 5.207   &17.49   &$\dots$   &$\dots$   & -0.057   &  0.486  \\
 PN~G326.0-06.5   & 4.562   &17.45   &$\dots$   &$\dots$   &$\dots$   &$\dots$  \\
 PN~G327.1-01.8   & 4.366   &17.29   & 4.114   &$\dots$   & -1.617   &  0.198  \\
 PN~G327.8-06.1   &13.910   &17.48   & 3.677   &$\dots$   &  0.553   &  2.362  \\
 PN~G327.9-04.3   &$\dots$   &$\dots$   & 3.308   &$\dots$   &$\dots$   &  5.062  \\
 PN~G329.4-02.7   & 7.433   &17.47   & 2.782   &$\dots$   &  1.702   &  4.270  \\
 PN~G331.0-02.7   & 4.750   &17.34   & 4.339   &$\dots$   & -0.703   &  0.666  \\
 PN~G334.8-07.4   &$\dots$   &$\dots$   &$\dots$   &$\dots$   & -0.794   &  0.688  \\
 PN~G336.3-05.6   & 4.326   &17.36   & 3.722   & 56.0   &  1.120   &  7.424  \\
 PN~G336.9+08.3   &$\dots$   &$\dots$   &$\dots$   &$\dots$   &$\dots$   &  4.415  \\
 PN~G340.9-04.6   &$\dots$   &$\dots$   &$\dots$   &$\dots$   &  1.902   &  6.322  \\
 PN~G341.5-09.1   & 2.659   &17.50   & 3.206   &$\dots$   &  1.889   &  4.689  \\
 PN~G343.4+11.9   & 8.748   &17.51   &$\dots$   & 19.5   &$\dots$   &  5.401  \\
 PN~G344.2+04.7   &$\dots$   &$\dots$   & 3.619   &$\dots$   & -0.020   &  1.809  \\
 PN~G344.4+02.8   &$\dots$   &$\dots$   &$\dots$   &$\dots$   &$\dots$   &  7.628  \\
 PN~G344.8+03.4   &$\dots$   &$\dots$   & 3.306   &$\dots$   & -0.695   &  0.437  \\
 PN~G345.0+04.3   &$\dots$   &$\dots$   &$\dots$   &$\dots$   &$\dots$   &$\dots$  \\
 PN~G348.4-04.1   &$\dots$   &$\dots$   & 3.522   &$\dots$   &  1.427   &  6.867  \\
 PN~G348.8-09.0   &$\dots$   &$\dots$   &$\dots$   &$\dots$   &  0.753   &$\dots$  \\
 PN~G350.8-02.4   &$\dots$   &$\dots$   & 4.250   &$\dots$   &  0.380   &  1.737  \\
 PN~G351.3+07.6   &$\dots$   &$\dots$   &$\dots$   &$\dots$   &  1.004   &  2.871  \\
 PN~G351.9-01.9   &$\dots$   &$\dots$   & 3.449   &$\dots$   &  1.220   &  5.665  \\
 PN~G352.6+03.0   & 1.464   &17.36   & 3.779   &$\dots$   &  1.098   &  4.838  \\
 PN~G354.2+04.3   & 2.556   &17.49   & 3.533   &$\dots$   &  0.142   &  0.661  \\
 PN~G354.9+03.5   &$\dots$   &$\dots$   & 3.141   &$\dots$   & -0.027   &  1.120  \\
 PN~G355.2-02.5   & 3.563   &17.53   & 3.895   &$\dots$   &  0.963   &  5.252  \\
 PN~G355.7-03.0   & 4.252   &17.41   & 3.531   &$\dots$   &  1.351   &  4.671  \\
 PN~G355.9+02.7   & 4.518   &17.27   & 4.101   &$\dots$   &  1.257   &  6.313  \\
 PN~G356.2-04.4   & 0.540   &17.31   &$\dots$   &  4.5   &  1.677   &  2.439  \\
 PN~G356.5+01.5   & 2.253   &17.56   &$\dots$   &$\dots$   &  0.433   &  1.854  \\
 PN~G356.5-03.6   &$\dots$   &$\dots$   & 3.408   &$\dots$   &  1.173   &  3.672  \\
 PN~G356.8+03.3   &11.350   &17.42   &$\dots$   &$\dots$   & -1.377   &  0.211  \\
 PN~G357.1+01.9   &$\dots$   &$\dots$   & 2.413   &$\dots$   &  1.324   &  2.259  \\
 PN~G357.1-06.1   &$\dots$   &$\dots$   & 2.879   & 74.0   &  0.998   &  8.421  \\
 PN~G357.2+02.0   &$\dots$   &$\dots$   & 3.762   &$\dots$   &  1.536   &  9.364  \\
 PN~G357.6+01.7   & 1.896   &17.32   & 4.060   &$\dots$   &  1.390   &  5.787  \\
 PN~G357.6+02.6   & 5.437   &17.68   &$\dots$   &$\dots$   &  1.359   &  5.126  \\
 PN~G358.2+03.6   & 1.449   &17.35   & 3.583   & 14.5   &  1.675   &  5.124  \\
 PN~G358.3+01.2   &$\dots$   &$\dots$   & 3.839   &$\dots$   &  1.058   & 11.772  \\
 PN~G358.5+02.9   &$\dots$   &$\dots$   &$\dots$   &$\dots$   &$\dots$   &  8.496  \\
 PN~G358.5-04.2   & 0.237   &16.86   & 4.114   &$\dots$   &  0.167   &  2.430  \\
 PN~G358.6+01.8   &$\dots$   &$\dots$   &$\dots$   &$\dots$   &  1.605   & 12.150  \\
 PN~G358.7+05.2   &$\dots$   &$\dots$   &$\dots$   &$\dots$   &$\dots$   &  6.066  \\
 PN~G358.7-02.7   & 3.723   &17.36   & 3.792   &$\dots$   &$\dots$   &$\dots$  \\
 PN~G358.9+03.4   & 4.051   &17.10   & 3.073   &$\dots$   &  0.089   &  2.020  \\
 PN~G358.9-03.7   &$\dots$   &$\dots$   & 3.058   &$\dots$   & -0.375   &  0.445  \\
 PN~G359.3+03.6   &$\dots$   &$\dots$   &$\dots$   &$\dots$   &  1.337   &  7.758  \\
 PN~G359.4+02.3   &$\dots$   &$\dots$   & 4.114   &$\dots$   &$\dots$   &$\dots$  \\

\enddata
\end{deluxetable}

\begin{deluxetable}{lcccccccc}

\tablecaption{Morphological distribution among dust types}

\tablehead{ 
\colhead{dust type} &  \colhead{N}& \colhead{R}& \colhead{E}& \colhead{Symm.}& \colhead{BC}& \colhead{B}& \colhead{P}& \colhead{Asymm.}\\
&& $\%$& $\%$&$\%$&$\%$&$\%$&$\%$&$\%$\\
(1)&(2)&(3)&(4)&(5)&(6)&(7)&(8)&(9)\\}

\startdata

F&  		5& \dots&  80& 80& 20& \dots& \dots& 20\\
CRD& 	24& 13&61&74&4.2& 17&4.2& 26\\
ORD&      17&12& 35.3& 47& \dots& 35.3& 18& 53\\
MCD&      12& \dots& 50& 50& \dots& 25& 25& 50\\ 	
\enddata
\end{deluxetable}

\begin{deluxetable}{lrrr}

\tablecaption{Galactic bulge, disk, and Magellanic Cloud PN dust class distribution}

\tablehead{ 
\colhead{dust type} &  \colhead{Bulge, $\%$}& \colhead{Galactic, $\%$}& \colhead{Magellanic Clouds, $\%$}\\
(1)&(2)&(3)&(4)\\}

\startdata

F&  		4&20&41\\
CRD& 	11&28&52\\
ORD&      38&28&7\\
MCD&       46&24&N.A.\\	
        \enddata
\end{deluxetable}

\end{document}